\documentclass[aps, prx, reprint, superscriptaddress, longbibliography,floatfix]{revtex4-2}

\usepackage{graphicx}% Include figure files
\usepackage{xcolor} % Enable \definecolor
\usepackage{braket} % For ket notation
\usepackage{amsmath}
\usepackage{amssymb} % For filled in triangles
\usepackage{hhline}
\usepackage{siunitx}
\usepackage{csquotes}
\usepackage[version=4]{mhchem}

\DeclareMathOperator{\sinc}{sinc}

\definecolor{NewBlue}{rgb}{0, 0, 0.41}
\definecolor{NewRed}{rgb}{0.6, 0.07, 0.07}
\usepackage[colorlinks,
    linkcolor=NewBlue,
    citecolor=NewBlue,
    urlcolor=NewRed]{hyperref}
\usepackage{orcidlink}

\newcommand{\EqualContrib}{These authors contributed equally to this work}
\begin{document}
\title{Control of solid-state nuclear spin qubits using an electron spin-1/2}

\author{Hans~K.~C.~\surname{Beukers}\orcidlink{0000-0001-9934-1099}}
\thanks{\EqualContrib}
\author{Christopher~\surname{Waas}\orcidlink{0009-0008-1878-2051}}
\thanks{\EqualContrib}
\author{Matteo~\surname{Pasini}\orcidlink{0009-0005-1358-7896}}
\author{Hendrik~B.~\surname{van~Ommen}\orcidlink{0009-0004-1219-9413}}
\author{Zarije~\surname{Ademi}\orcidlink{0009-0005-5093-6784}}
\author{Mariagrazia~\surname{Iuliano}\orcidlink{0009-0003-4859-0521}}
\author{Nina~\surname{Codreanu}\orcidlink{0009-0006-6646-8396}}
\author{Julia~M.~\surname{Brevoord}\orcidlink{0000-0002-8801-9616}}
\author{Tim~\surname{Turan}\orcidlink{0009-0003-9908-7985}}
\author{Tim~H.~\surname{Taminiau}\orcidlink{0000-0002-2355-727X}}
\author{Ronald~\surname{Hanson}\orcidlink{0000-0001-8938-2137}}
\email{R.Hanson@TUDelft.nl}
 
\affiliation{QuTech and Kavli Institute of Nanoscience, Delft University of Technology, P.O. Box 5046, 2600 GA Delft, The Netherlands}

\begin{abstract}
Solid-state quantum registers consisting of optically active electron spins with nearby nuclear spins are promising building blocks for future quantum technologies. For electron spin-1 registers, dynamical decoupling (DD) quantum gates have been developed that enable the precise control of multiple nuclear spin qubits. However, for the important class of electron spin-1/2 systems, this control method suffers from intrinsic selectivity limitations, resulting in reduced nuclear spin gate fidelities. Here we demonstrate improved control of single nuclear spins by an electron spin-1/2 using Dynamically Decoupled Radio Frequency (DDRF) gates. We make use of the electron spin-1/2 of a diamond tin-vacancy center, showing high-fidelity single-qubit gates, single-shot readout, and spin coherence beyond a millisecond. The DD control is used as a benchmark to observe and control a single \ce{^13C} nuclear spin. Using the DDRF control method, we demonstrate improved control on that spin. In addition, we find and control an additional nuclear spin that is insensitive to the DD control method. Using these DDRF gates, we show entanglement between the electron and the nuclear spin with \qty{72(3)}{\percent} state fidelity. Our extensive simulations indicate that DDRF gate fidelities well in excess are feasible. Finally, we employ time-resolved photon detection during readout to quantify the hyperfine coupling for the electron's optically excited state. Our work provides key insights into the challenges and opportunities for nuclear spin control in electron spin-1/2 systems, opening the door to multi-qubit experiments on these promising qubit platforms.
\end{abstract}

\maketitle

\begin{figure}
    \centering
    \includegraphics{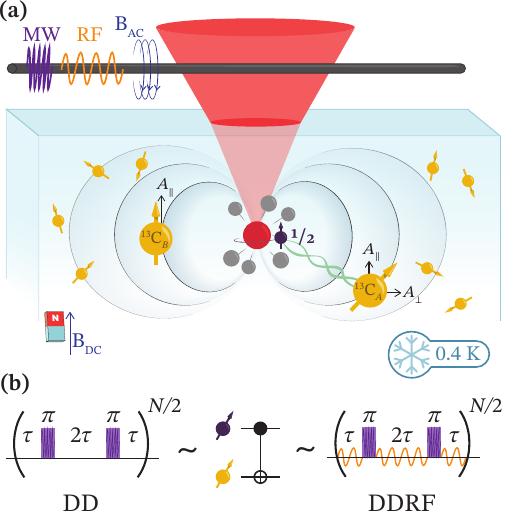}
    \caption{\textbf{Nuclear spin control with electron spin-1/2.}
    (a) The electron spin-1/2 of a negatively charged tin-vacancy center in a $\langle100\rangle$ surface-oriented diamond (purple spin) is initialized and read out by a red laser. The electron spin is controlled using the AC magnetic field of microwave (MW) radiation (purple sine) through a wire (gray line) spanned over the diamond. The surrounding nuclear spins (yellow spins) have a unique parallel and perpendicular hyperfine coupling ($A_\parallel$ and $A_\perp$) to the electron spin. This allows conditional control with the electron spin and generation of entanglement (green link). The nuclear spins can be directly driven using radiofrequency (RF) radiation (yellow sine).
    (b) Nuclear spin control is achieved by dynamical decoupling (DD) and dynamically decoupled radio frequency (DDRF) gates. For DD gates, interpulse delays $\tau$ resonant with the nuclear spin dynamics cause a rotation conditioned on the electron spin state. For DDRF gates, the interpulse delays do not need to follow the dynamics of the target nuclear spin: direct spin-state selective radio frequency driving with tailored phase updating enables a conditional rotation of the nuclear spin. Both techniques perform $\mathrm{CNOT}$-equivalent gates.}
    \label{fig:idea}
\end{figure} 

\section{Introduction} 
Optically interfaced electron spins in the solid state are promising platforms for quantum networking, computing, and sensing \cite{awschalom2018quantum}. Prominent examples are color centers \cite{maze2008nanoscale, abobeih2019atomicscale, stolk2024metropolitanscale, lukin2020integrated, ruf2021quantum, knaut2024entanglement} and single rare-earth ions \cite{dibos2018rei_telecom, ruskuc2022nuclear, uysal2023rei_nuclear, ruskuc2024scalable}. These electron spins offer fast control \cite{fuchs2009gigahertz, christle2015isolated, sukachev2017siliconvacancy} and high-fidelity readout \cite{robledo2011highfidelity, kindem2020control, bhaskar2020communication}. Moreover, their spin-photon interface enables remote entanglement generation \cite{beukers2024remoteentanglement}, while their solid-state nature facilitates on-chip integration \cite{wan2020largescale}. The use of surrounding nuclear spins as long-lived memory qubits further enhances the functionality of the electron spins. For example, these electron-nuclear registers have enabled recent demonstrations of fault-tolerant quantum computing \cite{abobeih2022faulttolerant}, memory-assisted quantum communication \cite{bhaskar2020communication} and a multi-node quantum network \cite{pompili2021multinode, hermans2023entangling}.

A key requirement for establishing coherent quantum gates between the electron and nuclear spins is the protection of the electron spin coherence during the gate~\cite{vandersar2012decoherenceprotected}. For this, different control methods have been developed, such as dynamical decoupling (DD) control \cite{taminiau2012detection} and dynamically decoupled radio frequency (DDRF) control \cite{bradley2019tenqubit}. Nuclear spin control using the DD method has been shown in different platforms including color centers in diamond \cite{cramer2016error, maity2022nuclear, nguyen2019prl}, silicon carbide \cite{babin2022fabrication} and single rare-earth ions \cite{uysal2023rei_nuclear}. In these decoherence-protected control methods, the selectivity of the control critically depends on the electron spin magnitude. Notably, the important class of electron spin-1/2 systems, which includes the silicon T-centers \cite{higginbottom2022optical, photonicinc2024distributed} and the diamond group-IV color centers \cite{nguyen2019prb, senkalla2024germanium, rosenthal2023microwave, wang2024transformlimited}, has an intrinsically reduced selectivity for the DD control method compared to spin systems with a higher magnitude \cite{zahedian2024blueprint, maity2018strain, nguyen2019prl, uysal2023rei_nuclear}. It has been hypothesized that the DDRF control method improves the selectivity for these interesting electron spin-1/2 systems \cite{bradley2019tenqubit}. 

In this work, we experimentally explore and investigate the control of two nuclear spins with the electron spin-1/2 of the negatively charged tin-vacancy (SnV) center in diamond (Fig.~\ref{fig:idea}(a)), using both the DD and DDRF control (Fig.~\ref{fig:idea}(b)). The SnV center has recently emerged as a highly promising quantum system because of its excellent optical and spin properties, compatibility with nanophotonic integration and operating temperature of above one Kelvin \cite{ditaliatchernij2017snv, iwasaki2017tinvacancy, trusheim2020transformlimited, gorlitz2020spectroscopic, rugar2020waveguide,  wan2020largescale, rugar2021cavity, debroux2021quantum, guo2023microwave, pasini2024nonlinear}. We provide a detailed selectivity comparison of spin-1/2 and spin-1 systems in simulation. Furthermore, we investigate the nuclear spins' coupling to the SnV center's excited state during the optical readout of the electron spin.

\begin{figure*}
    \centering
    \includegraphics{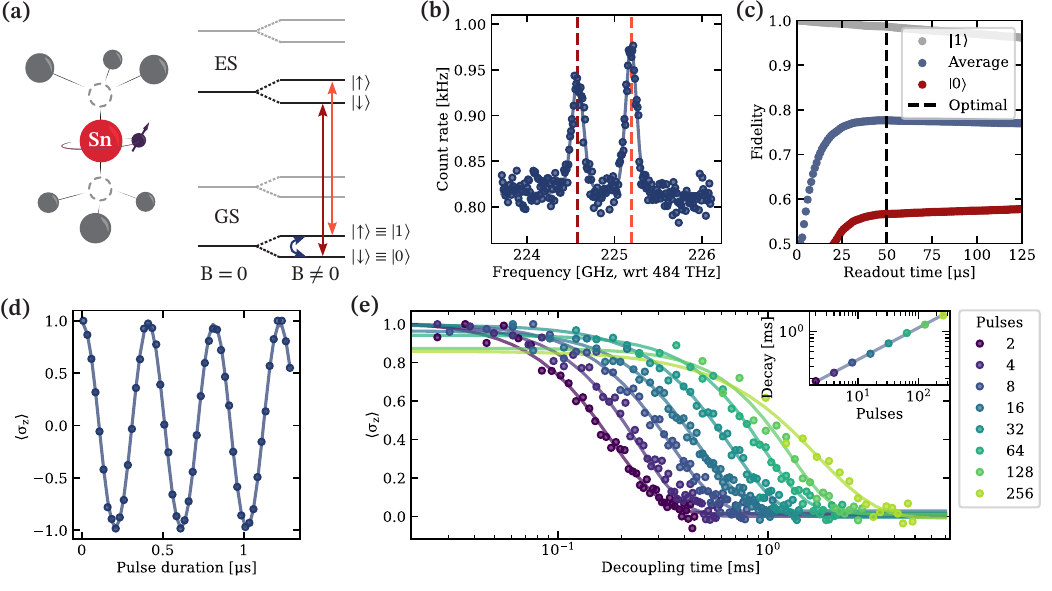}
    \caption{ \textbf{Electronic qubit characterization.}
    (a) Schematic lattice and level structure of the SnV center. The negatively charged SnV center consists of an interstitial tin atom with two carbon vacancies and an additional electron. The ground and excited state behave as an effective spin-1/2 system at low temperatures. The lowest two levels of the ground state form a spin-1/2 qubit and are connected to the excited state via an optical transition at \qty{619}{\nano\meter}.
    (b) Photoluminescence excitation measurement of the two spin-conserving transitions in (a).
    (c) Single-shot readout calibration curve. The splitting of the optical transitions allows for spin-selective readout via optical excitation. The qubit is initialized into an eigenstate. A single-shot readout result of $0 (1)$ is assigned when detecting at least one photon (no photon) during the excitation of the spin-down transition. The best average fidelity determines the optimal readout time.
    (d) Microwave control of the SnV center electron spin qubit. The fit of this Rabi oscillation is used to determine the duration of $\pi/2$ and $\pi$ pulses.
    (e) Electron coherence measurement with dynamical decoupling. We employ XY8 sequences with a varying amount of $\pi$ pulses. Increasing the number of decoupling pulses beyond 100 creates heating in this device which shows up as a lowered contrast.}
    \label{fig:electron}
\end{figure*} 

\section{Electron spin control}

Our experiments are performed on a chemical vapor deposition grown diamond that is implanted with \qty{5e10}{ions\per\centi\meter\squared} of \ce{^120Sn} at a target depth of \qty{88}{\nano\meter} and subsequently annealed at \qty{1100}{\degreeCelsius} \cite{pasini2024nonlinear}. A wire is spanned over the diamond sample to deliver the microwave and radio frequency signals. A \ce{^3He}-cryostat with a confocal optical microscope cools the sample to \qty{0.4}{\kelvin}. We align a bias magnetic field of \qty{0.1}{\tesla} with the symmetry axis of the SnV center, see Fig.~\ref{fig:idea}(a).

We have used the negatively charged tin-vacancy (SnV) center throughout this work. The electronic waveform of the SnV center is a spin and orbital doublet in the ground and excited state, as depicted in Fig.~\ref{fig:electron}(a). The spin-orbit interaction and lattice strain lift this degeneracy. At temperatures below \qty{1.5}{\kelvin} thermal occupation of the higher levels is negligible, resulting in an effective spin-1/2 system. The other levels are still observable in effects like an anisotropic g-factor, reduced microwave driving efficiency, and an optical cyclicity that depends on the magnetic field alignment \cite{rosenthal2023microwave}. We implanted the spinless \ce{^120Sn} isotope, followed by an annealing step to activate the SnV center. 

The electron spin qubit is initialized and read out optically with a \qty{619}{\nano\meter} laser. In Fig.~\ref{fig:electron}(b), a photoluminescent excitation (PLE) measurement shows the two spin-conserving transitions split by \qty{611(3)}{\mega\hertz} due to a difference in the $g$-factors of ground and excited state. The SnV center is initialized in the correct charge state and the optical lines are ensured to be on resonance with the laser by a charge-resonance check \cite{brevoord2024heralded}. The single-shot readout is implemented by spin-selective optical excitation of the spin-down transition \cite{rosenthal2024singleshot}. The bright $\ket{0}$ state is assigned in case at least one photon is recorded; otherwise the $\ket{1}$ state is assigned. A finite spin-flipping probability during optical cycling causes readout infidelity of the bright $\ket{0}$ state, whereas noise counts limit the readout fidelity of the dark $\ket{1}$ state. These two contributions are optimized in Fig.~\ref{fig:electron}(c), resulting in an average readout fidelity of \qty{77.7(3)}{\percent}. Note that single-shot readout can be achieved in this setup despite a low collection efficiency of $\approx \qty{0.2}{\percent}$, thanks to the high cyclicity of $\approx \num{1200}$ of this SnV center.

Initialization of the qubit is achieved by spin pumping, where a laser, on resonance with the spin-up transition, excites the electron. The finite spin-flipping probability during optical decay causes initialization into the spin-down state. The initialization fidelity of \qty{98.1(5)}{\percent} is deduced from the residual fluorescence at the end of spin pumping for \qty{300}{\micro\second} and can be readily improved by implementing a longer spin pumping time. 

The electron spin is controlled by microwave driving with a Rabi frequency of \qty{2.46}{\mega\hertz}, as shown in Fig.~\ref{fig:electron}(d). The average fidelity of the calibrated gates was measured using process tomography to be \qty{98(2)}{\percent}. The investigated SnV center has a spin dephasing time of $T_2^*=\qty{2.42(4)}{\micro\second}$, see Appendix~\ref{app:electron_ramsey}. The coherence is extended by XY8 dynamical decoupling of the electron spin to $T_2^\text{DD} = \qty{1.7(5)}{\milli\second}$ using \num{256} $\pi$ pulses \cite{delange2010universal}. The observed scaling of the coherence during DD in Fig.~\ref{fig:electron}(e) matches
\begin{equation}
    T_2^\text{DD} = T_2^\text{Echo} N ^ \chi,
\end{equation}
where $N$ is the number of decoupling pulses, $T_2^\text{Echo}=\qty{129(2)}{\micro\second}$ is the fitted constant equivalent to the single echo coherence, and $\chi=\num{0.47(1)}$ is the scaling factor.

To investigate the limiting effect for the $T_2^\text{Echo}$, we perform a double electron-electron resonance (DEER) measurement, see Appendix~\ref{app:deer}. In the DEER, we simultaneously apply an echo pulse on the SnV center electron spin and at the MW frequency of free electron spins with a $g$-factor of \num{2}. This DEER shows a reduced coherence time for the SnV electron spin, which points to the presence of an electron spin bath as a noise source in this device. Therefore, eliminating this bath in fabrication would improve the coherence of the electron spin to that achieved in recent reports \cite{karapatzakis2024microwave}.

\begin{figure*}
    \centering
    \includegraphics{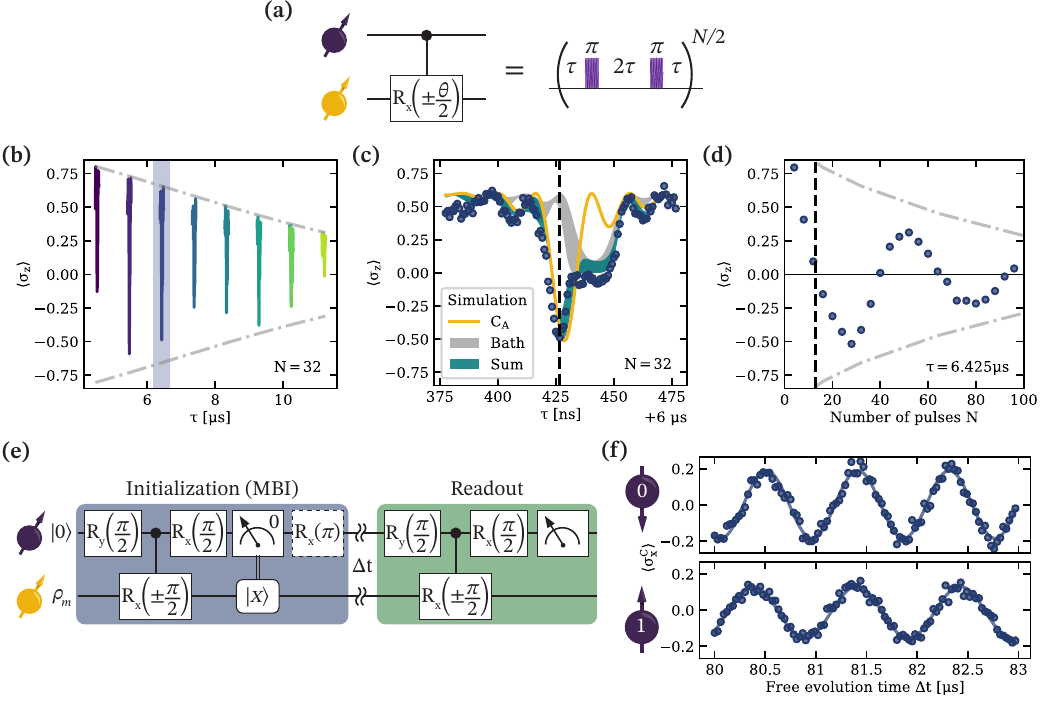}
    \caption{\textbf{Dynamical decoupling nuclear spin control.}
    (a) A dynamical decoupling sequence performs a conditional rotation if $\tau$ fulfills the resonance condition of the nuclear spin.
    (b) Dynamical decoupling sequence for varying interpulse delay $\tau$. The dips in the coherence indicate coupling to the nuclear spins around. The dashed line indicates the contrast limit due to the electron coherence.
    (c) Zoom into the shaded region of $\tau$ in (b). The gray area shows the effect of the spin bath on the electron spin. The spread stems from simulating many different configurations of the spin bath. The narrow dip stems from an individual nuclear spin $C_A$, where the yellow line is the simulated response. The blue area is the simulated combined effect of the bath and the nuclear spin $C_A$. Appendix~\ref{app:dd_simulation} describes the simulations.
    (d) Nuclear gate calibration. By fixing $\tau=\qty{6.425}{\micro\second}$ (dashed line in (b)) and varying the number of decoupling pulses $N$, coherent control of the nuclear spin can be observed by the oscillation of the electron coherence. The fully entangling gate is achieved when $\langle \sigma_z \rangle = 0$. The gray dashed line indicates the contrast limit due to the electron coherence.
    (e) Pulse sequence for nuclear Ramsey measurement. The nuclear spin is prepared via a measurement-based initialization (MBI). The controlled gate is achieved by the DD gate characterized in (c) and (d). A measurement of $\ket{0}$ of the electron at the end of the sequence projects the nuclear spin into $\ket{x}$. An optional $\pi$ pulse allows for measuring the evolution of the nuclear spin for both electron spin states. After a free evolution time $\Delta t$, the nuclear spin is measured in the $x$-basis through the electron spin.
    (f) Ramsey signal of the nuclear spin-dependent on the electron spin state. The precession frequency of the nuclear spin changes depending on whether the electron is kept in $\ket{0}$ or flipped to $\ket{1}$ after the MBI. The measured frequencies are $\omega_0/2\pi=\qty{1116.1(2)}{\kilo\hertz}$ and $\omega_1/2\pi=\qty{985.4(3)}{\kilo\hertz}$.}
    \label{fig:dd}
\end{figure*} 
\section{Dynamical Decoupling control}

The hyperfine coupling between the electron spin of the SnV center and the surrounding \ce{^13C}-spins is caused by the dipole-dipole interaction and the Fermi contact hyperfine interaction \cite{nizovtsev2018nonflipping}. Therefore, the coupling is dependent on their relative distance and orientation. A single nuclear spin can be controlled by targeting its unique hyperfine coupling. Strongly coupled \ce{^13C} can be observed directly in optically detected magnetic resonance (ODMR) of the electron spin \cite{neumann2008multipartite, grimm2024coherent}. However, more weakly coupled spins, where the coupling is smaller than $1/T_2^*$ of the electron spin, are not resolvable by ODMR. To observe and control them, one needs to extend the coherence of the electron spin at the same time. Dynamical decoupling of the electron spin brings the detection limit down to the inverse of the coherence time of the electron spin under dynamical decoupling $1/T_2^\text{DD}$ \cite{taminiau2012detection, kolkowitz2012sensing, zhao2012sensing}.

The Hamiltonian of a single \ce{^13C} spin, in the secular approximation, is
\begin{equation}
    H_\text{hf} = \omega_L I_z + A_\parallel S_z I_z + A_\perp S_z I_x,
\end{equation}
where $S_i$ and $I_i$ are the spin operators for the electron and nucleus, respectively, and $\omega_L = \gamma_c B_z$ is the Larmor frequency of the \ce{^13C}-spin with $\gamma_c/2\pi = \qty{10.71}{\mega\hertz\per\tesla}$. $A_\parallel$ and $A_\perp$ are the parallel and perpendicular hyperfine parameters with respect to the external magnetic field, depicted in Fig.~\ref{fig:idea}(a). As a result of this Hamiltonian, the precession axis of the nuclear spin depends on the state of the electron as 
\begin{equation}
    \vec{\omega}_i = (s_i A_\perp, 0, \omega_L + s_i A_\parallel),
    \label{eq:omega}
\end{equation} 
where $s_i$ is the spin projection of the electron spin for qubit state $\ket{i}$ \cite{zahedian2024blueprint}. 

Nuclear spins can be detected and controlled using the dynamical decoupling sequence of the electron. The nuclear spin precesses around a different axis depending on the electron spin state. Periodically changing between these two rotation axes can give rise to a conditional rotation of the nuclear spin \cite{taminiau2012detection, kolkowitz2012sensing, zhao2012sensing}, as depicted in Fig.~\ref{fig:dd}(a). To control a nuclear spin with DD, the $\pi$ pulses on the electron spin need to be applied in resonance with the dynamics of the nuclear spin. In the high magnetic field regime ($\omega_L \gg A_\parallel, A_\perp$), the resonant condition is met when the $\tau$ of the decoupling sequence is
\begin{equation}
    \tau_p \approx \frac{(2p+1)\pi}{2\bar{\omega}},
\end{equation}
where $p$ is the order of the resonance and $\bar{\omega} = (\omega_0 + \omega_1)/2$ is the average nuclear precession frequency. To target a specific nuclear spin, a unique $\tau_p$ is required to avoid cross-talk with other spins. Therefore, the selectivity of the DD control method depends on a difference in the average nuclear precession frequency $\bar\omega$. Expanding it in hyperfine parameters gives \cite{zahedian2024blueprint}
\begin{equation} \label{eq:omega_average}
    \bar{\omega} = \omega_L \left[1 + \frac{s_0+s_1}{2}\frac{A_\parallel}{\omega_L}+\frac{s_0^2+s_1^2}{4}\left(\frac{A_\perp}{\omega_L}\right)^2\right].
\end{equation}
The first-order term drops out of the expression if $s_0$ and $s_1$ have opposite signs and the same magnitude. In a spin-1/2 system, this is unavoidable, resulting in a second-order selectivity in $A_\perp$. In contrast, in systems with a larger spin magnitude, one can choose the spin projections to achieve a first-order sensitivity in $A_\parallel$.

Low magnetic fields will give a stronger influence of $A_\perp/\omega_L$, which benefits the nuclear spin selectivity in spin-1/2 systems. However, this usually results in a trade-off, as the magnetic field also influences other properties. For the SnV center, for example, a lower field comes at the cost of less separation between the optical transitions and, therefore, reduced optical initialization and readout fidelity for the electron spin qubit. 

In Fig.~\ref{fig:dd}(b), we first use the DD method to detect the nuclear spins around the SnV center by sweeping $\tau$ for 32 decoupling pulses. As this conditional interaction acts like $S_z I_x$, the electron spin can be used as a sensor by applying the conditional gate while the electron is prepared in $\ket{x}= (\ket{0}+\ket{1})/\sqrt{2}$ and read out in the $x$-basis. A conditional gate entangles the electron spin with the nuclear spin, which shows up as a loss of coherence of the electron spin in Fig.~\ref{fig:dd}(b). A drop in coherence to $\braket{\sigma_z} = \num{0}$ can be caused by coupling to many different spins. A value below that shows a coherent interaction with a single spin system \cite{taminiau2012detection}. The nuclear spin bath, composed of spins that have small coupling and cannot be resolved individually, shows up at $\tau \approx \frac{2p+1}{4} \tau_L$, where $\tau_L=2\pi/\omega_L$ is the Larmor period. In Fig.~\ref{fig:dd}(c), the dip in coherence around $\tau=\qty{6.45}{\micro\second}$ is shown. Next to the nuclear spin bath, we observe one dip separate from the bath, which goes well below $\braket{\sigma_z} = \num{0}$, indicating coherent coupling. We calibrate a two-qubit gate with this spin based on the DD spectrum by using the $\tau$ of the center of the dip.  The number of pulses in the DD sequence is swept in Fig.~\ref{fig:dd}(d), where $\braket{\sigma_z} = \num{0}$ indicates the number of pulses required for a maximally entangling gate (dashed line). The decay of the signal is caused by a combination of loss of coherence of the electron spin and residual cross-talk with the spin bath.

The control over the nuclear spin allows us to perform a Ramsey experiment using the sequence in Fig.~\ref{fig:dd}(e), where the nuclear spin is prepared in $\ket{x}$ and read out in the $x$-basis. The state of the nuclear spin can be read out by entangling it with the electron spin and reading out the electron spin. We initialize the nuclear spin using measurement-based initialization (MBI) \cite{abobeih2021thesis}, where measuring the electron spin to be $\ket{0}$  prepares $\ket{x}$ on the nuclear spin. A Ramsey measurement is performed in Fig.~\ref{fig:dd}(f) by varying the time between the MBI and measurement. After the MBI, the electron spin is in $\ket{0}$, and the Ramsey shows $\omega_0/2\pi=\qty{1116.1(2)}{\kilo\hertz}$. Adding a $\pi$ pulse after the MBI allows the measurement of $\omega_1/2\pi=\qty{ 985.4(3)}{\kilo\hertz}$. 

In the system of the electron spin of the SnV center and the \ce{^13C} nuclear spin, there is no direct measure of the magnetic field. The electronic $g$-factor depends on the unknown strain in the lattice. For the nuclear spin, the unknown hyperfine coupling always influences the precession frequency of the \ce{^13C}-spin, as there is no spin-0 projection of the electron. Therefore, we extract the Larmor frequency of the \ce{^13C}-spins from a fit to Fig.~\ref{fig:dd}(d), yielding $\omega_L / 2 \pi = \qty{1048.52(8)}{\kilo\hertz}$. From this we extract $A_{\parallel} / 2 \pi = \qty{-130.9(3)}{\kilo\hertz}$ and $A_{\perp} / 2 \pi = \qty{137(6)}{\kilo\hertz}$ for this \ce{^13C}-spin $C_A$. We have corrected for small changes in magnetic field strength between different measurements, see Appendix~\ref{app:magnetic_field}.  We can observe and control this nuclear spin using the DD control method because it has a significantly large $A_\perp$ that separates its resonance enough from the nuclear spin bath, whereas the resonances of other nuclear spins overlap with the nuclear spin bath. 

\begin{figure*}
    \centering
    \includegraphics{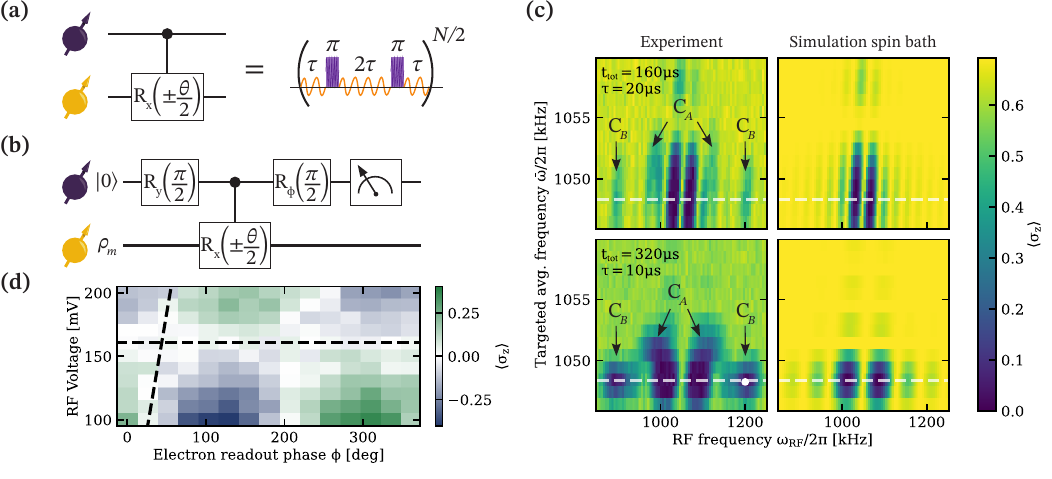}
    \caption{\textbf{Dynamical decoupled radio frequency (DDRF) control.}
    (a) A DDRF gate is realized by direct RF driving of the nuclear spin during the interpulse delay of the DD sequence. 
    (b) The sequence to calibrate the gate. The angle $\theta$ is calibrated with the applied RF power. The angle $\phi$ of the second $\pi/2$ gate is calibrated to counteract a phase that the electron picks up due to the RF driving.
    (c) DDRF spectrum. For each data point in the left panels, the sequence in (b) is measured for multiple electron readout $\pi/2$ pulse phases $\phi$ and fitted to a sine function to account for phases picked up during the RF pulses. The amplitude of the sine is a measure for $\langle \sigma_z \rangle$. The values for $\tau$ are chosen such that for both values of $N$, the expected electron coherence at the end of the sequence is similar. The right panels show a simulation of the Hamiltonian for a collection of weakly coupled spins to imitate the nuclear spin bath, which is described in Appendix~\ref{app:ddrf_simulation}. Additionally to the characteristic bath features, two coupled spins (indicated by the arrows) are observable. The white dashed line indicates $\omega_L$. The solid white circle is the used operating point in (d).
    (d) Nuclear spin gate calibration. The controlled gate can be calibrated by varying the electron readout $\pi/2$ pulse phase and the RF voltage. A fit of the full data with a heuristic model determines the correct voltage for the controlled gate as well as the additional phase picked up by the electron during the gate (dashed lines).
    }
    \label{fig:ddrf_spectrum}
\end{figure*} 
\section{Dynamically Decoupled Radio Frequency control}

Dynamical Decoupled Radio Frequency (DDRF) control has been suggested to yield a better selectivity for a spin-1/2 system than DD control \cite{bradley2019tenqubit}. In DDRF, a direct drive of the nuclear spin is combined with coherence protection of the electron spin using dynamical decoupling. The most direct way of driving the nuclear spin is with radio frequency (RF) radiation. The required RF driving frequency $\omega_i$ depends on the electron spin state $\ket{i}$ as described by Eq.~\ref{eq:omega}. The electron spin coherence must be extended with DD while driving the nuclear spin. To make it work in practice, we need to apply the RF driving pulses in phase with the nuclear spin evolution. Van Ommen et al. \cite{vanommen2024ddrf} extend the original work on DDRF \cite{bradley2019tenqubit} with a more precise and generalized analysis that includes the effects of the bandwidths of the RF pulses.

To achieve the desired nuclear spin evolution the phase of the RF pulses is updated after each $\pi$ pulse on the electron spin with
\begin{equation}
    \delta\phi = 2\tau\bar{\omega} + \pi,
\end{equation}
where $\bar{\omega}=(\omega_0+\omega_1)/2$ is the average nuclear spin precession frequency. The $2\tau\bar{\omega}$ is the phase update required to follow the spin evolution such that the RF drive induces rotations along a fixed axis in the rotation frame of the nuclear spin. The $\pi$ phase makes the rotation conditional as it inverts the rotation axis for the opposite electron spin state. As the phase update rule depends on $\bar\omega$, it has a similar selectivity as the DD control. However, the direct driving of the spin RF frequency, typical of DDRF control, increases its selectivity.

The Fourier transform of the pulse's temporal shape determines the frequency response of the RF drive \cite{vanommen2024ddrf}. For a square pulse of length $2\tau$ it is
\begin{equation}
    \Omega(\Delta, \tau) \propto \sinc(\Delta \tau),
\end{equation}
where $\Delta=\omega-\omega_\text{RF}$ is the detuning from the RF driving frequency, and $\sinc(x) = \sin(x)/x$. For a typical $\tau$ used in this work, around \qty{5}{\micro\second}, this sets an upper bound for the selectivity of the RF drive of around \qty{100}{\kilo\hertz}. In our experiments, the power broadening of the nuclear spin transition can be neglected as it is around \qty{2}{\kilo\hertz}.
% Distance between zeros of sinc(x*w): width_w = 2\pi / x = 2 pi / tau --> width_f = 1 / tau

In previous work \cite{bradley2019tenqubit}, only a single driving frequency was used, effectively reducing the nuclear spins' driving to half the time. Here, we drive the nuclear spin at both $\omega_0$ and $\omega_1$, making the driving more efficient and reducing heating. Simulations of our experiments show that a double drive with a driving strength of $\Omega/2$ for both transitions is equivalent to a single drive with $\Omega$. For simplicity and consistency, we will describe the experiments and theory in this work as if a single drive was used.

As the DDRF control method also gives an effective $S_z I_x$ interaction, like the DD method, similar detection and control circuit diagrams can be used \cite{taminiau2012detection}. The DDRF control method now performs the conditional gate, as depicted in Fig.~\ref{fig:ddrf_spectrum}(a). The DDRF gate depends both on the used RF frequency $\omega_\text{RF}$, and the targeted average precession frequency $\bar\omega$ through the phase update rule, resulting in a two-dimensional DDRF spectrum. This spectrum is achieved by the gates in Fig.~\ref{fig:ddrf_spectrum}(b).

In Fig.~\ref{fig:ddrf_spectrum}(c), we show the resulting spectra for two different decoupling sequences, both using a Rabi frequency of \qty{1.64}{\kilo\hertz}. We compare the spectra to simulations of the nuclear spin bath. In this way, we can see which resonances are caused by the bath and which ones show the presence of controllable single nuclear spins. Our measurements show the known nuclear spin $C_A$, as indicated in Fig.~\ref{fig:ddrf_spectrum}(c). However, there are also resonances belonging to another nuclear spin $C_B$. This nuclear spin has an average precession frequency close to $\omega_L$ (white dashed line), indicating a small $A_\perp$. The splitting between the dips of $\omega_0$ and $\omega_1$ for $C_B$ indicates $A_\parallel/2\pi \approx \qty{300}{\kilo\hertz}$. This nuclear spin was not visible in the DD measurements as that method requires a high $A_\perp$ to make it stand out from the bath.

In the spectra, the width of the resonances in targeted $\bar\omega$ (vertical) is inversely proportional to the total gate duration as this is the time over which a difference in $\bar\omega$ can build up a phase difference. The width of the resonances in $\omega_\text{RF}$ (horizontal) is inversely proportional to the $\tau$ used, as this corresponds to the frequency width of the RF pulse \cite{vanommen2024ddrf}. 

Nuclear spins have different combinations of $A_\parallel$ and $A_\perp$, resulting in various combinations of $\bar{\omega}$ and $\omega_\text{RF}$ for the resonances. This means that the two parameters need to be searched independently to find the different nuclear spins. This is different in a spin-1 system, such as the nitrogen-vacancy center in diamond, when using $s_0=0$ and $s_1=\pm1$, where the hyperfine parameters only influence $\omega_1$, with $\omega_0$ being constant. For a spin-1, the search space can be reduced to a one-dimensional slice as $\omega_0$ is known in advance.

The high symmetry of the DD sequence, in combination with the symmetric spin projections of a spin-1/2 system, causes the loss of first-order selectivity in the phase update rule. To circumvent this, we explored the use of the less symmetric Uhrig Dynamical Decoupling (UDD) sequence \cite{uhrig2007udd} for the DDRF gate, which is discussed in Appendix~\ref{app:udd}. The UDD gives differently shaped resonances, which, in our case, causes separation of the nuclear spin $C_A$ from the bath. However, in our system, UDD decoupling is inferior to the XY8 decoupling sequence \cite{delange2010universal}, resulting in lower electron coherence, and was therefore not further pursued. 

The dips in the coherence of the electron spin show which combination of $\omega_\text{RF}$ and $\bar{\omega}$ can be used to control the nuclear spin. We use the right dip of $C_B$ (white dot) to control it. At this operating point, the RF amplitude is swept in the sequence of Fig.~\ref{fig:ddrf_spectrum}(b) to calibrate the DDRF gate. A fully entangling gate is created by setting the RF amplitude to the point where $\braket{\sigma_z}=0$ (Fig.~\ref{fig:ddrf_spectrum}(d)). The phase of the electron spin is also calibrated as the magnetic part of the RF radiation adds an extra phase to the electron spin, which needs to be compensated. We measure the nuclear spin precession frequencies using a nuclear magnetic resonance (NMR, see Appendix~\ref{app:nmr}) experiment extracting $\omega_0/2\pi = \qty{896.02(3)}{\kilo\hertz}$ and $\omega_1/2\pi = \qty{1200.49(3)}{\kilo\hertz}$ for \ce{^13C}-spin $C_B$, which translates to $A_{\parallel} / 2 \pi = \qty{304.45(5)}{\kilo\hertz}$ and $A_{\perp} / 2 \pi = \qty{0(13)}{\kilo\hertz}$. For $C_A$, we observe an improved contrast compared to the DD method, indicating that we can better separate this spin from the bath. For the best gate settings, we find a contrast of $\qty{0.23}{}$ ($\qty{0.23}{}$) for a Ramsey measurement with the electron spin in state $\ket{0}$ ($\ket{1}$) compared to the contrast of $\qty{0.19}{}$ ($\qty{0.14}{}$) via DD control.

\begin{figure*}
    \centering
    \includegraphics{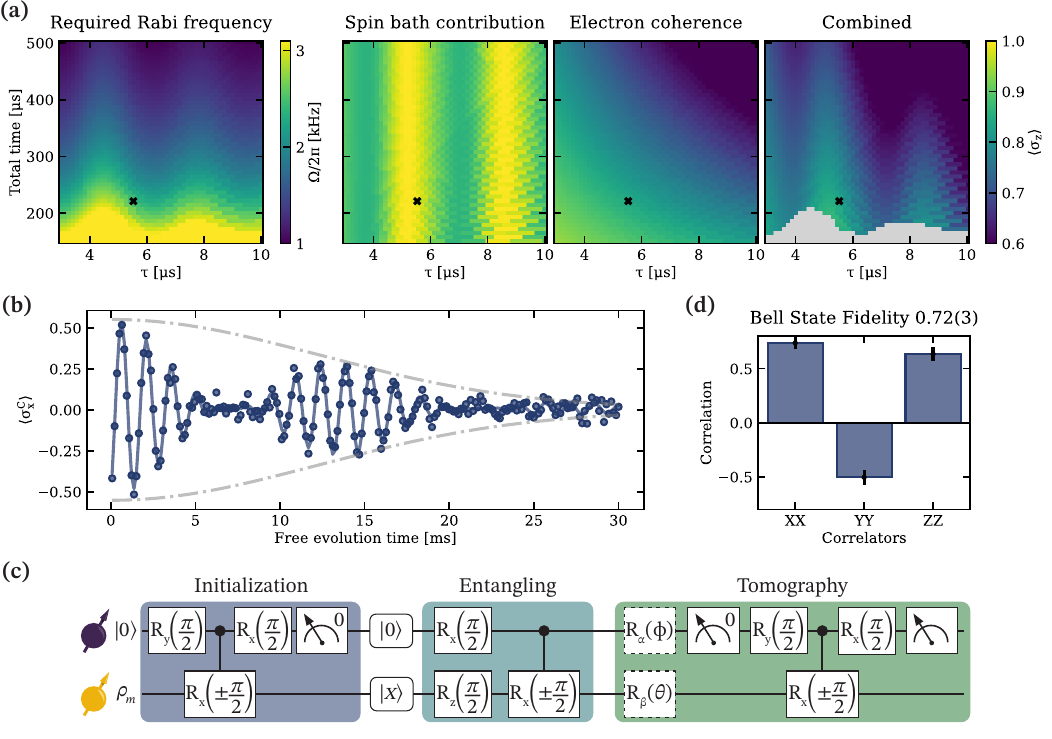}
    \caption{\textbf{Gate optimization and electron-nuclear entanglement.}
    (a) Gate parameter simulation. A set of $\tau$ and the total gate duration puts a requirement on the Rabi frequency that needs to be achieved for a gate with $\theta=\pi/2$. For specific values of $\tau$, the spectral shape of the RF tone also drives the other nuclear spin transition (i.e. driving $\omega_0$ while targeting $\omega_1$), resulting in a higher required Rabi frequency $\Omega$ to compensate for this off-resonant driving \cite{vanommen2024ddrf}. The gate fidelity is determined by the combination of electron coherence during DD and extra coherence loss due to the applied DDRF gate. Considering the fidelity and excluding gates with a Rabi frequency higher than the maximally achievable \qty{3.1}{\kilo\hertz} (gray region), we choose the working point indicated by the cross.
    (b) Measurement of the nuclear $T_2^*$ via a Ramsey sequence. The fit accounts for two coupled spins that induce a beating of the Ramsey signal. The envelope (gray dashed line) has the form $A\exp\left(-(t/T_2^*)^n\right)$ with $n=\qty{2.0(2)}{}$ and decay time $T_2^*=\qty{17.2(6)}{\milli\second}$.
    (c) Gate sequence for electron-nuclear entanglement. The nuclear spin is prepared into $\ket{x}$ via MBI. The entangling block creates the Bell state $\ket{\Phi^+}$, which is consecutively measured with different correlators. The tomography pulses on the nuclear spin are realized by Larmor precession ($z$-rotations) or direct RF driving.
    (d) Fidelity of the $\ket{\Phi^+}$ state. The readout of the nuclear spin in the tomography block is corrected for errors in the two-qubit gate, resulting in an entangled state fidelity of $\mathcal{F}_\text{Bell} = \num{0.72(3)}$. 
    }
    \label{fig:bell}
\end{figure*} 

\section{Electron-nuclear entanglement}

To use the DDRF gate on $C_B$ for entangling the electron and nuclear spin, we first optimize the gate through simulation in Fig.~\ref{fig:bell}(a), using the hyperfine parameters determined in the previous section. For this, we simulate the required RF Rabi frequencies for fully entangling gates with different total lengths and $\tau$. The effective Rabi frequency depends on the total length of the gate and the amount of off-resonant driving of the other nuclear transition. This off-resonant driving occurs because the RF pulse has a finite sinc-shaped frequency bandwidth \cite{vanommen2024ddrf}. Taking the RF Rabi frequency, total gate time, and $\tau$, we simulate the effect of the presence of a nuclear spin bath. Furthermore, the coherence of the electron needs to be preserved for a high-fidelity gate. Therefore, we calculate the coherence of the electron for different DD sequences determined in Fig.\ref{fig:electron}(b). By combining these two contributions to the infidelity, we get an estimate for the fidelity of a full two-qubit gate. We choose an operating point that gives high fidelity within the region of achievable Rabi frequencies ($\Omega/2\pi<\qty{3.1}{\kilo\hertz}$). 

For the optimization, there is a trade-off between preserving electron coherence and gate performance. Shortening the total gate time improves electron coherence but requires a higher Rabi driving frequency. Increasing $\tau$ improves the RF selectivity as the pulse bandwidth is reduced but lowers the electron coherence. Besides this, higher-order resonances with the nuclear spin bath can be avoided by carefully selecting appropriate values for $\tau$.

In Fig.~\ref{fig:bell}(b), the optimized gate is used in an under-sampled Ramsey measurement on $C_B$. The observed beating pattern is explained by a coupling to two different nuclear spins with \qty{67(4)}{\hertz} and \qty{71(4)}{\hertz}. The $T_2^*=\qty{17.2(6)}{\milli\second}$ of this nuclear spin is comparable to high values measured using the nitrogen-vacancy center in diamond \cite{bradley2019tenqubit}. We extract a two-qubit gate fidelity of $\mathcal{F}_\text{gate} = \qty{0.874(4)}{}$ (see Appendix \ref{app:readout_gate} for details), which is slightly less than the simulated fidelity $\mathcal{F}_\text{sim} = \qty{0.915}{}$. We attribute the difference to the decoherence of the electron spin to coupling to undetected spins, which are not captured by the spin bath simulation.

With this two-qubit gate, we have the tools to entangle the SnV electron spin with the \ce{^13C} spin. For this, we initialize the \ce{^13C}-spin, prepare the electron in a superposition state, and use the two-qubit entangling gate to create the Bell state $\ket{\Phi^+} = \frac{1}{\sqrt{2}}\left(\ket{00}+\ket{11}\right)$, see Fig. \ref{fig:bell}(c). The correlators are measured with tomography, from which the fidelity to the Bell state $\ket{\Phi^+}$ is calculated as $\mathcal{F}=(\braket{XX}-\braket{YY}+\braket{ZZ}+1)/4$. By correcting for known tomography errors, as explained in Appendix~\ref{app:readout_gate}, we find a fidelity to the Bell state $\ket{\Phi^+}$ of \qty{72(3)}{\percent}. As discussed in Appendix~\ref{app:error_budget}, this is mainly limited by the electron coherence loss during the nuclear spin gates.

\label{sec:comparison}
\begin{figure}
    \centering
    \includegraphics{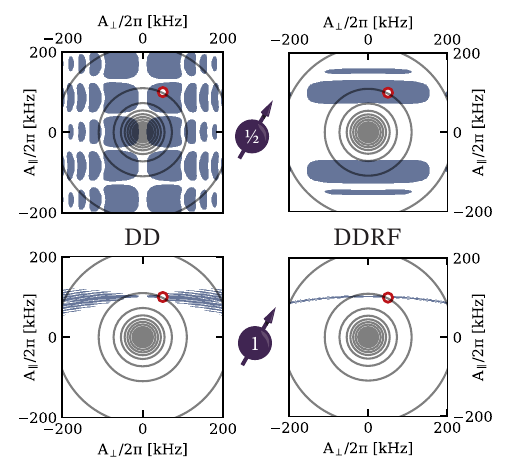}
    \caption{\textbf{Comparison of nuclear spin control with DD and DDRF methods.} Simulated cross-talk regimes when controlling a target nuclear spin with $A_\parallel/2\pi = \qty{100}{\kilo\hertz}$ and $A_\perp/2\pi = \qty{50}{\kilo\hertz}$ (red circle). DD and DDRF control methods are compared for electron spin-1/2 and spin-1. The blue areas correspond to hyperfine parameters for which a bystander nuclear spin would cause more than \qty{5}{\percent} decoherence of the electron spin, during a gate performed on the target nuclear spin. The regions between the gray circles contain, for a natural \ce{^13C} abundance of \qty{1.1}{\percent}, on average one bystander \ce{^13C} nuclear spin. For simulation details, see Appendix~\ref{app:comparison_simulation}.}
    \label{fig:comparison}
\end{figure} 

\section{Comparison control methods}
In previous sections, we explored the experimental control of \ce{^13C}-spin with a spin-1/2 electron spin using the DD and DDRF control methods. In this section, we compare the selectivity of these two methods in simulation. Additionally, we analyze the difference for the case of an electron spin-1 system \cite{taminiau2012detection, vanommen2024ddrf}. The different nuclear spins surrounding the electron spin can be characterized by their specific $A_\parallel$ and $A_\perp$. Figure~\ref{fig:comparison} visualizes the selectivity of both control methods. We simulate a nuclear spin gate for a nuclear spin with $A_\parallel/2\pi=\qty{100}{\kilo\hertz}$ and $A_\perp/2\pi=\qty{50}{\kilo\hertz}$ (red circle ). The blue region indicates for which hyperfine parameters a bystander spin would cause more than \qty{5}{\percent} coupling to the electron during the same gate.

For each scenario, the most selective gate is found by optimizing the gate parameters under the boundary conditions that the electron has preserved more than \qty{99}{\percent} coherence and the gate does not exceed \qty{1}{\milli\second}. For the simulations, we assume an electron echo time of $T_2^\text{Echo} = \qty{1}{\milli\second}$, dynamical decoupling scaling factor $\chi=2/3$ and a maximum nuclear Rabi frequency of \qty{5}{\kilo\hertz}.

The likelihood of having a bystander spin in the blue areas depends on the distribution of hyperfine parameters of the nuclear spins surrounding the color center. The placement of \ce{^13C} atoms around a color center in the diamond lattice is probabilistic, resulting in a varying arrangement of the bystander spins for different centers. We simulate the distribution of nuclear spin hyperfine parameters by generating many different random spin configurations and calculating their hyperfine parameters. The distribution of hyperfine parameters of bystander spins is visualized by the regions between the gray lines, which contain on average one nuclear spin. More simulation details can be found in Appendix~\ref{app:comparison_simulation}.

We observe that the selectivity for DD in spin-1/2 systems is reduced compared to spin-1 \cite{nguyen2019prl}. Most notably, there is considerable cross-talk with the nuclear spin bath, i.e., the spins with a weak coupling. We note that the specific analysis will depend on the chosen target spin. For example, spins with lower $A_\perp$ will have more cross-talk for DD control with electron spin-1/2. The DDRF method has an enhanced selectivity compared to the DD as the RF driving has a selectivity that depends on $A_\parallel$, thereby reducing the cross-talk with the spin bath. DDRF for spin-1 systems has an even more enhanced selectivity as it can leverage the first-order sensitivity of $A_\parallel$ in the phase update rule. 

\begin{figure*}
    \centering
    \includegraphics{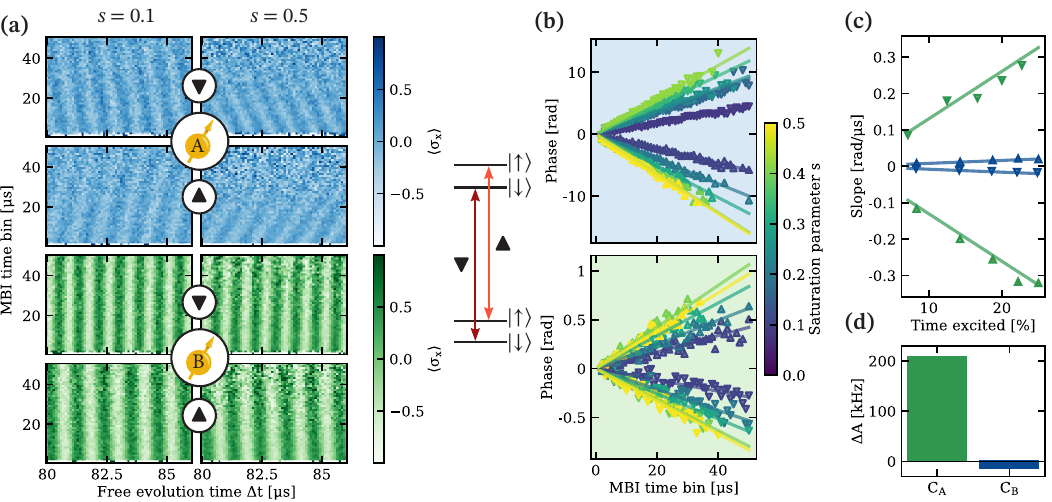}
    \caption{\textbf{Excited state hyperfine coupling.}
    (a) The Ramsey measurements of the nuclear spins are binned for the readout time during MBI. We find an additional phase shift depending on the duration of the readout. The blue (green) colored data correspond to $C_A$ ($C_B$).
    The rate of acquiring phase increases for higher laser powers (left and right column), i.e., longer occupation of the excited state. Changing the optical transition used for the readout during MBI ($\blacktriangle$ or $\blacktriangledown$) flips the direction of the phase shift. The difference in contrast stems from the better gate fidelities for $C_B$ compared to $C_A$.
    (b) Extracting the phase shift for different MBI durations and saturation numbers $s$ results in linear slopes.
    (c) The used power in (b) is converted to time spent in the excited state and shows a linear relation with the slope fitted in (b).
    (d) The difference in hyperfine coupling to the ground and excited state for both nuclear spins.
    }
    \label{fig:hyperfine}
\end{figure*} 

\section{Hyperfine coupling with the excited state}

So far, we have considered the hyperfine interaction with the ground state of the SnV electron spin. However, the performed Ramsey measurements also give us insight into the hyperfine interaction in the excited state. We observe that the nuclear spin acquires a phase depending on how much time the electron is in the excited state.

During measurement-based initialization (MBI), we use a readout of the electron that is stopped when a photon is detected. This yields the most reliable assignment of the post-measurement state but also results in a variable readout time. During the readout, the electron spin gets cycled between the ground and excited state. Therefore, the time the electron spends in the excited state depends on the optical power and the readout duration. 

In Fig.~\ref{fig:hyperfine}(a), Ramsey measurements for $C_A$ and $C_B$ have been binned by the readout duration during MBI. We observe that the Ramsey fringes are shifted for longer readout durations, indicating that the nuclear spin acquired an extra phase during readout. Furthermore, the shift depends on the readout power. We use the saturation parameter $s=P/P_\text{sat}$, to compare the readout power $P$ to the saturation power  $P_\text{sat}\approx \qty{88}{\nano\watt}$. The observed shift is more significant for a readout with $s=0.5$ in the right column than a readout with $s=0.1$ in the left column.  Lastly, we see that reading out using the spin-down transition ($\blacktriangledown$ in Fig.~\ref{fig:hyperfine}) results in the opposite shift compared to reading out using the spin-up transition ($\blacktriangle$ in Fig.~\ref{fig:hyperfine}). The more noisy data for longer readout duration is a consequence of the small amount of data resulting from a low probability of detecting the first photon at those times. 

We attribute the extra phase to a different hyperfine coupling of the \ce{^13C}-spin to the SnV electron in the ground and the excited state. Note that these effects can be avoided in integer-spin systems such as the diamond nitrogen-vacancy center by using an $m_S = 0$ optical transition. Intuitively, as the excited state exhibits a different electron wave function, the coupling to the \ce{^13C} can differ from that of the ground state. We can extract the acquired additional phase from the binned Ramsey measurements in the excited state. Fig.~\ref{fig:hyperfine}(b) shows the fitted phases for the binned Ramsey fringes in Fig.~\ref{fig:hyperfine}(a), which are well explained by a constant phase acquisition. Based on saturation measurements, the readout power can be converted to a fraction of the time spent in the excited state (see Section~1.7 in reference \cite{loudon2000quantum}):
\begin{equation}
    t_\text{exc}/t = \frac{s}{2s+1}.
\end{equation}
The factor 2 in the denominator stems from the readout being done resonantly, capturing the contribution of stimulated emission. Figure~\ref{fig:hyperfine}(c) shows the angular frequency corresponding to the additional phase acquisition as a function of the fraction of time spent in the excited state. Combining this occupancy with the additional phase acquisition in the excited state gives the difference in coupling. The observed shift in coupling is $\Delta A_A=\qty{208(17)}{\kilo\hertz}$ and $\Delta A_B=\qty{-13(2)}{\kilo\hertz}$ as displayed in Fig.~\ref{fig:hyperfine}(d). This shift can mainly be assigned to a difference in $A_\parallel$ between the ground and excited state, as a difference in $A_\perp$ has a small second-order effect on the precession frequencies. The contrast of the Ramsey measurements is mostly preserved for longer readout durations, even after hundreds of optical excitations, suggesting the nuclear spins are a viable storage of qubit states during optical operations on the electron spin.

\section{Discussion and outlook}
We have investigated the nuclear spin control using an electron spin-1/2 system. We have shown an improved selectivity in controlling nuclear spins using Dynamically Decoupled Radio Frequency (DDRF) control compared to the more commonly used Dynamical Decoupling (DD) control. Using these methods, we control two nuclear spins and show entanglement between the nuclear spin of a \ce{^13C} atom and the electron spin of a tin-vacancy (SnV) center in diamond.

These findings directly translate to other systems with an electron spin-1/2, such as rare-earth ions \cite{uysal2023rei_nuclear} and many other color centers like the silicon-vacancy in diamond \cite{knaut2024entanglement}, the T-center in silicon \cite{photonicinc2024distributed} and the vanadium center in silicon carbide \cite{astner2024vanadium}.
    
Decoupling with longer $\tau$ improves the RF selectivity. Longer gates allow for better phase selectivity.  Both these factors improve the selectivity for a single gate, allowing the control of more \ce{^13C}-spins. In this work, the electron coherence limits the electron-nuclear gate fidelity. This is not intrinsic to the SnV center or the nuclear spins but can be attributed to the presence of an electron spin bath. Recently, it has been shown that the SnV can indeed have longer coherence times during dynamical decoupling \cite{karapatzakis2024microwave}, which would improve the nuclear spin control as well.

The selectivity of the nuclear spin gates with DDRF is better than with DD for electron spin-1/2. However, it is still reduced compared to spin systems with a larger magnitude. To counteract this, less symmetric decoupling sequences \cite{casanova2015robust}, like the UDD \cite{uhrig2007udd}, can be explored to increase the selectivity of DDRF further.

The selectivity of the nuclear spin gates in electron spin-1/2 systems can be further improved by interleaving a single DD or DDRF gate by periods in which the electron spin is in an eigenstate. In these periods, the nuclear spins will precess with $\omega_i$ as opposed to $\bar\omega$ during decoupling, thereby recovering the first-order selectivity to $A_\parallel$ of spin-1 systems. To preserve the state of the electron spin during those periods, it needs to be temporarily stored in an easily accessible quantum memory \cite{stas2022robust}. In the SnV center, this role could be fulfilled by the nuclear spin of \ce{^117Sn} or \ce{^119Sn} \cite{harris2023hyperfine}.

Lastly, we investigated the hyperfine coupling of the nuclear spins with the excited state of the electron, showing a different coupling in the excited state compared to the ground state of the electron. Additionally, we can conclude that it also showed that the nuclear spins surrounding the SnV center stay coherent while the optical transition is cycled hundreds of times, making these memories robust against the optical readout of the electron spin.

The improved selectivity of the DDRF control and the insights in nuclear spin coupling during readout bring high-fidelity spin control in electron spin-1/2 systems closer.

The data and simulations that support this manuscript are available at 4TU.ResearchData \cite{beukers2024dataset4tu}.

\section*{Acknowledgements}

We thank H. P. Bartling, C. E. Bradley and V. V. Dobrovitski for helpful and important discussions. 
We acknowledge financial support from the joint research program “Modular quantum computers” by Fujitsu Limited and Delft University of Technology co-funded by the Netherlands Enterprise Agency under project number PPS2007, from the Dutch Research Council (NWO) through the Spinoza prize 2019 (project number SPI 63-264), from the Dutch Ministry of Economic Affairs and Climate Policy (EZK) as part of the Quantum Delta NL programme, from the Quantum Internet Alliance through the Horizon Europe program (grant agreement No. 101080128) and from the European Research Council (ERC) under the European Union’s Horizon 2020 research and innovation programme (grant agreement No. 852410).

\appendix

\begin{figure*}
    \centering
    \includegraphics{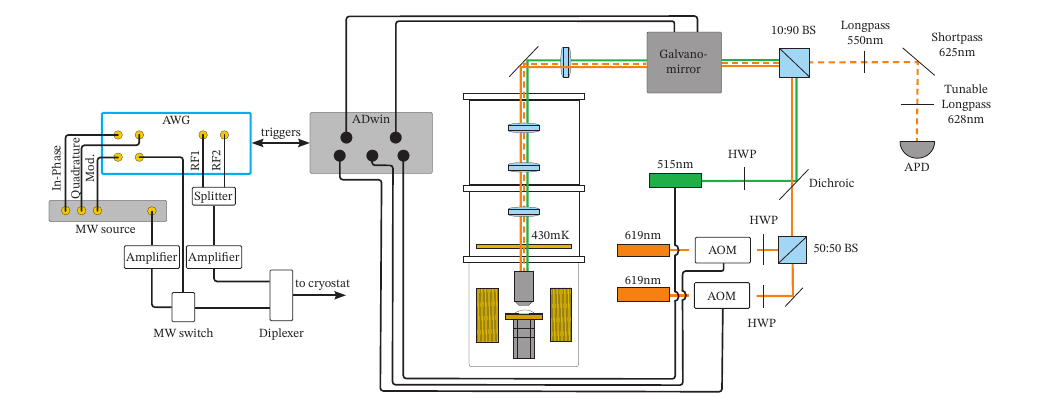}
    \caption{\textbf{Experimental Setup.} The setup consists of a cryostat with optical access, control electronics, and optical excitation and collection paths. The dashed line represents the fluorescence from the SnV center. A detailed description is provided in the text.}
    \label{fig:experimental_setup}
\end{figure*}

\section{Experimental setup}
\label{app:setup}

The setup is schematically shown in Fig.~\ref{fig:experimental_setup}.
The center of the setup is a BlueFors LD250He system with a base temperature of $\qty{430}{\milli\kelvin}$ with a 1-1-\qty{1}{\tesla} vector magnet. A positioner stack provides movement of the device.
Confocal microscopy experiments are performed with a high-NA objective and a double-4f system with a two-axis galvanomirror.
Microwaves and radiofrequency drives are supplied via superconducting coaxial cables (not shown in the figure) and a flex-cable to allow movement of the device (not shown in the figure). To minimize heating, the signal is guided via a custom PCB to a bondwire $\approx \qty{100}{\micro\meter}$ above the device.
The total microwave losses through the system at the relevant frequency $\qty{3.1}{\giga\hertz}$ are $\qty{5}{\decibel}$.

The experiment is controlled via a home-written software infrastructure based on QMI \cite{raa2023qmi}. A microcontroller (ADwin Pro II, Jäger Messtechnik) controls the experiment, and laser pulses are controlled via its DAC module.
For microwave and radio-frequency pulses, the ADwin triggers an arbitrary waveform generator (HDAWG, Zurich Instruments). The HDAWG modulates a microwave signal generator (SGS100A, Rohde \& Schwarz) with IQ-signals to generate phase-controlled microwave pulses.
The microwave pulses are amplified (Model 40S1G4, Amplifier Research) to $\qty[qualifier-mode = combine]{26}{\deci\bel\of{m}}$ for all measurements in the main text, which yields an estimated power of $\qty[qualifier-mode = combine]{20}{\deci\bel\of{m}}$ at the device location. To prevent the shot noise of the amplifier during free evolution periods from entering the system,
a microwave switch and high-pass filter (not in the figure) are employed.
The radio-frequency signals are directly generated by oscillators of the HDAWG and amplified to $\qty[qualifier-mode = combine]{18}{\deci\bel\of{m}}$. The transmission for these $\qty{}{\mega\hertz}$-signals is almost lossless.
We use two driving frequencies for the control with DDRF gates and drive both resonance frequencies of the nuclear spin, resulting in a $\sqrt{2}$-improvement of the Rabi frequency for the same power.
These two driving tones are combined with a directional coupler and then via a diplexer with the microwave signal. An additional DC-block (not in the figure) after the diplexer completes the chain.

The electron spin is addressed optically via a confocal microscopy setup. Two resonant excitation lasers (DLpro SHG-TA, Toptica) are modulated via the ADwin and AOMs (Fiber-Q 637nm, Gooch\&Housego) and combined via a beamsplitter.
An off-resonant repump laser at 515nm (Cobolt 06-03-MLD515, Hübner Photonics) is added with a dichroic mirror. The excitation light is coupled into the beampath of the cryostat through a 10:90 beamsplitter and the galvanomirror.
The collected light is spectrally filtered (FELH0600, Thorlabs; FF625-SDi01 and TLP01-628, Semrock) and detected by a single photon detector (COUNT, LaserComponents). A counter module of the ADwin stores the registered counts in $\qty{1}{\micro\second}$ time bins.

\begin{figure}
    \centering
    \includegraphics{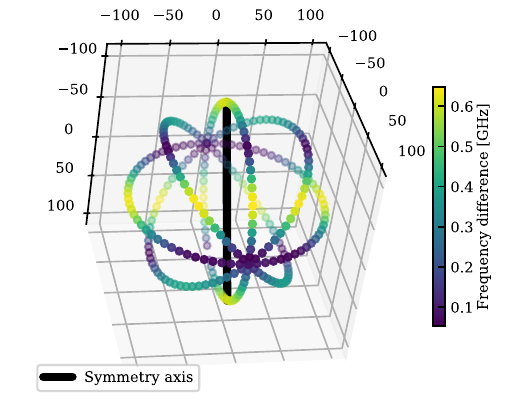}
    \caption{\textbf{Magnetic field sweep.} The frequency difference of the two spin-conserving optical transitions is shown as a function of the 3D magnetic field in the laboratory frame in units of \qty{}{\milli\tesla}. The solid black line shows the direction of the SnV center symmetry axis. In laboratory spherical coordinates, it is oriented along $(\theta, \phi) = (\qty{54.2(1)}{\degree}, \qty{-1.0(2)}{\degree})$.}
    \label{fig:magnetic_field_sweep}
\end{figure}

\section{Magnetic field sweep}
We investigate the behavior of the SnV center by sweeping the magnetic field along four great circles with magnetic field strength $\qty{100}{\milli\tesla}$ (see Fig.~\ref{fig:magnetic_field_sweep}). We monitor the frequencies of both spin-conserving transitions with PLE measurements.
We model the SnV center Hamiltonian as described in reference \cite{rosenthal2023microwave}, Eq.~B5.
To extract relevant parameters from the available data, we have to fix some Hamiltonian parameters. Specifically, our PLE measurements do not give us access to the ground-state splitting. 

Fitting the resulting data with the SnV center Hamiltonian with the strain and the laboratory orientation of the SnV center as free parameters, we find a laboratory direction of the SnV center symmetry axis of $(\theta, \phi) = (\qty{54.2(1)}{\degree}, \qty{-1.0(2)}{\degree})$. Since we use a $\langle 100 \rangle$-cut diamond with $\langle 110 \rangle$ side faces, these values match the expectation given the alignment of the diamond with respect to the magnet.
The strain values result in $\Upsilon_g = \qty{355(8)}{\giga\hertz}$ and $\Upsilon_e = \qty{689(20)}{\giga\hertz}$ for the ground and excited state, respectively. This moderate strain is smaller than the spin-orbit coupling in both manifolds, yet allows for direct driving of the microwave transition. The fit parameters are summarized in Table~\ref{tab:magnetic_field_fit_parameters}.

\begin{table}
    \centering
    \caption{Magnetic field fitting parameters.}
    \begin{tabular}{c c c}
        \hline{}
        Term & Value & Origin \\ \hline
        $\lambda_g/2\pi$ & $\qty{830}{\giga\hertz}$ & \cite{rosenthal2023microwave} \\
        $\lambda_e/2\pi$ & $\qty{2988}{\giga\hertz}$ & \cite{rosenthal2023microwave} \\
        $B$ & $\qty{97.8}{\milli\tesla}$ & Appendix~\ref{app:magnetic_field}\\
        $\theta_\text{SnV}$ & $\qty{54.2(1)}{\degree}$ & Fit \\
        $\phi_\text{SnV}$ & $\qty{-1.0(2)}{\degree}$ & Fit \\
        $\Upsilon_g$ & $\qty{352(7)}{\giga\hertz}$ & Fit \\
        $\Upsilon_e$ & $\qty{674(17)}{\giga\hertz}$ & Fit \\
        $f_g$ & $\qty{0.171}{}$ & \cite{rosenthal2023microwave} \\
        $f_e$ & $\qty{0.072}{}$ & \cite{rosenthal2023microwave} \\
        $\delta_g$ & $\qty{0.015}{}$ & \cite{rosenthal2023microwave} \\
        $\delta_e$ & $\qty{0.176}{}$ & \cite{rosenthal2023microwave} \\
        \hline
    \end{tabular}
    \label{tab:magnetic_field_fit_parameters}
\end{table}

\section{Electron coherence}
\subsection{Ramsey}
\label{app:electron_ramsey}
\begin{figure}
    \centering
    \includegraphics{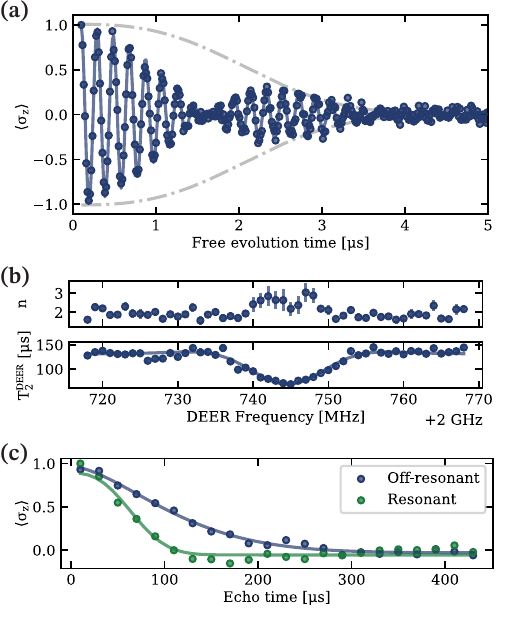}
    \caption{\textbf{Electron Ramsey and DEER measurements.} (a) Ramsey measurement. The Ramsey measurement is performed with an artificial detuning of \qty{5}{\mega\hertz}, implemented via an evolution time-dependent phase on the second $\pi/2$ gate. The beating frequency of $\qty{312(3)}{\kilo\hertz}$. (b) The frequency of the second MW pulse in the DEER measurement is varied to find the resonance of the electron spin bath. Both the echo time and the power of the power exponential fit in Eq.~\ref{eq:ramsey} are given. (c) Two examples of the Hahn-Echo measurements from (b), where a resonant MW pulse with the electron spin bath decreases the coherence time of the Hahn-Echo.}
    \label{fig:electron_coherence}
\end{figure}

To investigate the coherence of the electron spin, we conduct a Ramsey measurement. We perform the experiment on resonance and vary the phase of the second $\pi/2$ pulse to artificially introduce the signature of a detuning with $\Delta=\qty{5}{\mega\hertz}$. The resulting signal in Fig. \ref{fig:electron_coherence}(a) shows a slow beating in addition to the oscillation with $\Delta$. A fit of two sine functions with a power exponential envelope

\begin{align}
\label{eq:ramsey}
    \langle\sigma_z\rangle(t) = 
    A\exp\left(-\left(\frac{t}{T_2^*}\right)^n\right)\sum_{i=1,2}\left(\sin\left(\omega_it+\phi_i\right)\right)+c
\end{align}
reveals a frequency difference of $\qty{312(3)}{\kHz}$ of the sines, coinciding with the coupling strength of $C_B$ in the main text. The decay of the envelope with exponent $n=\qty{2.8(2)}{}$ leads to $T_2^*=\qty{2.42(4)}{\micro\second}$. This result is in accordance with the expected Ramsey coherence time of color centers in naturally abundant diamond \cite{taminiau2014universal}.

\subsection{DEER}
\label{app:deer}
The scaling of the coherence time $T_2^N = T_2^\text{Echo} N^\chi$ with $T_2^\text{Echo}=\qty{129(2)}{\micro\second}$ with the number of pulses $N$ via $\chi=\qty{0.47(1)}{}$ is less than we expect for a pure nuclear spin bath limitation \cite{wang2012comparison}. We conduct a double electron-electron resonance (DEER) measurement to verify electron bath noise. When measuring the Hahn-Echo decay time of the electron spin while applying a microwave pulse with varying frequency simultaneously to the decoupling pulse, we see a clear reduction of the coherence time when driving the surrounding with $\qty{2.745}{\giga\hertz}$, see Fig.~\ref{fig:electron_coherence}(b). We extract the decay constant $\tau$ and the exponent $n$ from the function
\begin{equation}
    T_2^\text{DEER} = A \exp\left(-(t/\tau)^n \right).
\end{equation}

We find a spin bath resonance frequency of $\qty{2.7446(2)}{\giga\hertz}$. This corresponds very well with electrons with a $g$-factor of 2 and a magnetic field of \qty{97.884}{\milli\tesla}, as determined in Appendix~\ref{app:magnetic_field} determined. The $T_2^\text{DEER} = \qty{76(4)}{\micro\second}$ has significantly decreased compared to a $T_2^\text{Echo}=\qty{128(5)}{\micro\second}$. The origin of this spin bath noise is unknown, it might stem from crystal damage due to insufficient annealing or residual tin ions in the surrounding.

\begin{figure}
    \centering
    \includegraphics{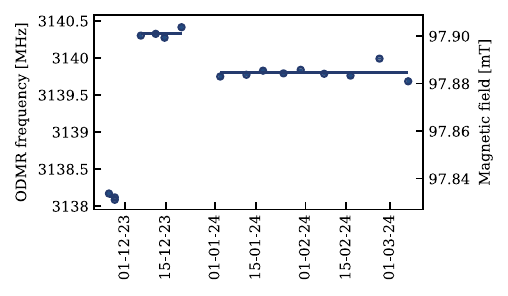}
    \caption{\textbf{Magnetic field strength over time.} The changes in the magnetic field strength are monitored using the MW frequency of the electron spin, through an ODMR measurement.}
    \label{fig:magnetic_field_strength}
\end{figure}

\section{Magnetic field strength changes}
\label{app:magnetic_field}

During the experiments, the magnetic field strength has not been constant as shown in Fig.~\ref{fig:magnetic_field_strength}, hampering direct comparison between measurements. As we have no direct measure of the magnetic field, we use the MW transition frequency of the electron spin, as measured by the ODMR measurements over time, to keep track of the relative magnitude of the magnetic field. The orientation of the magnetic field also influences the MW transition frequency, but as this is a second-order effect \cite{rosenthal2023microwave} we neglect this contribution. We observed three periods in which the magnetic field was stable. We suspect that switching the superconducting magnet to persistent mode caused the first jump and temporarily warming up to \qty{4}{\kelvin} caused the second one. Using these relative variations in the field strength allowed us to fit Fig.~\ref{fig:dd}(d) while using the $\omega_0$ and $\omega_1$ extracted from Fig.~\ref{fig:dd}(f) to determine the Larmor frequency. This Larmor frequency could then be converted to an absolute magnetic field strength by using the gyromagnetic ratio of the nuclear spin, as indicated in Fig.~\ref{fig:magnetic_field_strength}. The reported value of $\omega_L$ in the main text was based on the second time period.

\section{Simulations}
\label{app:simulation}

All the code to generate the figures in this paper, including the simulations are available in the repository supporting this manuscript \cite{beukers2024dataset4tu}.
\subsection{DD}
\label{app:dd_simulation}
The DD simulations are all based on the analytic formulas from reference \cite{zahedian2024blueprint}, which describe both electron spin-1/2 and spin-1 systems.

To simulate the nuclear spin bath, we first generate a random distribution of the nuclear spins. For this, we randomly place the nuclear spins in a sphere with radius \qty{30}{\nano\meter} around the electron spin with a density corresponding to \qty{1.1}{\percent} \ce{^13C} atoms. The hyperfine parameters are then calculated by assuming just the dipole-dipole interaction. For the DD spin bath simulations only spins with a hyperfine coupling $A/2\pi < \qty{100}{\kilo\hertz}$ are included, where $A = \sqrt{A_\parallel^2+A_\perp^2}$.

For Fig.~\ref{fig:dd}(c), 10 different realizations of the spin bath are generated and the region between the highest and lowest line are shown.

\subsection{DDRF}
\label{app:ddrf_simulation}

For the DDRF simulations in Fig.~\ref{fig:ddrf_spectrum}(b), Fig.~\ref{fig:bell}(a) and Fig.~\ref{fig:uddrf_spectra} numerical methods based on the Python framework QuTiP \cite{johansson2012qutip} were used to simulate the Hamiltonian of the RF drive from reference \cite{vanommen2024ddrf}. These simulations include the exact pulse shape and length as were used in our experiments. For the comparison in Fig.~\ref{fig:comparison} the analytic formulas from reference \cite{vanommen2024ddrf} were used.

The simulations of the nuclear spin bath in Fig.~\ref{fig:ddrf_spectrum}(b), Fig.~\ref{fig:bell}(a) and Fig.~\ref{fig:uddrf_spectra} are based on the spin bath simulations in reference \cite{vanommen2024ddrf}. Rather than simulating a randomly generated nuclear spin bath, we approximate it with a probability distribution of the hyperfine parameters. For the distribution of $A_\parallel$ we take the heuristic distribution from the supplement of reference \cite{vandestolpe2024mapping}. We assume that the nuclear spin bath has $A_\perp=\qty{0}{\kilo\hertz}$, as the bath has weak coupling and $A_\perp$ has only a second-order effect on $\omega_0$ and $\omega_1$, which are the parameters relevant for the DDRF simulation. In this work, we consider spins with a $|A_\parallel|/2\pi < \qty{20}{\kilo\hertz}$ to be part of the spin bath for the DDRF simulations as these spins are below our detection limit.  We then subdivide the distribution of the bath in bins of $\Delta A_\parallel=\qty{2}{\kilo\hertz}$ and weigh their effect by the number of expected spins in that bin. Finally, we combine the effect of the different bins on the electron to get the effect of the whole nuclear spin bath.

\subsection{Comparison}
\label{app:comparison_simulation}

The comparison in Fig.~\ref{fig:comparison} is performed by finding the coherent and fast two-qubit gate on the target spin that minimizes the cross-talk with bystander spins. We considered it fast and coherent if the electron coherence after the gate was above \qty{99}{\percent} and the total gate duration was below \qty{1}{\milli\second}. A bystander spin is considered to cause cross-talk for a gate if the loss in the coherence of the electron due to this bystander spin is bigger than \qty{5}{\percent}. The gate is optimized to cause cross-talk with the fewest possible bystander spins, i.e., minimizing the blue region in Fig.~\ref{fig:comparison}. This metric was not weighted for the bystander spin density. 

For the DD method, the gate with the smallest cross talk was found by sweeping the interpulse delay $\tau$ and the number of decoupling pulses $N$. For the DDRF method, the interpulse delay $\tau$, the number of decoupling pulses $N$, and the RF driving strength $\Omega$ were swept to find the most selective gate.

The circles indicating the bystander spin density are based on the average of \num{100} randomly generated spin bath configurations. The spin baths are generated in the same way as for the DD spin bath simulations, see Appendix~\ref{app:dd_simulation}.

\begin{figure}
    \centering
    \includegraphics{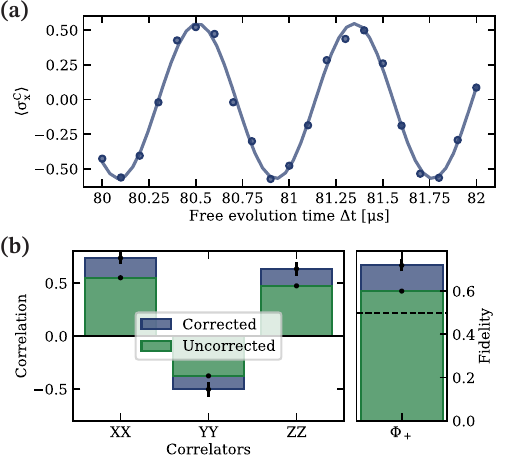}
    \caption{\textbf{Readout correction for Bell state measurement.} (a) Ramsey measurement to determine two-qubit gate fidelity. The contrast of the Ramsey measurement is reduced due to imperfect two-qubit gates. For the measurement, the two-qubit gate is performed to initialize and read out the nuclear spin. A model based on imperfect state preparation and POVM measurements yields $\mathcal{F}_\text{gate} = \qty{0.874(4)}{}$. (b) Additional uncorrected data for the Bell state measurement in Fig.~\ref{fig:bell}(d).}
    \label{fig:bell_correction}
\end{figure}

\section{Bell state}
\subsection{Readout gate correction}
\label{app:readout_gate}
When determining the fidelity of the Bell state $\ket{\Phi^+}$ in the main text, we apply a readout correction to the nuclear spin readout via MBI. To this end, we estimate the two-qubit gate fidelity $\mathcal{F}_\text{gate}$ via a Ramsey measurement. The fit of a sinusoidal function to Fig.~\ref{fig:bell_correction}(a) shows an amplitude of $\langle\sigma_x\rangle=\qty{0.559(12)}{}$. We model the imperfect initialization and readout of the nuclear spin in the $x$-basis with the preparation of state
\begin{equation}
    \rho = \begin{pmatrix}
        \mathcal{F}_\text{gate} & 0 \\
        0 & 1-\mathcal{F}_\text{gate}
    \end{pmatrix},
\end{equation}
and readout via a POVM measurement with the matrix set
\begin{equation}
    E_x = \begin{pmatrix}
        \mathcal{F}_\text{gate} & 0 \\
        0 & 1-\mathcal{F}_\text{gate}
    \end{pmatrix}, \quad
    E_{-x} = \begin{pmatrix}
        1-\mathcal{F}_\text{gate} & 0 \\
        0 & \mathcal{F}_\text{gate}
    \end{pmatrix}.
\end{equation}
The expectation value of this model is 
\begin{equation}
    \langle\sigma_x\rangle = 1 \cdot p_x - 1 \cdot p_{-x} = 1 - 4\mathcal{F}_\text{gate} + 4\mathcal{F}_\text{gate}^2,
\end{equation}
with $p_i = \mathrm{Tr}\left(\rho E_i\right)$. Solving for the gate fidelity results in $\mathcal{F}_\text{gate} = \qty{0.874(4)}{}$. This value is slightly below the simulated gate fidelity shown in Fig.~\ref{fig:bell}(a) of $\qty{0.915}{}$. We attribute this difference to the coupling of the electron spin to unknown spins.

We use the estimated fidelity to perform a readout correction following the formalism outlined in reference \cite{pompili2021thesis} Appendix~A. The uncorrected and corrected data of the Bell state measurement is shown in Fig.~\ref{fig:bell_correction}(b).

\subsection{Error budget estimation}
\label{app:error_budget}
The measured Bell state does not have perfect fidelity. This section provides an overview of the known and unknown sources of infidelity to estimate the expected fidelity for the Bell state measurements.

The experimental gate sequence, as shown in Fig.~\ref{fig:bell}(c) is comprised of three logic blocks with single- and two-qubit gates.

In the \enquote{Initialization}-block, both the electron spin and nuclear spin are prepared into known states $\ket{0}$ and $\ket{x}$. The probability of correctly measuring the electron in state $\ket{0}$ after the two-qubit gate is $\qty{98}{\percent}$, mainly limited by the spin pump fidelity. The correct initialization of the nuclear spin is limited by the two-qubit gate fidelity of $\qty{87.4}{\percent}$. 

The \enquote{Entangling}-block consists of a $\pi/2$ gate on the electron spin, a $z$-rotation of the nuclear spin (implemented by a waiting time, assumed to be perfect) and a two-qubit gate identical to the initialization. We determine the average electron spin gate fidelity to $\qty{0.98(2)}{}$ via process tomography. 

The last block, \enquote{Tomography}, consists of another single qubit rotation on the electron spin as well as the nuclear spin, followed by a single-shot readout of both qubits individually. To prepare the nuclear spin for a readout in the $x$- or $y$-basis, a $z$-rotation is performed via a waiting time. The error of this gate is determined by the uncertainty of the Ramsey frequency, which we determine to be less than $\qty{1e-4}{}$ relative error in the main text.
For readout in the $z$-basis, a direct RF gate is implemented as reported in Appendix~\ref{app:nmr}. This gate is calibrated via a Rabi experiment. However, the fidelity of the gate has not been determined.
We correct for the readout processes following the description of Appendix~A of reference \cite{pompili2021thesis} for the electron spin and Appendix~\ref{app:readout_gate} for the nuclear spin. 

All known and unknown fidelities are listed in table \ref{tab:fidelities}.

\begin{table}[]
    \centering
    \caption{Gate fidelites of electron (e) and nuclear (n) spin.}
    \label{tab:fidelities}
    \begin{tabular}{c c c}
        \hline{}
        Gate  &  Fidelity &  Method\\\hline
        Init. (e) & 0.981(5) & Calibration \\
        Init. (n) & 0.874(4) & Ramsey contrast \\
        $R(\pi/2)$ (e) & 0.98(2) & Process tomography \\
        $R_z(\pi/2)$ (n) & 1 & Calibration \\
        $cR_x(\pm\pi/2)$ & 0.874(4) & Ramsey contrast \\
        $R_{x,y}(\pi/2)$ (n) & - & Undetermined \\\hline
        
    \end{tabular}
\end{table}

To estimate the expected fidelity of the Bell state measurement, we use a simplified model in which we assume all gates except the $cR_x(\pm\pi/2)$ to be perfect for simplicity, as their infidelity is small compared to the two-qubit gate. We describe the two-qubit gate as an ideal gate with probability $p = \mathcal{F}_\text{gate}$ and an ideal gate with an additional phase error on the electron with probability $1-p$. With this phase error, we take the loss of coherence of the electron spin during the gate into account, which is the main reason for the gate infidelity, as detailed in Fig.~\ref{fig:bell}(a). For the modeling, we exclude the tomography pulses and hence compare the calculated value to the readout-corrected measured fidelity.

The estimated fidelity of the created Bell state with the sequence shown in Fig.~\ref{fig:bell}(c) is $\mathcal{F}_\text{ideal} = \qty{0.76}{}$ with respect to the target state $\ket{\Phi}^+$. This is in good agreement with the measured fidelity of $\mathcal{F}_\text{Bell} = \qty{0.72(3)}{}$. We attribute the additional infidelity in the measurement to the single-qubit errors summarized in this section.

\begin{figure}
    \centering
    \includegraphics{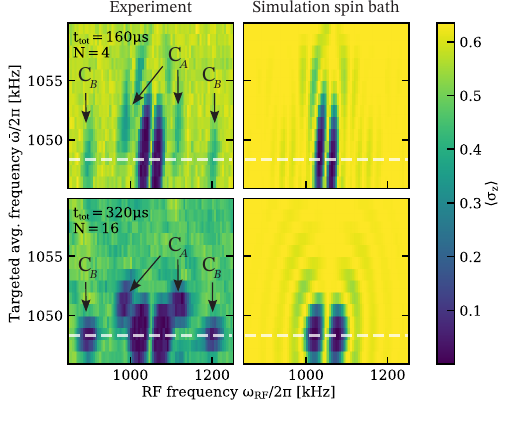}
    \caption{\textbf{DDRF Spectrum using the UDD sequence.} The same DDRF spectrum  as in Fig.~\ref{fig:ddrf_spectrum}(c) but now implemented with UDD. The upper row is using 4, while the bottom row is using 16 decoupling pulses. }
    \label{fig:uddrf_spectra}
\end{figure}

\section{DDRF using UDD}
\label{app:udd}
To see if we could improve on the selectivity of DDRF with the symmetric XY8 decoupling sequence, we used Uhrig Dynamical Decoupling \cite{uhrig2007udd}. In this sequence, the time between the pulses varies as
\begin{equation}
    \delta_j= \sin^2\left(\frac{j \pi}{2N + 2}\right),
\end{equation}
where $j$ is the index of the interpulse delay and $N$ is the total number of pulses. The DDRF can be implemented in the same way as for an XY8 decoupling sequence. However, the phase update for the RF driving field is now different between each pulse and needs to be calculated explicitly by keeping track of the phase of the nuclear spin.

In Fig.~\ref{fig:ddrf_spectrum}, a spectrum similar to Fig.~\ref{fig:ddrf_spectrum}(c) is shown but now implemented with UDD. The top row shows UDD with $N=4$, and the bottom row shows $N=16$. The same two nuclear spins can be observed as with the DDRF with XY8. However, the nuclear spin bath has a different fingerprint, revealing $C_A$ better in the spectrum of $N=4$.

\begin{figure}
    \centering
    \includegraphics{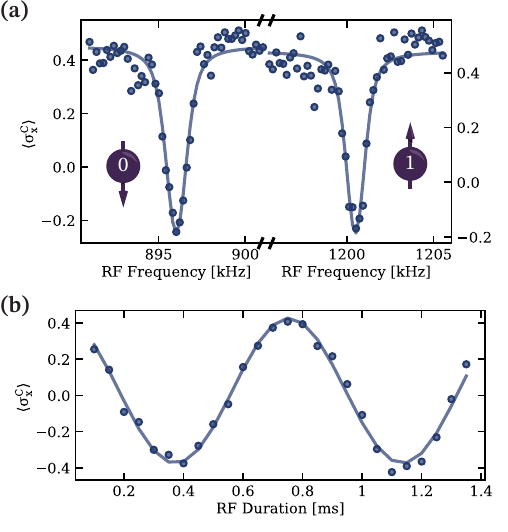}
    \caption{\textbf{Nuclear spin magnetic resonance and Rabi drive.} (a) Nuclear magnetic resonance experiment. The two resonance frequencies appear as a dip in the readout signal and depend on the state of the electron spin. The frequency matches well with the frequencies determined by Ramsey measurements. (b) Direct RF drive Rabi experiment. The measured Rabi frequency of $\qty{1.31(1)}{\kilo\hertz}$ matches the expected frequency determined by simulation.}
    \label{fig:nmr}
\end{figure}

\section{Nuclear spin NMR and Rabi}
\label{app:nmr}
In addition to Ramsey measurements to determine the nuclear precession frequencies, we use a direct RF drive on $C_B$ to find the two resonance frequencies. We measure the nuclear magnetic resonance (NMR) spectra by initializing the nuclear spin into state $\ket{x}$, applying the RF signal for $\qty{300}{\micro\second}$ with varying frequency, and reading out in the $x$-basis. The resonances are shown in Fig.~\ref{fig:nmr}(a) for both electron spin states during the experiment. We determine $\omega_0/2\pi = \qty{896.02(3)}{\kilo\hertz}$ and $\omega_1/2\pi = \qty{1200.49(3)}{\kilo\hertz}$. 

For the tomography of the Bell state, we need to employ an $R_x(\pi/2)$-gate on the nuclear spin. We implement this gate via direct RF drive at $\omega_0$ while the electron is in state $\ket{0}$. In Fig. \ref{fig:nmr}(b), a Rabi experiment shows a Rabi frequency $\Omega/2\pi = \qty{1.31(1)}{\kilo\hertz}$ of the nuclear spin. From this, we calibrate the $\pi/2$ gate employed in the experiments.  Furthermore, this measurement allowed us to covert the RF power used in the DDRF spectra in Fig.~\ref{fig:ddrf_spectrum} and Fig.~\ref{fig:uddrf_spectra} to a Rabi frequency of $\Omega/2\pi = \qty{1.64}{\kilo\hertz}$, which was used in the nuclear spin bath simulations.

\bibliography{library.bib}% Produces the bibliography via BibTeX.

\begin{thebibliography}{70}%
\makeatletter
\providecommand \@ifxundefined [1]{%
 \@ifx{#1\undefined}
}%
\providecommand \@ifnum [1]{%
 \ifnum #1\expandafter \@firstoftwo
 \else \expandafter \@secondoftwo
 \fi
}%
\providecommand \@ifx [1]{%
 \ifx #1\expandafter \@firstoftwo
 \else \expandafter \@secondoftwo
 \fi
}%
\providecommand \natexlab [1]{#1}%
\providecommand \enquote  [1]{``#1''}%
\providecommand \bibnamefont  [1]{#1}%
\providecommand \bibfnamefont [1]{#1}%
\providecommand \citenamefont [1]{#1}%
\providecommand \href@noop [0]{\@secondoftwo}%
\providecommand \href [0]{\begingroup \@sanitize@url \@href}%
\providecommand \@href[1]{\@@startlink{#1}\@@href}%
\providecommand \@@href[1]{\endgroup#1\@@endlink}%
\providecommand \@sanitize@url [0]{\catcode `\\12\catcode `\$12\catcode `\&12\catcode `\#12\catcode `\^12\catcode `\_12\catcode `\%12\relax}%
\providecommand \@@startlink[1]{}%
\providecommand \@@endlink[0]{}%
\providecommand \url  [0]{\begingroup\@sanitize@url \@url }%
\providecommand \@url [1]{\endgroup\@href {#1}{\urlprefix }}%
\providecommand \urlprefix  [0]{URL }%
\providecommand \Eprint [0]{\href }%
\providecommand \doibase [0]{https://doi.org/}%
\providecommand \selectlanguage [0]{\@gobble}%
\providecommand \bibinfo  [0]{\@secondoftwo}%
\providecommand \bibfield  [0]{\@secondoftwo}%
\providecommand \translation [1]{[#1]}%
\providecommand \BibitemOpen [0]{}%
\providecommand \bibitemStop [0]{}%
\providecommand \bibitemNoStop [0]{.\EOS\space}%
\providecommand \EOS [0]{\spacefactor3000\relax}%
\providecommand \BibitemShut  [1]{\csname bibitem#1\endcsname}%
\let\auto@bib@innerbib\@empty
%</preamble>
\bibitem [{\citenamefont {Awschalom}\ \emph {et~al.}(2018)\citenamefont {Awschalom}, \citenamefont {Hanson}, \citenamefont {Wrachtrup},\ and\ \citenamefont {Zhou}}]{awschalom2018quantum}%
  \BibitemOpen
  \bibfield  {author} {\bibinfo {author} {\bibfnamefont {D.~D.}\ \bibnamefont {Awschalom}}, \bibinfo {author} {\bibfnamefont {R.}~\bibnamefont {Hanson}}, \bibinfo {author} {\bibfnamefont {J.}~\bibnamefont {Wrachtrup}},\ and\ \bibinfo {author} {\bibfnamefont {B.~B.}\ \bibnamefont {Zhou}},\ }\bibfield  {title} {\bibinfo {title} {Quantum technologies with optically interfaced solid-state spins},\ }\href {https://doi.org/10.1038/s41566-018-0232-2} {\bibfield  {journal} {\bibinfo  {journal} {Nature Photonics}\ }\textbf {\bibinfo {volume} {12}},\ \bibinfo {pages} {516} (\bibinfo {year} {2018})}\BibitemShut {NoStop}%
\bibitem [{\citenamefont {Maze}\ \emph {et~al.}(2008)\citenamefont {Maze}, \citenamefont {Stanwix}, \citenamefont {Hodges}, \citenamefont {Hong}, \citenamefont {Taylor}, \citenamefont {Cappellaro}, \citenamefont {Jiang}, \citenamefont {Dutt}, \citenamefont {Togan}, \citenamefont {Zibrov}, \citenamefont {Yacoby}, \citenamefont {Walsworth},\ and\ \citenamefont {Lukin}}]{maze2008nanoscale}%
  \BibitemOpen
  \bibfield  {author} {\bibinfo {author} {\bibfnamefont {J.~R.}\ \bibnamefont {Maze}}, \bibinfo {author} {\bibfnamefont {P.~L.}\ \bibnamefont {Stanwix}}, \bibinfo {author} {\bibfnamefont {J.~S.}\ \bibnamefont {Hodges}}, \bibinfo {author} {\bibfnamefont {S.}~\bibnamefont {Hong}}, \bibinfo {author} {\bibfnamefont {J.~M.}\ \bibnamefont {Taylor}}, \bibinfo {author} {\bibfnamefont {P.}~\bibnamefont {Cappellaro}}, \bibinfo {author} {\bibfnamefont {L.}~\bibnamefont {Jiang}}, \bibinfo {author} {\bibfnamefont {M.~V.~G.}\ \bibnamefont {Dutt}}, \bibinfo {author} {\bibfnamefont {E.}~\bibnamefont {Togan}}, \bibinfo {author} {\bibfnamefont {A.~S.}\ \bibnamefont {Zibrov}}, \bibinfo {author} {\bibfnamefont {A.}~\bibnamefont {Yacoby}}, \bibinfo {author} {\bibfnamefont {R.~L.}\ \bibnamefont {Walsworth}},\ and\ \bibinfo {author} {\bibfnamefont {M.~D.}\ \bibnamefont {Lukin}},\ }\bibfield  {title} {\bibinfo {title} {Nanoscale magnetic sensing with an individual electronic spin in diamond},\ }\href
  {https://doi.org/10.1038/nature07279} {\bibfield  {journal} {\bibinfo  {journal} {Nature}\ }\textbf {\bibinfo {volume} {455}},\ \bibinfo {pages} {644} (\bibinfo {year} {2008})}\BibitemShut {NoStop}%
\bibitem [{\citenamefont {Abobeih}\ \emph {et~al.}(2019)\citenamefont {Abobeih}, \citenamefont {Randall}, \citenamefont {Bradley}, \citenamefont {Bartling}, \citenamefont {Bakker}, \citenamefont {Degen}, \citenamefont {Markham}, \citenamefont {Twitchen},\ and\ \citenamefont {Taminiau}}]{abobeih2019atomicscale}%
  \BibitemOpen
  \bibfield  {author} {\bibinfo {author} {\bibfnamefont {M.~H.}\ \bibnamefont {Abobeih}}, \bibinfo {author} {\bibfnamefont {J.}~\bibnamefont {Randall}}, \bibinfo {author} {\bibfnamefont {C.~E.}\ \bibnamefont {Bradley}}, \bibinfo {author} {\bibfnamefont {H.~P.}\ \bibnamefont {Bartling}}, \bibinfo {author} {\bibfnamefont {M.~A.}\ \bibnamefont {Bakker}}, \bibinfo {author} {\bibfnamefont {M.~J.}\ \bibnamefont {Degen}}, \bibinfo {author} {\bibfnamefont {M.}~\bibnamefont {Markham}}, \bibinfo {author} {\bibfnamefont {D.~J.}\ \bibnamefont {Twitchen}},\ and\ \bibinfo {author} {\bibfnamefont {T.~H.}\ \bibnamefont {Taminiau}},\ }\bibfield  {title} {\bibinfo {title} {Atomic-scale imaging of a 27-nuclear-spin cluster using a quantum sensor},\ }\href {https://doi.org/10.1038/s41586-019-1834-7} {\bibfield  {journal} {\bibinfo  {journal} {Nature}\ }\textbf {\bibinfo {volume} {576}},\ \bibinfo {pages} {411} (\bibinfo {year} {2019})}\BibitemShut {NoStop}%
\bibitem [{\citenamefont {Stolk}\ \emph {et~al.}(2024)\citenamefont {Stolk}, \citenamefont {{van der Enden}}, \citenamefont {Slater}, \citenamefont {te~{Raa-Derckx}}, \citenamefont {Botma}, \citenamefont {{van Rantwijk}}, \citenamefont {Biemond}, \citenamefont {Hagen}, \citenamefont {Herfst}, \citenamefont {Koek}, \citenamefont {Meskers}, \citenamefont {Vollmer}, \citenamefont {{van Zwet}}, \citenamefont {Markham}, \citenamefont {Edmonds}, \citenamefont {Geus}, \citenamefont {Elsen}, \citenamefont {Jungbluth}, \citenamefont {Haefner}, \citenamefont {Tresp}, \citenamefont {Stuhler}, \citenamefont {Ritter},\ and\ \citenamefont {Hanson}}]{stolk2024metropolitanscale}%
  \BibitemOpen
  \bibfield  {author} {\bibinfo {author} {\bibfnamefont {A.~J.}\ \bibnamefont {Stolk}}, \bibinfo {author} {\bibfnamefont {K.~L.}\ \bibnamefont {{van der Enden}}}, \bibinfo {author} {\bibfnamefont {M.-C.}\ \bibnamefont {Slater}}, \bibinfo {author} {\bibfnamefont {I.}~\bibnamefont {te~{Raa-Derckx}}}, \bibinfo {author} {\bibfnamefont {P.}~\bibnamefont {Botma}}, \bibinfo {author} {\bibfnamefont {J.}~\bibnamefont {{van Rantwijk}}}, \bibinfo {author} {\bibfnamefont {B.}~\bibnamefont {Biemond}}, \bibinfo {author} {\bibfnamefont {R.~A.~J.}\ \bibnamefont {Hagen}}, \bibinfo {author} {\bibfnamefont {R.~W.}\ \bibnamefont {Herfst}}, \bibinfo {author} {\bibfnamefont {W.~D.}\ \bibnamefont {Koek}}, \bibinfo {author} {\bibfnamefont {A.~J.~H.}\ \bibnamefont {Meskers}}, \bibinfo {author} {\bibfnamefont {R.}~\bibnamefont {Vollmer}}, \bibinfo {author} {\bibfnamefont {E.~J.}\ \bibnamefont {{van Zwet}}}, \bibinfo {author} {\bibfnamefont {M.}~\bibnamefont {Markham}}, \bibinfo {author} {\bibfnamefont {A.~M.}\ \bibnamefont {Edmonds}},
  \bibinfo {author} {\bibfnamefont {J.~F.}\ \bibnamefont {Geus}}, \bibinfo {author} {\bibfnamefont {F.}~\bibnamefont {Elsen}}, \bibinfo {author} {\bibfnamefont {B.}~\bibnamefont {Jungbluth}}, \bibinfo {author} {\bibfnamefont {C.}~\bibnamefont {Haefner}}, \bibinfo {author} {\bibfnamefont {C.}~\bibnamefont {Tresp}}, \bibinfo {author} {\bibfnamefont {J.}~\bibnamefont {Stuhler}}, \bibinfo {author} {\bibfnamefont {S.}~\bibnamefont {Ritter}},\ and\ \bibinfo {author} {\bibfnamefont {R.}~\bibnamefont {Hanson}},\ }\href@noop {} {\bibinfo {title} {Metropolitan-scale heralded entanglement of solid-state qubits}} (\bibinfo {year} {2024}),\ \Eprint {https://arxiv.org/abs/2404.03723} {arXiv:2404.03723} \BibitemShut {NoStop}%
\bibitem [{\citenamefont {Lukin}\ \emph {et~al.}(2020)\citenamefont {Lukin}, \citenamefont {Guidry},\ and\ \citenamefont {Vu{\v c}kovi{\'c}}}]{lukin2020integrated}%
  \BibitemOpen
  \bibfield  {author} {\bibinfo {author} {\bibfnamefont {D.~M.}\ \bibnamefont {Lukin}}, \bibinfo {author} {\bibfnamefont {M.~A.}\ \bibnamefont {Guidry}},\ and\ \bibinfo {author} {\bibfnamefont {J.}~\bibnamefont {Vu{\v c}kovi{\'c}}},\ }\bibfield  {title} {\bibinfo {title} {Integrated {{Quantum Photonics}} with {{Silicon Carbide}}: {{Challenges}} and {{Prospects}}},\ }\href {https://doi.org/10.1103/PRXQuantum.1.020102} {\bibfield  {journal} {\bibinfo  {journal} {PRX Quantum}\ }\textbf {\bibinfo {volume} {1}},\ \bibinfo {pages} {020102} (\bibinfo {year} {2020})}\BibitemShut {NoStop}%
\bibitem [{\citenamefont {Ruf}\ \emph {et~al.}(2021)\citenamefont {Ruf}, \citenamefont {Wan}, \citenamefont {Choi}, \citenamefont {Englund},\ and\ \citenamefont {Hanson}}]{ruf2021quantum}%
  \BibitemOpen
  \bibfield  {author} {\bibinfo {author} {\bibfnamefont {M.}~\bibnamefont {Ruf}}, \bibinfo {author} {\bibfnamefont {N.~H.}\ \bibnamefont {Wan}}, \bibinfo {author} {\bibfnamefont {H.}~\bibnamefont {Choi}}, \bibinfo {author} {\bibfnamefont {D.}~\bibnamefont {Englund}},\ and\ \bibinfo {author} {\bibfnamefont {R.}~\bibnamefont {Hanson}},\ }\bibfield  {title} {\bibinfo {title} {Quantum networks based on color centers in diamond},\ }\href {https://doi.org/10.1063/5.0056534} {\bibfield  {journal} {\bibinfo  {journal} {Journal of Applied Physics}\ }\textbf {\bibinfo {volume} {130}},\ \bibinfo {pages} {070901} (\bibinfo {year} {2021})}\BibitemShut {NoStop}%
\bibitem [{\citenamefont {Knaut}\ \emph {et~al.}(2024)\citenamefont {Knaut}, \citenamefont {Suleymanzade}, \citenamefont {Wei}, \citenamefont {Assumpcao}, \citenamefont {Stas}, \citenamefont {Huan}, \citenamefont {Machielse}, \citenamefont {Knall}, \citenamefont {Sutula}, \citenamefont {Baranes}, \citenamefont {Sinclair}, \citenamefont {{De-Eknamkul}}, \citenamefont {Levonian}, \citenamefont {Bhaskar}, \citenamefont {Park}, \citenamefont {Lon{\v c}ar},\ and\ \citenamefont {Lukin}}]{knaut2024entanglement}%
  \BibitemOpen
  \bibfield  {author} {\bibinfo {author} {\bibfnamefont {C.~M.}\ \bibnamefont {Knaut}}, \bibinfo {author} {\bibfnamefont {A.}~\bibnamefont {Suleymanzade}}, \bibinfo {author} {\bibfnamefont {Y.-C.}\ \bibnamefont {Wei}}, \bibinfo {author} {\bibfnamefont {D.~R.}\ \bibnamefont {Assumpcao}}, \bibinfo {author} {\bibfnamefont {P.-J.}\ \bibnamefont {Stas}}, \bibinfo {author} {\bibfnamefont {Y.~Q.}\ \bibnamefont {Huan}}, \bibinfo {author} {\bibfnamefont {B.}~\bibnamefont {Machielse}}, \bibinfo {author} {\bibfnamefont {E.~N.}\ \bibnamefont {Knall}}, \bibinfo {author} {\bibfnamefont {M.}~\bibnamefont {Sutula}}, \bibinfo {author} {\bibfnamefont {G.}~\bibnamefont {Baranes}}, \bibinfo {author} {\bibfnamefont {N.}~\bibnamefont {Sinclair}}, \bibinfo {author} {\bibfnamefont {C.}~\bibnamefont {{De-Eknamkul}}}, \bibinfo {author} {\bibfnamefont {D.~S.}\ \bibnamefont {Levonian}}, \bibinfo {author} {\bibfnamefont {M.~K.}\ \bibnamefont {Bhaskar}}, \bibinfo {author} {\bibfnamefont {H.}~\bibnamefont {Park}}, \bibinfo {author}
  {\bibfnamefont {M.}~\bibnamefont {Lon{\v c}ar}},\ and\ \bibinfo {author} {\bibfnamefont {M.~D.}\ \bibnamefont {Lukin}},\ }\bibfield  {title} {\bibinfo {title} {Entanglement of nanophotonic quantum memory nodes in a telecom network},\ }\href {https://doi.org/10.1038/s41586-024-07252-z} {\bibfield  {journal} {\bibinfo  {journal} {Nature}\ }\textbf {\bibinfo {volume} {629}},\ \bibinfo {pages} {573} (\bibinfo {year} {2024})}\BibitemShut {NoStop}%
\bibitem [{\citenamefont {Dibos}\ \emph {et~al.}(2018)\citenamefont {Dibos}, \citenamefont {Raha}, \citenamefont {Phenicie},\ and\ \citenamefont {Thompson}}]{dibos2018rei_telecom}%
  \BibitemOpen
  \bibfield  {author} {\bibinfo {author} {\bibfnamefont {A.~M.}\ \bibnamefont {Dibos}}, \bibinfo {author} {\bibfnamefont {M.}~\bibnamefont {Raha}}, \bibinfo {author} {\bibfnamefont {C.~M.}\ \bibnamefont {Phenicie}},\ and\ \bibinfo {author} {\bibfnamefont {J.~D.}\ \bibnamefont {Thompson}},\ }\bibfield  {title} {\bibinfo {title} {Atomic {{Source}} of {{Single Photons}} in the {{Telecom Band}}},\ }\href {https://doi.org/10.1103/PhysRevLett.120.243601} {\bibfield  {journal} {\bibinfo  {journal} {Physical Review Letters}\ }\textbf {\bibinfo {volume} {120}},\ \bibinfo {pages} {243601} (\bibinfo {year} {2018})}\BibitemShut {NoStop}%
\bibitem [{\citenamefont {Ruskuc}\ \emph {et~al.}(2022)\citenamefont {Ruskuc}, \citenamefont {Wu}, \citenamefont {Rochman}, \citenamefont {Choi},\ and\ \citenamefont {Faraon}}]{ruskuc2022nuclear}%
  \BibitemOpen
  \bibfield  {author} {\bibinfo {author} {\bibfnamefont {A.}~\bibnamefont {Ruskuc}}, \bibinfo {author} {\bibfnamefont {C.-J.}\ \bibnamefont {Wu}}, \bibinfo {author} {\bibfnamefont {J.}~\bibnamefont {Rochman}}, \bibinfo {author} {\bibfnamefont {J.}~\bibnamefont {Choi}},\ and\ \bibinfo {author} {\bibfnamefont {A.}~\bibnamefont {Faraon}},\ }\bibfield  {title} {\bibinfo {title} {Nuclear spin-wave quantum register for a solid-state qubit},\ }\href {https://doi.org/10.1038/s41586-021-04293-6} {\bibfield  {journal} {\bibinfo  {journal} {Nature}\ }\textbf {\bibinfo {volume} {602}},\ \bibinfo {pages} {408} (\bibinfo {year} {2022})}\BibitemShut {NoStop}%
\bibitem [{\citenamefont {Uysal}\ \emph {et~al.}(2023)\citenamefont {Uysal}, \citenamefont {Raha}, \citenamefont {Chen}, \citenamefont {Phenicie}, \citenamefont {Ourari}, \citenamefont {Wang}, \citenamefont {{Van de Walle}}, \citenamefont {Dobrovitski},\ and\ \citenamefont {Thompson}}]{uysal2023rei_nuclear}%
  \BibitemOpen
  \bibfield  {author} {\bibinfo {author} {\bibfnamefont {M.~T.}\ \bibnamefont {Uysal}}, \bibinfo {author} {\bibfnamefont {M.}~\bibnamefont {Raha}}, \bibinfo {author} {\bibfnamefont {S.}~\bibnamefont {Chen}}, \bibinfo {author} {\bibfnamefont {C.~M.}\ \bibnamefont {Phenicie}}, \bibinfo {author} {\bibfnamefont {S.}~\bibnamefont {Ourari}}, \bibinfo {author} {\bibfnamefont {M.}~\bibnamefont {Wang}}, \bibinfo {author} {\bibfnamefont {C.~G.}\ \bibnamefont {{Van de Walle}}}, \bibinfo {author} {\bibfnamefont {V.~V.}\ \bibnamefont {Dobrovitski}},\ and\ \bibinfo {author} {\bibfnamefont {J.~D.}\ \bibnamefont {Thompson}},\ }\bibfield  {title} {\bibinfo {title} {Coherent {{Control}} of a {{Nuclear Spin}} via {{Interactions}} with a {{Rare-Earth Ion}} in the {{Solid State}}},\ }\href {https://doi.org/10.1103/PRXQuantum.4.010323} {\bibfield  {journal} {\bibinfo  {journal} {PRX Quantum}\ }\textbf {\bibinfo {volume} {4}},\ \bibinfo {pages} {010323} (\bibinfo {year} {2023})}\BibitemShut {NoStop}%
\bibitem [{\citenamefont {Ruskuc}\ \emph {et~al.}(2024)\citenamefont {Ruskuc}, \citenamefont {Wu}, \citenamefont {Green}, \citenamefont {Hermans}, \citenamefont {Choi},\ and\ \citenamefont {Faraon}}]{ruskuc2024scalable}%
  \BibitemOpen
  \bibfield  {author} {\bibinfo {author} {\bibfnamefont {A.}~\bibnamefont {Ruskuc}}, \bibinfo {author} {\bibfnamefont {C.-J.}\ \bibnamefont {Wu}}, \bibinfo {author} {\bibfnamefont {E.}~\bibnamefont {Green}}, \bibinfo {author} {\bibfnamefont {S.~L.~N.}\ \bibnamefont {Hermans}}, \bibinfo {author} {\bibfnamefont {J.}~\bibnamefont {Choi}},\ and\ \bibinfo {author} {\bibfnamefont {A.}~\bibnamefont {Faraon}},\ }\href@noop {} {\bibinfo {title} {Scalable {{Multipartite Entanglement}} of {{Remote Rare-earth Ion Qubits}}}} (\bibinfo {year} {2024}),\ \Eprint {https://arxiv.org/abs/2402.16224} {arXiv:2402.16224} \BibitemShut {NoStop}%
\bibitem [{\citenamefont {Fuchs}\ \emph {et~al.}(2009)\citenamefont {Fuchs}, \citenamefont {Dobrovitski}, \citenamefont {Toyli}, \citenamefont {Heremans},\ and\ \citenamefont {Awschalom}}]{fuchs2009gigahertz}%
  \BibitemOpen
  \bibfield  {author} {\bibinfo {author} {\bibfnamefont {G.~D.}\ \bibnamefont {Fuchs}}, \bibinfo {author} {\bibfnamefont {V.~V.}\ \bibnamefont {Dobrovitski}}, \bibinfo {author} {\bibfnamefont {D.~M.}\ \bibnamefont {Toyli}}, \bibinfo {author} {\bibfnamefont {F.~J.}\ \bibnamefont {Heremans}},\ and\ \bibinfo {author} {\bibfnamefont {D.~D.}\ \bibnamefont {Awschalom}},\ }\bibfield  {title} {\bibinfo {title} {Gigahertz {{Dynamics}} of a {{Strongly Driven Single Quantum Spin}}},\ }\href {https://doi.org/10.1126/science.1181193} {\bibfield  {journal} {\bibinfo  {journal} {Science}\ }\textbf {\bibinfo {volume} {326}},\ \bibinfo {pages} {1520} (\bibinfo {year} {2009})}\BibitemShut {NoStop}%
\bibitem [{\citenamefont {Christle}\ \emph {et~al.}(2015)\citenamefont {Christle}, \citenamefont {Falk}, \citenamefont {Andrich}, \citenamefont {Klimov}, \citenamefont {Hassan}, \citenamefont {Son}, \citenamefont {Janz{\'e}n}, \citenamefont {Ohshima},\ and\ \citenamefont {Awschalom}}]{christle2015isolated}%
  \BibitemOpen
  \bibfield  {author} {\bibinfo {author} {\bibfnamefont {D.~J.}\ \bibnamefont {Christle}}, \bibinfo {author} {\bibfnamefont {A.~L.}\ \bibnamefont {Falk}}, \bibinfo {author} {\bibfnamefont {P.}~\bibnamefont {Andrich}}, \bibinfo {author} {\bibfnamefont {P.~V.}\ \bibnamefont {Klimov}}, \bibinfo {author} {\bibfnamefont {J.~U.}\ \bibnamefont {Hassan}}, \bibinfo {author} {\bibfnamefont {N.~T.}\ \bibnamefont {Son}}, \bibinfo {author} {\bibfnamefont {E.}~\bibnamefont {Janz{\'e}n}}, \bibinfo {author} {\bibfnamefont {T.}~\bibnamefont {Ohshima}},\ and\ \bibinfo {author} {\bibfnamefont {D.~D.}\ \bibnamefont {Awschalom}},\ }\bibfield  {title} {\bibinfo {title} {Isolated electron spins in silicon carbide with millisecond coherence times},\ }\href {https://doi.org/10.1038/nmat4144} {\bibfield  {journal} {\bibinfo  {journal} {Nature Materials}\ }\textbf {\bibinfo {volume} {14}},\ \bibinfo {pages} {160} (\bibinfo {year} {2015})}\BibitemShut {NoStop}%
\bibitem [{\citenamefont {Sukachev}\ \emph {et~al.}(2017)\citenamefont {Sukachev}, \citenamefont {Sipahigil}, \citenamefont {Nguyen}, \citenamefont {Bhaskar}, \citenamefont {Evans}, \citenamefont {Jelezko},\ and\ \citenamefont {Lukin}}]{sukachev2017siliconvacancy}%
  \BibitemOpen
  \bibfield  {author} {\bibinfo {author} {\bibfnamefont {D.~D.}\ \bibnamefont {Sukachev}}, \bibinfo {author} {\bibfnamefont {A.}~\bibnamefont {Sipahigil}}, \bibinfo {author} {\bibfnamefont {C.~T.}\ \bibnamefont {Nguyen}}, \bibinfo {author} {\bibfnamefont {M.~K.}\ \bibnamefont {Bhaskar}}, \bibinfo {author} {\bibfnamefont {R.~E.}\ \bibnamefont {Evans}}, \bibinfo {author} {\bibfnamefont {F.}~\bibnamefont {Jelezko}},\ and\ \bibinfo {author} {\bibfnamefont {M.~D.}\ \bibnamefont {Lukin}},\ }\bibfield  {title} {\bibinfo {title} {Silicon-{{Vacancy Spin Qubit}} in {{Diamond}}: {{A Quantum Memory Exceeding}} 10 ms with {{Single-Shot State Readout}}},\ }\href {https://doi.org/10.1103/PhysRevLett.119.223602} {\bibfield  {journal} {\bibinfo  {journal} {Physical Review Letters}\ }\textbf {\bibinfo {volume} {119}},\ \bibinfo {pages} {223602} (\bibinfo {year} {2017})}\BibitemShut {NoStop}%
\bibitem [{\citenamefont {Robledo}\ \emph {et~al.}(2011)\citenamefont {Robledo}, \citenamefont {Childress}, \citenamefont {Bernien}, \citenamefont {Hensen}, \citenamefont {Alkemade},\ and\ \citenamefont {Hanson}}]{robledo2011highfidelity}%
  \BibitemOpen
  \bibfield  {author} {\bibinfo {author} {\bibfnamefont {L.}~\bibnamefont {Robledo}}, \bibinfo {author} {\bibfnamefont {L.}~\bibnamefont {Childress}}, \bibinfo {author} {\bibfnamefont {H.}~\bibnamefont {Bernien}}, \bibinfo {author} {\bibfnamefont {B.}~\bibnamefont {Hensen}}, \bibinfo {author} {\bibfnamefont {P.~F.~A.}\ \bibnamefont {Alkemade}},\ and\ \bibinfo {author} {\bibfnamefont {R.}~\bibnamefont {Hanson}},\ }\bibfield  {title} {\bibinfo {title} {High-fidelity projective read-out of a solid-state spin quantum register},\ }\href {https://doi.org/10.1038/nature10401} {\bibfield  {journal} {\bibinfo  {journal} {Nature}\ }\textbf {\bibinfo {volume} {477}},\ \bibinfo {pages} {574} (\bibinfo {year} {2011})}\BibitemShut {NoStop}%
\bibitem [{\citenamefont {Kindem}\ \emph {et~al.}(2020)\citenamefont {Kindem}, \citenamefont {Ruskuc}, \citenamefont {Bartholomew}, \citenamefont {Rochman}, \citenamefont {Huan},\ and\ \citenamefont {Faraon}}]{kindem2020control}%
  \BibitemOpen
  \bibfield  {author} {\bibinfo {author} {\bibfnamefont {J.~M.}\ \bibnamefont {Kindem}}, \bibinfo {author} {\bibfnamefont {A.}~\bibnamefont {Ruskuc}}, \bibinfo {author} {\bibfnamefont {J.~G.}\ \bibnamefont {Bartholomew}}, \bibinfo {author} {\bibfnamefont {J.}~\bibnamefont {Rochman}}, \bibinfo {author} {\bibfnamefont {Y.~Q.}\ \bibnamefont {Huan}},\ and\ \bibinfo {author} {\bibfnamefont {A.}~\bibnamefont {Faraon}},\ }\bibfield  {title} {\bibinfo {title} {Control and single-shot readout of an ion embedded in a nanophotonic cavity},\ }\href {https://doi.org/10.1038/s41586-020-2160-9} {\bibfield  {journal} {\bibinfo  {journal} {Nature}\ }\textbf {\bibinfo {volume} {580}},\ \bibinfo {pages} {201} (\bibinfo {year} {2020})}\BibitemShut {NoStop}%
\bibitem [{\citenamefont {Bhaskar}\ \emph {et~al.}(2020)\citenamefont {Bhaskar}, \citenamefont {Riedinger}, \citenamefont {Machielse}, \citenamefont {Levonian}, \citenamefont {Nguyen}, \citenamefont {Knall}, \citenamefont {Park}, \citenamefont {Englund}, \citenamefont {Lon{\v c}ar}, \citenamefont {Sukachev},\ and\ \citenamefont {Lukin}}]{bhaskar2020communication}%
  \BibitemOpen
  \bibfield  {author} {\bibinfo {author} {\bibfnamefont {M.~K.}\ \bibnamefont {Bhaskar}}, \bibinfo {author} {\bibfnamefont {R.}~\bibnamefont {Riedinger}}, \bibinfo {author} {\bibfnamefont {B.}~\bibnamefont {Machielse}}, \bibinfo {author} {\bibfnamefont {D.~S.}\ \bibnamefont {Levonian}}, \bibinfo {author} {\bibfnamefont {C.~T.}\ \bibnamefont {Nguyen}}, \bibinfo {author} {\bibfnamefont {E.~N.}\ \bibnamefont {Knall}}, \bibinfo {author} {\bibfnamefont {H.}~\bibnamefont {Park}}, \bibinfo {author} {\bibfnamefont {D.}~\bibnamefont {Englund}}, \bibinfo {author} {\bibfnamefont {M.}~\bibnamefont {Lon{\v c}ar}}, \bibinfo {author} {\bibfnamefont {D.~D.}\ \bibnamefont {Sukachev}},\ and\ \bibinfo {author} {\bibfnamefont {M.~D.}\ \bibnamefont {Lukin}},\ }\bibfield  {title} {\bibinfo {title} {Experimental demonstration of memory-enhanced quantum communication},\ }\href {https://doi.org/10.1038/s41586-020-2103-5} {\bibfield  {journal} {\bibinfo  {journal} {Nature}\ }\textbf {\bibinfo {volume} {580}},\ \bibinfo {pages} {60}
  (\bibinfo {year} {2020})}\BibitemShut {NoStop}%
\bibitem [{\citenamefont {Beukers}\ \emph {et~al.}(2024{\natexlab{a}})\citenamefont {Beukers}, \citenamefont {Pasini}, \citenamefont {Choi}, \citenamefont {Englund}, \citenamefont {Hanson},\ and\ \citenamefont {Borregaard}}]{beukers2024remoteentanglement}%
  \BibitemOpen
  \bibfield  {author} {\bibinfo {author} {\bibfnamefont {H.~K.~C.}\ \bibnamefont {Beukers}}, \bibinfo {author} {\bibfnamefont {M.}~\bibnamefont {Pasini}}, \bibinfo {author} {\bibfnamefont {H.}~\bibnamefont {Choi}}, \bibinfo {author} {\bibfnamefont {D.}~\bibnamefont {Englund}}, \bibinfo {author} {\bibfnamefont {R.}~\bibnamefont {Hanson}},\ and\ \bibinfo {author} {\bibfnamefont {J.}~\bibnamefont {Borregaard}},\ }\bibfield  {title} {\bibinfo {title} {Remote-{{Entanglement Protocols}} for {{Stationary Qubits}} with {{Photonic Interfaces}}},\ }\href {https://doi.org/10.1103/PRXQuantum.5.010202} {\bibfield  {journal} {\bibinfo  {journal} {PRX Quantum}\ }\textbf {\bibinfo {volume} {5}},\ \bibinfo {pages} {010202} (\bibinfo {year} {2024}{\natexlab{a}})}\BibitemShut {NoStop}%
\bibitem [{\citenamefont {Wan}\ \emph {et~al.}(2020)\citenamefont {Wan}, \citenamefont {Lu}, \citenamefont {Chen}, \citenamefont {Walsh}, \citenamefont {Trusheim}, \citenamefont {De~Santis}, \citenamefont {Bersin}, \citenamefont {Harris}, \citenamefont {Mouradian}, \citenamefont {Christen}, \citenamefont {Bielejec},\ and\ \citenamefont {Englund}}]{wan2020largescale}%
  \BibitemOpen
  \bibfield  {author} {\bibinfo {author} {\bibfnamefont {N.~H.}\ \bibnamefont {Wan}}, \bibinfo {author} {\bibfnamefont {T.-J.}\ \bibnamefont {Lu}}, \bibinfo {author} {\bibfnamefont {K.~C.}\ \bibnamefont {Chen}}, \bibinfo {author} {\bibfnamefont {M.~P.}\ \bibnamefont {Walsh}}, \bibinfo {author} {\bibfnamefont {M.~E.}\ \bibnamefont {Trusheim}}, \bibinfo {author} {\bibfnamefont {L.}~\bibnamefont {De~Santis}}, \bibinfo {author} {\bibfnamefont {E.~A.}\ \bibnamefont {Bersin}}, \bibinfo {author} {\bibfnamefont {I.~B.}\ \bibnamefont {Harris}}, \bibinfo {author} {\bibfnamefont {S.~L.}\ \bibnamefont {Mouradian}}, \bibinfo {author} {\bibfnamefont {I.~R.}\ \bibnamefont {Christen}}, \bibinfo {author} {\bibfnamefont {E.~S.}\ \bibnamefont {Bielejec}},\ and\ \bibinfo {author} {\bibfnamefont {D.}~\bibnamefont {Englund}},\ }\bibfield  {title} {\bibinfo {title} {Large-scale integration of artificial atoms in hybrid photonic circuits},\ }\href {https://doi.org/10.1038/s41586-020-2441-3} {\bibfield  {journal} {\bibinfo  {journal}
  {Nature}\ }\textbf {\bibinfo {volume} {583}},\ \bibinfo {pages} {226} (\bibinfo {year} {2020})}\BibitemShut {NoStop}%
\bibitem [{\citenamefont {Abobeih}\ \emph {et~al.}(2022)\citenamefont {Abobeih}, \citenamefont {Wang}, \citenamefont {Randall}, \citenamefont {Loenen}, \citenamefont {Bradley}, \citenamefont {Markham}, \citenamefont {Twitchen}, \citenamefont {Terhal},\ and\ \citenamefont {Taminiau}}]{abobeih2022faulttolerant}%
  \BibitemOpen
  \bibfield  {author} {\bibinfo {author} {\bibfnamefont {M.~H.}\ \bibnamefont {Abobeih}}, \bibinfo {author} {\bibfnamefont {Y.}~\bibnamefont {Wang}}, \bibinfo {author} {\bibfnamefont {J.}~\bibnamefont {Randall}}, \bibinfo {author} {\bibfnamefont {S.~J.~H.}\ \bibnamefont {Loenen}}, \bibinfo {author} {\bibfnamefont {C.~E.}\ \bibnamefont {Bradley}}, \bibinfo {author} {\bibfnamefont {M.}~\bibnamefont {Markham}}, \bibinfo {author} {\bibfnamefont {D.~J.}\ \bibnamefont {Twitchen}}, \bibinfo {author} {\bibfnamefont {B.~M.}\ \bibnamefont {Terhal}},\ and\ \bibinfo {author} {\bibfnamefont {T.~H.}\ \bibnamefont {Taminiau}},\ }\bibfield  {title} {\bibinfo {title} {Fault-tolerant operation of a logical qubit in a diamond quantum processor},\ }\href {https://doi.org/10.1038/s41586-022-04819-6} {\bibfield  {journal} {\bibinfo  {journal} {Nature}\ }\textbf {\bibinfo {volume} {606}},\ \bibinfo {pages} {884} (\bibinfo {year} {2022})}\BibitemShut {NoStop}%
\bibitem [{\citenamefont {Pompili}\ \emph {et~al.}(2021)\citenamefont {Pompili}, \citenamefont {Hermans}, \citenamefont {Baier}, \citenamefont {Beukers}, \citenamefont {Humphreys}, \citenamefont {Schouten}, \citenamefont {Vermeulen}, \citenamefont {Tiggelman}, \citenamefont {Martins}, \citenamefont {Dirkse}, \citenamefont {Wehner},\ and\ \citenamefont {Hanson}}]{pompili2021multinode}%
  \BibitemOpen
  \bibfield  {author} {\bibinfo {author} {\bibfnamefont {M.}~\bibnamefont {Pompili}}, \bibinfo {author} {\bibfnamefont {S.~L.~N.}\ \bibnamefont {Hermans}}, \bibinfo {author} {\bibfnamefont {S.}~\bibnamefont {Baier}}, \bibinfo {author} {\bibfnamefont {H.~K.~C.}\ \bibnamefont {Beukers}}, \bibinfo {author} {\bibfnamefont {P.~C.}\ \bibnamefont {Humphreys}}, \bibinfo {author} {\bibfnamefont {R.~N.}\ \bibnamefont {Schouten}}, \bibinfo {author} {\bibfnamefont {R.~F.~L.}\ \bibnamefont {Vermeulen}}, \bibinfo {author} {\bibfnamefont {M.~J.}\ \bibnamefont {Tiggelman}}, \bibinfo {author} {\bibfnamefont {L.~d.~S.}\ \bibnamefont {Martins}}, \bibinfo {author} {\bibfnamefont {B.}~\bibnamefont {Dirkse}}, \bibinfo {author} {\bibfnamefont {S.}~\bibnamefont {Wehner}},\ and\ \bibinfo {author} {\bibfnamefont {R.}~\bibnamefont {Hanson}},\ }\bibfield  {title} {\bibinfo {title} {Realization of a multinode quantum network of remote solid-state qubits},\ }\href {https://doi.org/10.1126/science.abg1919} {\bibfield  {journal} {\bibinfo
  {journal} {Science}\ }\textbf {\bibinfo {volume} {372}},\ \bibinfo {pages} {259} (\bibinfo {year} {2021})}\BibitemShut {NoStop}%
\bibitem [{\citenamefont {Hermans}\ \emph {et~al.}(2023)\citenamefont {Hermans}, \citenamefont {Pompili}, \citenamefont {Martins}, \citenamefont {Montblanch}, \citenamefont {Beukers}, \citenamefont {Baier}, \citenamefont {Borregaard},\ and\ \citenamefont {Hanson}}]{hermans2023entangling}%
  \BibitemOpen
  \bibfield  {author} {\bibinfo {author} {\bibfnamefont {S.~L.~N.}\ \bibnamefont {Hermans}}, \bibinfo {author} {\bibfnamefont {M.}~\bibnamefont {Pompili}}, \bibinfo {author} {\bibfnamefont {L.~D.~S.}\ \bibnamefont {Martins}}, \bibinfo {author} {\bibfnamefont {A.~R.-P.}\ \bibnamefont {Montblanch}}, \bibinfo {author} {\bibfnamefont {H.~K.~C.}\ \bibnamefont {Beukers}}, \bibinfo {author} {\bibfnamefont {S.}~\bibnamefont {Baier}}, \bibinfo {author} {\bibfnamefont {J.}~\bibnamefont {Borregaard}},\ and\ \bibinfo {author} {\bibfnamefont {R.}~\bibnamefont {Hanson}},\ }\bibfield  {title} {\bibinfo {title} {Entangling remote qubits using the single-photon protocol: An in-depth theoretical and experimental study},\ }\href {https://doi.org/10.1088/1367-2630/acb004} {\bibfield  {journal} {\bibinfo  {journal} {New Journal of Physics}\ }\textbf {\bibinfo {volume} {25}},\ \bibinfo {pages} {013011} (\bibinfo {year} {2023})}\BibitemShut {NoStop}%
\bibitem [{\citenamefont {{van der Sar}}\ \emph {et~al.}(2012)\citenamefont {{van der Sar}}, \citenamefont {Wang}, \citenamefont {Blok}, \citenamefont {Bernien}, \citenamefont {Taminiau}, \citenamefont {Toyli}, \citenamefont {Lidar}, \citenamefont {Awschalom}, \citenamefont {Hanson},\ and\ \citenamefont {Dobrovitski}}]{vandersar2012decoherenceprotected}%
  \BibitemOpen
  \bibfield  {author} {\bibinfo {author} {\bibfnamefont {T.}~\bibnamefont {{van der Sar}}}, \bibinfo {author} {\bibfnamefont {Z.~H.}\ \bibnamefont {Wang}}, \bibinfo {author} {\bibfnamefont {M.~S.}\ \bibnamefont {Blok}}, \bibinfo {author} {\bibfnamefont {H.}~\bibnamefont {Bernien}}, \bibinfo {author} {\bibfnamefont {T.~H.}\ \bibnamefont {Taminiau}}, \bibinfo {author} {\bibfnamefont {D.~M.}\ \bibnamefont {Toyli}}, \bibinfo {author} {\bibfnamefont {D.~A.}\ \bibnamefont {Lidar}}, \bibinfo {author} {\bibfnamefont {D.~D.}\ \bibnamefont {Awschalom}}, \bibinfo {author} {\bibfnamefont {R.}~\bibnamefont {Hanson}},\ and\ \bibinfo {author} {\bibfnamefont {V.~V.}\ \bibnamefont {Dobrovitski}},\ }\bibfield  {title} {\bibinfo {title} {Decoherence-protected quantum gates for a hybrid solid-state spin register},\ }\href {https://doi.org/10.1038/nature10900} {\bibfield  {journal} {\bibinfo  {journal} {Nature}\ }\textbf {\bibinfo {volume} {484}},\ \bibinfo {pages} {82} (\bibinfo {year} {2012})}\BibitemShut {NoStop}%
\bibitem [{\citenamefont {Taminiau}\ \emph {et~al.}(2012)\citenamefont {Taminiau}, \citenamefont {Wagenaar}, \citenamefont {{van der Sar}}, \citenamefont {Jelezko}, \citenamefont {Dobrovitski},\ and\ \citenamefont {Hanson}}]{taminiau2012detection}%
  \BibitemOpen
  \bibfield  {author} {\bibinfo {author} {\bibfnamefont {T.~H.}\ \bibnamefont {Taminiau}}, \bibinfo {author} {\bibfnamefont {J.~J.~T.}\ \bibnamefont {Wagenaar}}, \bibinfo {author} {\bibfnamefont {T.}~\bibnamefont {{van der Sar}}}, \bibinfo {author} {\bibfnamefont {F.}~\bibnamefont {Jelezko}}, \bibinfo {author} {\bibfnamefont {V.~V.}\ \bibnamefont {Dobrovitski}},\ and\ \bibinfo {author} {\bibfnamefont {R.}~\bibnamefont {Hanson}},\ }\bibfield  {title} {\bibinfo {title} {Detection and {{Control}} of {{Individual Nuclear Spins Using}} a {{Weakly Coupled Electron Spin}}},\ }\href {https://doi.org/10.1103/PhysRevLett.109.137602} {\bibfield  {journal} {\bibinfo  {journal} {Physical Review Letters}\ }\textbf {\bibinfo {volume} {109}},\ \bibinfo {pages} {137602} (\bibinfo {year} {2012})}\BibitemShut {NoStop}%
\bibitem [{\citenamefont {Bradley}\ \emph {et~al.}(2019)\citenamefont {Bradley}, \citenamefont {Randall}, \citenamefont {Abobeih}, \citenamefont {Berrevoets}, \citenamefont {Degen}, \citenamefont {Bakker}, \citenamefont {Markham}, \citenamefont {Twitchen},\ and\ \citenamefont {Taminiau}}]{bradley2019tenqubit}%
  \BibitemOpen
  \bibfield  {author} {\bibinfo {author} {\bibfnamefont {C.~E.}\ \bibnamefont {Bradley}}, \bibinfo {author} {\bibfnamefont {J.}~\bibnamefont {Randall}}, \bibinfo {author} {\bibfnamefont {M.~H.}\ \bibnamefont {Abobeih}}, \bibinfo {author} {\bibfnamefont {R.~C.}\ \bibnamefont {Berrevoets}}, \bibinfo {author} {\bibfnamefont {M.~J.}\ \bibnamefont {Degen}}, \bibinfo {author} {\bibfnamefont {M.~A.}\ \bibnamefont {Bakker}}, \bibinfo {author} {\bibfnamefont {M.}~\bibnamefont {Markham}}, \bibinfo {author} {\bibfnamefont {D.~J.}\ \bibnamefont {Twitchen}},\ and\ \bibinfo {author} {\bibfnamefont {T.~H.}\ \bibnamefont {Taminiau}},\ }\bibfield  {title} {\bibinfo {title} {A {{Ten-Qubit Solid-State Spin Register}} with {{Quantum Memory}} up to {{One Minute}}},\ }\href {https://doi.org/10.1103/PhysRevX.9.031045} {\bibfield  {journal} {\bibinfo  {journal} {Physical Review X}\ }\textbf {\bibinfo {volume} {9}},\ \bibinfo {pages} {031045} (\bibinfo {year} {2019})}\BibitemShut {NoStop}%
\bibitem [{\citenamefont {Cramer}\ \emph {et~al.}(2016)\citenamefont {Cramer}, \citenamefont {Kalb}, \citenamefont {Rol}, \citenamefont {Hensen}, \citenamefont {Blok}, \citenamefont {Markham}, \citenamefont {Twitchen}, \citenamefont {Hanson},\ and\ \citenamefont {Taminiau}}]{cramer2016error}%
  \BibitemOpen
  \bibfield  {author} {\bibinfo {author} {\bibfnamefont {J.}~\bibnamefont {Cramer}}, \bibinfo {author} {\bibfnamefont {N.}~\bibnamefont {Kalb}}, \bibinfo {author} {\bibfnamefont {M.~A.}\ \bibnamefont {Rol}}, \bibinfo {author} {\bibfnamefont {B.}~\bibnamefont {Hensen}}, \bibinfo {author} {\bibfnamefont {M.~S.}\ \bibnamefont {Blok}}, \bibinfo {author} {\bibfnamefont {M.}~\bibnamefont {Markham}}, \bibinfo {author} {\bibfnamefont {D.~J.}\ \bibnamefont {Twitchen}}, \bibinfo {author} {\bibfnamefont {R.}~\bibnamefont {Hanson}},\ and\ \bibinfo {author} {\bibfnamefont {T.~H.}\ \bibnamefont {Taminiau}},\ }\bibfield  {title} {\bibinfo {title} {Repeated quantum error correction on a continuously encoded qubit by real-time feedback},\ }\href {https://doi.org/10.1038/ncomms11526} {\bibfield  {journal} {\bibinfo  {journal} {Nature Communications}\ }\textbf {\bibinfo {volume} {7}},\ \bibinfo {pages} {11526} (\bibinfo {year} {2016})}\BibitemShut {NoStop}%
\bibitem [{\citenamefont {Maity}\ \emph {et~al.}(2022)\citenamefont {Maity}, \citenamefont {Pingault}, \citenamefont {Joe}, \citenamefont {Chalupnik}, \citenamefont {Assump{\c c}{\~a}o}, \citenamefont {Cornell}, \citenamefont {Shao},\ and\ \citenamefont {Lon{\v c}ar}}]{maity2022nuclear}%
  \BibitemOpen
  \bibfield  {author} {\bibinfo {author} {\bibfnamefont {S.}~\bibnamefont {Maity}}, \bibinfo {author} {\bibfnamefont {B.}~\bibnamefont {Pingault}}, \bibinfo {author} {\bibfnamefont {G.}~\bibnamefont {Joe}}, \bibinfo {author} {\bibfnamefont {M.}~\bibnamefont {Chalupnik}}, \bibinfo {author} {\bibfnamefont {D.}~\bibnamefont {Assump{\c c}{\~a}o}}, \bibinfo {author} {\bibfnamefont {E.}~\bibnamefont {Cornell}}, \bibinfo {author} {\bibfnamefont {L.}~\bibnamefont {Shao}},\ and\ \bibinfo {author} {\bibfnamefont {M.}~\bibnamefont {Lon{\v c}ar}},\ }\bibfield  {title} {\bibinfo {title} {Mechanical {{Control}} of a {{Single Nuclear Spin}}},\ }\href {https://doi.org/10.1103/PhysRevX.12.011056} {\bibfield  {journal} {\bibinfo  {journal} {Physical Review X}\ }\textbf {\bibinfo {volume} {12}},\ \bibinfo {pages} {011056} (\bibinfo {year} {2022})}\BibitemShut {NoStop}%
\bibitem [{\citenamefont {Nguyen}\ \emph {et~al.}(2019{\natexlab{a}})\citenamefont {Nguyen}, \citenamefont {Sukachev}, \citenamefont {Bhaskar}, \citenamefont {Machielse}, \citenamefont {Levonian}, \citenamefont {Knall}, \citenamefont {Stroganov}, \citenamefont {Riedinger}, \citenamefont {Park}, \citenamefont {Lon{\v c}ar},\ and\ \citenamefont {Lukin}}]{nguyen2019prl}%
  \BibitemOpen
  \bibfield  {author} {\bibinfo {author} {\bibfnamefont {C.~T.}\ \bibnamefont {Nguyen}}, \bibinfo {author} {\bibfnamefont {D.~D.}\ \bibnamefont {Sukachev}}, \bibinfo {author} {\bibfnamefont {M.~K.}\ \bibnamefont {Bhaskar}}, \bibinfo {author} {\bibfnamefont {B.}~\bibnamefont {Machielse}}, \bibinfo {author} {\bibfnamefont {D.~S.}\ \bibnamefont {Levonian}}, \bibinfo {author} {\bibfnamefont {E.~N.}\ \bibnamefont {Knall}}, \bibinfo {author} {\bibfnamefont {P.}~\bibnamefont {Stroganov}}, \bibinfo {author} {\bibfnamefont {R.}~\bibnamefont {Riedinger}}, \bibinfo {author} {\bibfnamefont {H.}~\bibnamefont {Park}}, \bibinfo {author} {\bibfnamefont {M.}~\bibnamefont {Lon{\v c}ar}},\ and\ \bibinfo {author} {\bibfnamefont {M.~D.}\ \bibnamefont {Lukin}},\ }\bibfield  {title} {\bibinfo {title} {Quantum {{Network Nodes Based}} on {{Diamond Qubits}} with an {{Efficient Nanophotonic Interface}}},\ }\href {https://doi.org/10.1103/PhysRevLett.123.183602} {\bibfield  {journal} {\bibinfo  {journal} {Physical Review Letters}\
  }\textbf {\bibinfo {volume} {123}},\ \bibinfo {pages} {183602} (\bibinfo {year} {2019}{\natexlab{a}})}\BibitemShut {NoStop}%
\bibitem [{\citenamefont {Babin}\ \emph {et~al.}(2022)\citenamefont {Babin}, \citenamefont {St{\"o}hr}, \citenamefont {Morioka}, \citenamefont {Linkewitz}, \citenamefont {Steidl}, \citenamefont {W{\"o}rnle}, \citenamefont {Liu}, \citenamefont {Hesselmeier}, \citenamefont {Vorobyov}, \citenamefont {Denisenko}, \citenamefont {Hentschel}, \citenamefont {Gobert}, \citenamefont {Berwian}, \citenamefont {Astakhov}, \citenamefont {Knolle}, \citenamefont {Majety}, \citenamefont {Saha}, \citenamefont {Radulaski}, \citenamefont {Son}, \citenamefont {{Ul-Hassan}}, \citenamefont {Kaiser},\ and\ \citenamefont {Wrachtrup}}]{babin2022fabrication}%
  \BibitemOpen
  \bibfield  {author} {\bibinfo {author} {\bibfnamefont {C.}~\bibnamefont {Babin}}, \bibinfo {author} {\bibfnamefont {R.}~\bibnamefont {St{\"o}hr}}, \bibinfo {author} {\bibfnamefont {N.}~\bibnamefont {Morioka}}, \bibinfo {author} {\bibfnamefont {T.}~\bibnamefont {Linkewitz}}, \bibinfo {author} {\bibfnamefont {T.}~\bibnamefont {Steidl}}, \bibinfo {author} {\bibfnamefont {R.}~\bibnamefont {W{\"o}rnle}}, \bibinfo {author} {\bibfnamefont {D.}~\bibnamefont {Liu}}, \bibinfo {author} {\bibfnamefont {E.}~\bibnamefont {Hesselmeier}}, \bibinfo {author} {\bibfnamefont {V.}~\bibnamefont {Vorobyov}}, \bibinfo {author} {\bibfnamefont {A.}~\bibnamefont {Denisenko}}, \bibinfo {author} {\bibfnamefont {M.}~\bibnamefont {Hentschel}}, \bibinfo {author} {\bibfnamefont {C.}~\bibnamefont {Gobert}}, \bibinfo {author} {\bibfnamefont {P.}~\bibnamefont {Berwian}}, \bibinfo {author} {\bibfnamefont {G.~V.}\ \bibnamefont {Astakhov}}, \bibinfo {author} {\bibfnamefont {W.}~\bibnamefont {Knolle}}, \bibinfo {author} {\bibfnamefont
  {S.}~\bibnamefont {Majety}}, \bibinfo {author} {\bibfnamefont {P.}~\bibnamefont {Saha}}, \bibinfo {author} {\bibfnamefont {M.}~\bibnamefont {Radulaski}}, \bibinfo {author} {\bibfnamefont {N.~T.}\ \bibnamefont {Son}}, \bibinfo {author} {\bibfnamefont {J.}~\bibnamefont {{Ul-Hassan}}}, \bibinfo {author} {\bibfnamefont {F.}~\bibnamefont {Kaiser}},\ and\ \bibinfo {author} {\bibfnamefont {J.}~\bibnamefont {Wrachtrup}},\ }\bibfield  {title} {\bibinfo {title} {Fabrication and nanophotonic waveguide integration of silicon carbide colour centres with preserved spin-optical coherence},\ }\href {https://doi.org/10.1038/s41563-021-01148-3} {\bibfield  {journal} {\bibinfo  {journal} {Nature Materials}\ }\textbf {\bibinfo {volume} {21}},\ \bibinfo {pages} {67} (\bibinfo {year} {2022})}\BibitemShut {NoStop}%
\bibitem [{\citenamefont {Higginbottom}\ \emph {et~al.}(2022)\citenamefont {Higginbottom}, \citenamefont {Kurkjian}, \citenamefont {Chartrand}, \citenamefont {Kazemi}, \citenamefont {Brunelle}, \citenamefont {MacQuarrie}, \citenamefont {Klein}, \citenamefont {{Lee-Hone}}, \citenamefont {Stacho}, \citenamefont {Ruether}, \citenamefont {Bowness}, \citenamefont {Bergeron}, \citenamefont {DeAbreu}, \citenamefont {Harrigan}, \citenamefont {Kanaganayagam}, \citenamefont {Marsden}, \citenamefont {Richards}, \citenamefont {Stott}, \citenamefont {Roorda}, \citenamefont {Morse}, \citenamefont {Thewalt},\ and\ \citenamefont {Simmons}}]{higginbottom2022optical}%
  \BibitemOpen
  \bibfield  {author} {\bibinfo {author} {\bibfnamefont {D.~B.}\ \bibnamefont {Higginbottom}}, \bibinfo {author} {\bibfnamefont {A.~T.~K.}\ \bibnamefont {Kurkjian}}, \bibinfo {author} {\bibfnamefont {C.}~\bibnamefont {Chartrand}}, \bibinfo {author} {\bibfnamefont {M.}~\bibnamefont {Kazemi}}, \bibinfo {author} {\bibfnamefont {N.~A.}\ \bibnamefont {Brunelle}}, \bibinfo {author} {\bibfnamefont {E.~R.}\ \bibnamefont {MacQuarrie}}, \bibinfo {author} {\bibfnamefont {J.~R.}\ \bibnamefont {Klein}}, \bibinfo {author} {\bibfnamefont {N.~R.}\ \bibnamefont {{Lee-Hone}}}, \bibinfo {author} {\bibfnamefont {J.}~\bibnamefont {Stacho}}, \bibinfo {author} {\bibfnamefont {M.}~\bibnamefont {Ruether}}, \bibinfo {author} {\bibfnamefont {C.}~\bibnamefont {Bowness}}, \bibinfo {author} {\bibfnamefont {L.}~\bibnamefont {Bergeron}}, \bibinfo {author} {\bibfnamefont {A.}~\bibnamefont {DeAbreu}}, \bibinfo {author} {\bibfnamefont {S.~R.}\ \bibnamefont {Harrigan}}, \bibinfo {author} {\bibfnamefont {J.}~\bibnamefont {Kanaganayagam}},
  \bibinfo {author} {\bibfnamefont {D.~W.}\ \bibnamefont {Marsden}}, \bibinfo {author} {\bibfnamefont {T.~S.}\ \bibnamefont {Richards}}, \bibinfo {author} {\bibfnamefont {L.~A.}\ \bibnamefont {Stott}}, \bibinfo {author} {\bibfnamefont {S.}~\bibnamefont {Roorda}}, \bibinfo {author} {\bibfnamefont {K.~J.}\ \bibnamefont {Morse}}, \bibinfo {author} {\bibfnamefont {M.~L.~W.}\ \bibnamefont {Thewalt}},\ and\ \bibinfo {author} {\bibfnamefont {S.}~\bibnamefont {Simmons}},\ }\bibfield  {title} {\bibinfo {title} {Optical observation of single spins in silicon},\ }\href {https://doi.org/10.1038/s41586-022-04821-y} {\bibfield  {journal} {\bibinfo  {journal} {Nature}\ }\textbf {\bibinfo {volume} {607}},\ \bibinfo {pages} {266} (\bibinfo {year} {2022})}\BibitemShut {NoStop}%
\bibitem [{\citenamefont {{Photonic Inc}}\ \emph {et~al.}(2024)\citenamefont {{Photonic Inc}}, \citenamefont {Afzal}, \citenamefont {Akhlaghi}, \citenamefont {Beale}, \citenamefont {Bedroya}, \citenamefont {Bell}, \citenamefont {Bergeron}, \citenamefont {{Bonsma-Fisher}}, \citenamefont {Bychkova}, \citenamefont {Chaisson}, \citenamefont {Chartrand}, \citenamefont {Clear}, \citenamefont {Darcie}, \citenamefont {DeAbreu}, \citenamefont {DeLisle},\ and\ \citenamefont {{and Others}}}]{photonicinc2024distributed}%
  \BibitemOpen
  \bibfield  {author} {\bibinfo {author} {\bibnamefont {{Photonic Inc}}}, \bibinfo {author} {\bibfnamefont {F.}~\bibnamefont {Afzal}}, \bibinfo {author} {\bibfnamefont {M.}~\bibnamefont {Akhlaghi}}, \bibinfo {author} {\bibfnamefont {S.~J.}\ \bibnamefont {Beale}}, \bibinfo {author} {\bibfnamefont {O.}~\bibnamefont {Bedroya}}, \bibinfo {author} {\bibfnamefont {K.}~\bibnamefont {Bell}}, \bibinfo {author} {\bibfnamefont {L.}~\bibnamefont {Bergeron}}, \bibinfo {author} {\bibfnamefont {K.}~\bibnamefont {{Bonsma-Fisher}}}, \bibinfo {author} {\bibfnamefont {P.}~\bibnamefont {Bychkova}}, \bibinfo {author} {\bibfnamefont {Z.~M.~E.}\ \bibnamefont {Chaisson}}, \bibinfo {author} {\bibfnamefont {C.}~\bibnamefont {Chartrand}}, \bibinfo {author} {\bibfnamefont {C.}~\bibnamefont {Clear}}, \bibinfo {author} {\bibfnamefont {A.}~\bibnamefont {Darcie}}, \bibinfo {author} {\bibfnamefont {A.}~\bibnamefont {DeAbreu}}, \bibinfo {author} {\bibfnamefont {C.}~\bibnamefont {DeLisle}},\ and\ \bibinfo {author} {\bibnamefont {{and
  Others}}},\ }\href@noop {} {\bibinfo {title} {Distributed {{Quantum Computing}} in {{Silicon}}}} (\bibinfo {year} {2024}),\ \Eprint {https://arxiv.org/abs/2406.01704} {arXiv:2406.01704} \BibitemShut {NoStop}%
\bibitem [{\citenamefont {Nguyen}\ \emph {et~al.}(2019{\natexlab{b}})\citenamefont {Nguyen}, \citenamefont {Sukachev}, \citenamefont {Bhaskar}, \citenamefont {Machielse}, \citenamefont {Levonian}, \citenamefont {Knall}, \citenamefont {Stroganov}, \citenamefont {Chia}, \citenamefont {Burek}, \citenamefont {Riedinger}, \citenamefont {Park}, \citenamefont {Lon{\v c}ar},\ and\ \citenamefont {Lukin}}]{nguyen2019prb}%
  \BibitemOpen
  \bibfield  {author} {\bibinfo {author} {\bibfnamefont {C.~T.}\ \bibnamefont {Nguyen}}, \bibinfo {author} {\bibfnamefont {D.~D.}\ \bibnamefont {Sukachev}}, \bibinfo {author} {\bibfnamefont {M.~K.}\ \bibnamefont {Bhaskar}}, \bibinfo {author} {\bibfnamefont {B.}~\bibnamefont {Machielse}}, \bibinfo {author} {\bibfnamefont {D.~S.}\ \bibnamefont {Levonian}}, \bibinfo {author} {\bibfnamefont {E.~N.}\ \bibnamefont {Knall}}, \bibinfo {author} {\bibfnamefont {P.}~\bibnamefont {Stroganov}}, \bibinfo {author} {\bibfnamefont {C.}~\bibnamefont {Chia}}, \bibinfo {author} {\bibfnamefont {M.~J.}\ \bibnamefont {Burek}}, \bibinfo {author} {\bibfnamefont {R.}~\bibnamefont {Riedinger}}, \bibinfo {author} {\bibfnamefont {H.}~\bibnamefont {Park}}, \bibinfo {author} {\bibfnamefont {M.}~\bibnamefont {Lon{\v c}ar}},\ and\ \bibinfo {author} {\bibfnamefont {M.~D.}\ \bibnamefont {Lukin}},\ }\bibfield  {title} {\bibinfo {title} {An integrated nanophotonic quantum register based on silicon-vacancy spins in diamond},\ }\href
  {https://doi.org/10.1103/PhysRevB.100.165428} {\bibfield  {journal} {\bibinfo  {journal} {Physical Review B}\ }\textbf {\bibinfo {volume} {100}},\ \bibinfo {pages} {165428} (\bibinfo {year} {2019}{\natexlab{b}})}\BibitemShut {NoStop}%
\bibitem [{\citenamefont {Senkalla}\ \emph {et~al.}(2024)\citenamefont {Senkalla}, \citenamefont {Genov}, \citenamefont {Metsch}, \citenamefont {Siyushev},\ and\ \citenamefont {Jelezko}}]{senkalla2024germanium}%
  \BibitemOpen
  \bibfield  {author} {\bibinfo {author} {\bibfnamefont {K.}~\bibnamefont {Senkalla}}, \bibinfo {author} {\bibfnamefont {G.}~\bibnamefont {Genov}}, \bibinfo {author} {\bibfnamefont {M.~H.}\ \bibnamefont {Metsch}}, \bibinfo {author} {\bibfnamefont {P.}~\bibnamefont {Siyushev}},\ and\ \bibinfo {author} {\bibfnamefont {F.}~\bibnamefont {Jelezko}},\ }\bibfield  {title} {\bibinfo {title} {Germanium {{Vacancy}} in {{Diamond Quantum Memory Exceeding}} 20 ms},\ }\href {https://doi.org/10.1103/PhysRevLett.132.026901} {\bibfield  {journal} {\bibinfo  {journal} {Physical Review Letters}\ }\textbf {\bibinfo {volume} {132}},\ \bibinfo {pages} {026901} (\bibinfo {year} {2024})}\BibitemShut {NoStop}%
\bibitem [{\citenamefont {Rosenthal}\ \emph {et~al.}(2023)\citenamefont {Rosenthal}, \citenamefont {Anderson}, \citenamefont {Kleidermacher}, \citenamefont {Stein}, \citenamefont {Lee}, \citenamefont {Grzesik}, \citenamefont {Scuri}, \citenamefont {Rugar}, \citenamefont {Riedel}, \citenamefont {Aghaeimeibodi}, \citenamefont {Ahn}, \citenamefont {Van~Gasse},\ and\ \citenamefont {Vu{\v c}kovi{\'c}}}]{rosenthal2023microwave}%
  \BibitemOpen
  \bibfield  {author} {\bibinfo {author} {\bibfnamefont {E.~I.}\ \bibnamefont {Rosenthal}}, \bibinfo {author} {\bibfnamefont {C.~P.}\ \bibnamefont {Anderson}}, \bibinfo {author} {\bibfnamefont {H.~C.}\ \bibnamefont {Kleidermacher}}, \bibinfo {author} {\bibfnamefont {A.~J.}\ \bibnamefont {Stein}}, \bibinfo {author} {\bibfnamefont {H.}~\bibnamefont {Lee}}, \bibinfo {author} {\bibfnamefont {J.}~\bibnamefont {Grzesik}}, \bibinfo {author} {\bibfnamefont {G.}~\bibnamefont {Scuri}}, \bibinfo {author} {\bibfnamefont {A.~E.}\ \bibnamefont {Rugar}}, \bibinfo {author} {\bibfnamefont {D.}~\bibnamefont {Riedel}}, \bibinfo {author} {\bibfnamefont {S.}~\bibnamefont {Aghaeimeibodi}}, \bibinfo {author} {\bibfnamefont {G.~H.}\ \bibnamefont {Ahn}}, \bibinfo {author} {\bibfnamefont {K.}~\bibnamefont {Van~Gasse}},\ and\ \bibinfo {author} {\bibfnamefont {J.}~\bibnamefont {Vu{\v c}kovi{\'c}}},\ }\bibfield  {title} {\bibinfo {title} {Microwave {{Spin Control}} of a {{Tin-Vacancy Qubit}} in {{Diamond}}},\ }\href
  {https://doi.org/10.1103/PhysRevX.13.031022} {\bibfield  {journal} {\bibinfo  {journal} {Physical Review X}\ }\textbf {\bibinfo {volume} {13}},\ \bibinfo {pages} {031022} (\bibinfo {year} {2023})}\BibitemShut {NoStop}%
\bibitem [{\citenamefont {Wang}\ \emph {et~al.}(2024)\citenamefont {Wang}, \citenamefont {Kazak}, \citenamefont {Senkalla}, \citenamefont {Siyushev}, \citenamefont {Abe}, \citenamefont {Taniguchi}, \citenamefont {Onoda}, \citenamefont {Kato}, \citenamefont {Makino}, \citenamefont {Hatano}, \citenamefont {Jelezko},\ and\ \citenamefont {Iwasaki}}]{wang2024transformlimited}%
  \BibitemOpen
  \bibfield  {author} {\bibinfo {author} {\bibfnamefont {P.}~\bibnamefont {Wang}}, \bibinfo {author} {\bibfnamefont {L.}~\bibnamefont {Kazak}}, \bibinfo {author} {\bibfnamefont {K.}~\bibnamefont {Senkalla}}, \bibinfo {author} {\bibfnamefont {P.}~\bibnamefont {Siyushev}}, \bibinfo {author} {\bibfnamefont {R.}~\bibnamefont {Abe}}, \bibinfo {author} {\bibfnamefont {T.}~\bibnamefont {Taniguchi}}, \bibinfo {author} {\bibfnamefont {S.}~\bibnamefont {Onoda}}, \bibinfo {author} {\bibfnamefont {H.}~\bibnamefont {Kato}}, \bibinfo {author} {\bibfnamefont {T.}~\bibnamefont {Makino}}, \bibinfo {author} {\bibfnamefont {M.}~\bibnamefont {Hatano}}, \bibinfo {author} {\bibfnamefont {F.}~\bibnamefont {Jelezko}},\ and\ \bibinfo {author} {\bibfnamefont {T.}~\bibnamefont {Iwasaki}},\ }\bibfield  {title} {\bibinfo {title} {Transform-{{Limited Photon Emission}} from a {{Lead-Vacancy Center}} in {{Diamond}} above 10 {{K}}},\ }\href {https://doi.org/10.1103/PhysRevLett.132.073601} {\bibfield  {journal} {\bibinfo  {journal} {Physical
  Review Letters}\ }\textbf {\bibinfo {volume} {132}},\ \bibinfo {pages} {073601} (\bibinfo {year} {2024})}\BibitemShut {NoStop}%
\bibitem [{\citenamefont {Zahedian}\ \emph {et~al.}(2024)\citenamefont {Zahedian}, \citenamefont {Vorobyov},\ and\ \citenamefont {Wrachtrup}}]{zahedian2024blueprint}%
  \BibitemOpen
  \bibfield  {author} {\bibinfo {author} {\bibfnamefont {M.}~\bibnamefont {Zahedian}}, \bibinfo {author} {\bibfnamefont {V.}~\bibnamefont {Vorobyov}},\ and\ \bibinfo {author} {\bibfnamefont {J.}~\bibnamefont {Wrachtrup}},\ }\bibfield  {title} {\bibinfo {title} {Blueprint for efficient nuclear spin characterization with color centers},\ }\href {https://doi.org/10.1103/PhysRevB.109.214111} {\bibfield  {journal} {\bibinfo  {journal} {Physical Review B}\ }\textbf {\bibinfo {volume} {109}},\ \bibinfo {pages} {214111} (\bibinfo {year} {2024})}\BibitemShut {NoStop}%
\bibitem [{\citenamefont {Maity}\ \emph {et~al.}(2018)\citenamefont {Maity}, \citenamefont {Shao}, \citenamefont {Sohn}, \citenamefont {Meesala}, \citenamefont {Machielse}, \citenamefont {Bielejec}, \citenamefont {Markham},\ and\ \citenamefont {Lon{\v c}ar}}]{maity2018strain}%
  \BibitemOpen
  \bibfield  {author} {\bibinfo {author} {\bibfnamefont {S.}~\bibnamefont {Maity}}, \bibinfo {author} {\bibfnamefont {L.}~\bibnamefont {Shao}}, \bibinfo {author} {\bibfnamefont {Y.-I.}\ \bibnamefont {Sohn}}, \bibinfo {author} {\bibfnamefont {S.}~\bibnamefont {Meesala}}, \bibinfo {author} {\bibfnamefont {B.}~\bibnamefont {Machielse}}, \bibinfo {author} {\bibfnamefont {E.}~\bibnamefont {Bielejec}}, \bibinfo {author} {\bibfnamefont {M.}~\bibnamefont {Markham}},\ and\ \bibinfo {author} {\bibfnamefont {M.}~\bibnamefont {Lon{\v c}ar}},\ }\bibfield  {title} {\bibinfo {title} {Spectral {{Alignment}} of {{Single-Photon Emitters}} in {{Diamond}} using {{Strain Gradient}}},\ }\href {https://doi.org/10.1103/PhysRevApplied.10.024050} {\bibfield  {journal} {\bibinfo  {journal} {Physical Review Applied}\ }\textbf {\bibinfo {volume} {10}},\ \bibinfo {pages} {024050} (\bibinfo {year} {2018})}\BibitemShut {NoStop}%
\bibitem [{\citenamefont {Ditalia~Tchernij}\ \emph {et~al.}(2017)\citenamefont {Ditalia~Tchernij}, \citenamefont {Herzig}, \citenamefont {Forneris}, \citenamefont {K{\"u}pper}, \citenamefont {Pezzagna}, \citenamefont {Traina}, \citenamefont {Moreva}, \citenamefont {Degiovanni}, \citenamefont {Brida}, \citenamefont {Skukan}, \citenamefont {Genovese}, \citenamefont {Jak{\v s}i{\'c}}, \citenamefont {Meijer},\ and\ \citenamefont {Olivero}}]{ditaliatchernij2017snv}%
  \BibitemOpen
  \bibfield  {author} {\bibinfo {author} {\bibfnamefont {S.}~\bibnamefont {Ditalia~Tchernij}}, \bibinfo {author} {\bibfnamefont {T.}~\bibnamefont {Herzig}}, \bibinfo {author} {\bibfnamefont {J.}~\bibnamefont {Forneris}}, \bibinfo {author} {\bibfnamefont {J.}~\bibnamefont {K{\"u}pper}}, \bibinfo {author} {\bibfnamefont {S.}~\bibnamefont {Pezzagna}}, \bibinfo {author} {\bibfnamefont {P.}~\bibnamefont {Traina}}, \bibinfo {author} {\bibfnamefont {E.}~\bibnamefont {Moreva}}, \bibinfo {author} {\bibfnamefont {I.~P.}\ \bibnamefont {Degiovanni}}, \bibinfo {author} {\bibfnamefont {G.}~\bibnamefont {Brida}}, \bibinfo {author} {\bibfnamefont {N.}~\bibnamefont {Skukan}}, \bibinfo {author} {\bibfnamefont {M.}~\bibnamefont {Genovese}}, \bibinfo {author} {\bibfnamefont {M.}~\bibnamefont {Jak{\v s}i{\'c}}}, \bibinfo {author} {\bibfnamefont {J.}~\bibnamefont {Meijer}},\ and\ \bibinfo {author} {\bibfnamefont {P.}~\bibnamefont {Olivero}},\ }\bibfield  {title} {\bibinfo {title} {Single-{{Photon-Emitting Optical Centers}} in
  {{Diamond Fabricated}} upon {{Sn Implantation}}},\ }\href {https://doi.org/10.1021/acsphotonics.7b00904} {\bibfield  {journal} {\bibinfo  {journal} {ACS Photonics}\ }\textbf {\bibinfo {volume} {4}},\ \bibinfo {pages} {2580} (\bibinfo {year} {2017})}\BibitemShut {NoStop}%
\bibitem [{\citenamefont {Iwasaki}\ \emph {et~al.}(2017)\citenamefont {Iwasaki}, \citenamefont {Miyamoto}, \citenamefont {Taniguchi}, \citenamefont {Siyushev}, \citenamefont {Metsch}, \citenamefont {Jelezko},\ and\ \citenamefont {Hatano}}]{iwasaki2017tinvacancy}%
  \BibitemOpen
  \bibfield  {author} {\bibinfo {author} {\bibfnamefont {T.}~\bibnamefont {Iwasaki}}, \bibinfo {author} {\bibfnamefont {Y.}~\bibnamefont {Miyamoto}}, \bibinfo {author} {\bibfnamefont {T.}~\bibnamefont {Taniguchi}}, \bibinfo {author} {\bibfnamefont {P.}~\bibnamefont {Siyushev}}, \bibinfo {author} {\bibfnamefont {M.~H.}\ \bibnamefont {Metsch}}, \bibinfo {author} {\bibfnamefont {F.}~\bibnamefont {Jelezko}},\ and\ \bibinfo {author} {\bibfnamefont {M.}~\bibnamefont {Hatano}},\ }\bibfield  {title} {\bibinfo {title} {Tin-{{Vacancy Quantum Emitters}} in {{Diamond}}},\ }\href {https://doi.org/10.1103/PhysRevLett.119.253601} {\bibfield  {journal} {\bibinfo  {journal} {Physical Review Letters}\ }\textbf {\bibinfo {volume} {119}},\ \bibinfo {pages} {253601} (\bibinfo {year} {2017})}\BibitemShut {NoStop}%
\bibitem [{\citenamefont {Trusheim}\ \emph {et~al.}(2020)\citenamefont {Trusheim}, \citenamefont {Pingault}, \citenamefont {Wan}, \citenamefont {G{\"u}ndo{\u g}an}, \citenamefont {De~Santis}, \citenamefont {Debroux}, \citenamefont {Gangloff}, \citenamefont {Purser}, \citenamefont {Chen}, \citenamefont {Walsh}, \citenamefont {Rose}, \citenamefont {Becker}, \citenamefont {Lienhard}, \citenamefont {Bersin}, \citenamefont {Paradeisanos}, \citenamefont {Wang}, \citenamefont {Lyzwa}, \citenamefont {Montblanch}, \citenamefont {Malladi}, \citenamefont {Bakhru}, \citenamefont {Ferrari}, \citenamefont {Walmsley}, \citenamefont {Atat{\"u}re},\ and\ \citenamefont {Englund}}]{trusheim2020transformlimited}%
  \BibitemOpen
  \bibfield  {author} {\bibinfo {author} {\bibfnamefont {M.~E.}\ \bibnamefont {Trusheim}}, \bibinfo {author} {\bibfnamefont {B.}~\bibnamefont {Pingault}}, \bibinfo {author} {\bibfnamefont {N.~H.}\ \bibnamefont {Wan}}, \bibinfo {author} {\bibfnamefont {M.}~\bibnamefont {G{\"u}ndo{\u g}an}}, \bibinfo {author} {\bibfnamefont {L.}~\bibnamefont {De~Santis}}, \bibinfo {author} {\bibfnamefont {R.}~\bibnamefont {Debroux}}, \bibinfo {author} {\bibfnamefont {D.}~\bibnamefont {Gangloff}}, \bibinfo {author} {\bibfnamefont {C.}~\bibnamefont {Purser}}, \bibinfo {author} {\bibfnamefont {K.~C.}\ \bibnamefont {Chen}}, \bibinfo {author} {\bibfnamefont {M.}~\bibnamefont {Walsh}}, \bibinfo {author} {\bibfnamefont {J.~J.}\ \bibnamefont {Rose}}, \bibinfo {author} {\bibfnamefont {J.~N.}\ \bibnamefont {Becker}}, \bibinfo {author} {\bibfnamefont {B.}~\bibnamefont {Lienhard}}, \bibinfo {author} {\bibfnamefont {E.}~\bibnamefont {Bersin}}, \bibinfo {author} {\bibfnamefont {I.}~\bibnamefont {Paradeisanos}}, \bibinfo {author}
  {\bibfnamefont {G.}~\bibnamefont {Wang}}, \bibinfo {author} {\bibfnamefont {D.}~\bibnamefont {Lyzwa}}, \bibinfo {author} {\bibfnamefont {A.~R.-P.}\ \bibnamefont {Montblanch}}, \bibinfo {author} {\bibfnamefont {G.}~\bibnamefont {Malladi}}, \bibinfo {author} {\bibfnamefont {H.}~\bibnamefont {Bakhru}}, \bibinfo {author} {\bibfnamefont {A.~C.}\ \bibnamefont {Ferrari}}, \bibinfo {author} {\bibfnamefont {I.~A.}\ \bibnamefont {Walmsley}}, \bibinfo {author} {\bibfnamefont {M.}~\bibnamefont {Atat{\"u}re}},\ and\ \bibinfo {author} {\bibfnamefont {D.}~\bibnamefont {Englund}},\ }\bibfield  {title} {\bibinfo {title} {Transform-{{Limited Photons From}} a {{Coherent Tin-Vacancy Spin}} in {{Diamond}}},\ }\href {https://doi.org/10.1103/PhysRevLett.124.023602} {\bibfield  {journal} {\bibinfo  {journal} {Physical Review Letters}\ }\textbf {\bibinfo {volume} {124}},\ \bibinfo {pages} {023602} (\bibinfo {year} {2020})}\BibitemShut {NoStop}%
\bibitem [{\citenamefont {G{\"o}rlitz}\ \emph {et~al.}(2020)\citenamefont {G{\"o}rlitz}, \citenamefont {Herrmann}, \citenamefont {Thiering}, \citenamefont {Fuchs}, \citenamefont {Gandil}, \citenamefont {Iwasaki}, \citenamefont {Taniguchi}, \citenamefont {Kieschnick}, \citenamefont {Meijer}, \citenamefont {Hatano}, \citenamefont {Gali},\ and\ \citenamefont {Becher}}]{gorlitz2020spectroscopic}%
  \BibitemOpen
  \bibfield  {author} {\bibinfo {author} {\bibfnamefont {J.}~\bibnamefont {G{\"o}rlitz}}, \bibinfo {author} {\bibfnamefont {D.}~\bibnamefont {Herrmann}}, \bibinfo {author} {\bibfnamefont {G.}~\bibnamefont {Thiering}}, \bibinfo {author} {\bibfnamefont {P.}~\bibnamefont {Fuchs}}, \bibinfo {author} {\bibfnamefont {M.}~\bibnamefont {Gandil}}, \bibinfo {author} {\bibfnamefont {T.}~\bibnamefont {Iwasaki}}, \bibinfo {author} {\bibfnamefont {T.}~\bibnamefont {Taniguchi}}, \bibinfo {author} {\bibfnamefont {M.}~\bibnamefont {Kieschnick}}, \bibinfo {author} {\bibfnamefont {J.}~\bibnamefont {Meijer}}, \bibinfo {author} {\bibfnamefont {M.}~\bibnamefont {Hatano}}, \bibinfo {author} {\bibfnamefont {A.}~\bibnamefont {Gali}},\ and\ \bibinfo {author} {\bibfnamefont {C.}~\bibnamefont {Becher}},\ }\bibfield  {title} {\bibinfo {title} {Spectroscopic investigations of negatively charged tin-vacancy centres in diamond},\ }\href {https://doi.org/10.1088/1367-2630/ab6631} {\bibfield  {journal} {\bibinfo  {journal} {New Journal of
  Physics}\ }\textbf {\bibinfo {volume} {22}},\ \bibinfo {pages} {013048} (\bibinfo {year} {2020})}\BibitemShut {NoStop}%
\bibitem [{\citenamefont {Rugar}\ \emph {et~al.}(2020)\citenamefont {Rugar}, \citenamefont {Dory}, \citenamefont {Aghaeimeibodi}, \citenamefont {Lu}, \citenamefont {Sun}, \citenamefont {Mishra}, \citenamefont {Shen}, \citenamefont {Melosh},\ and\ \citenamefont {Vu{\v c}kovi{\'c}}}]{rugar2020waveguide}%
  \BibitemOpen
  \bibfield  {author} {\bibinfo {author} {\bibfnamefont {A.~E.}\ \bibnamefont {Rugar}}, \bibinfo {author} {\bibfnamefont {C.}~\bibnamefont {Dory}}, \bibinfo {author} {\bibfnamefont {S.}~\bibnamefont {Aghaeimeibodi}}, \bibinfo {author} {\bibfnamefont {H.}~\bibnamefont {Lu}}, \bibinfo {author} {\bibfnamefont {S.}~\bibnamefont {Sun}}, \bibinfo {author} {\bibfnamefont {S.~D.}\ \bibnamefont {Mishra}}, \bibinfo {author} {\bibfnamefont {Z.-X.}\ \bibnamefont {Shen}}, \bibinfo {author} {\bibfnamefont {N.~A.}\ \bibnamefont {Melosh}},\ and\ \bibinfo {author} {\bibfnamefont {J.}~\bibnamefont {Vu{\v c}kovi{\'c}}},\ }\bibfield  {title} {\bibinfo {title} {Narrow-{{Linewidth Tin-Vacancy Centers}} in a {{Diamond Waveguide}}},\ }\href {https://doi.org/10.1021/acsphotonics.0c00833} {\bibfield  {journal} {\bibinfo  {journal} {ACS Photonics}\ }\textbf {\bibinfo {volume} {7}},\ \bibinfo {pages} {2356} (\bibinfo {year} {2020})},\ \Eprint {https://arxiv.org/abs/2005.10385} {arXiv:2005.10385} \BibitemShut {NoStop}%
\bibitem [{\citenamefont {Rugar}\ \emph {et~al.}(2021)\citenamefont {Rugar}, \citenamefont {Aghaeimeibodi}, \citenamefont {Riedel}, \citenamefont {Dory}, \citenamefont {Lu}, \citenamefont {McQuade}, \citenamefont {Shen}, \citenamefont {Melosh},\ and\ \citenamefont {Vu{\v c}kovi{\'c}}}]{rugar2021cavity}%
  \BibitemOpen
  \bibfield  {author} {\bibinfo {author} {\bibfnamefont {A.~E.}\ \bibnamefont {Rugar}}, \bibinfo {author} {\bibfnamefont {S.}~\bibnamefont {Aghaeimeibodi}}, \bibinfo {author} {\bibfnamefont {D.}~\bibnamefont {Riedel}}, \bibinfo {author} {\bibfnamefont {C.}~\bibnamefont {Dory}}, \bibinfo {author} {\bibfnamefont {H.}~\bibnamefont {Lu}}, \bibinfo {author} {\bibfnamefont {P.~J.}\ \bibnamefont {McQuade}}, \bibinfo {author} {\bibfnamefont {Z.-X.}\ \bibnamefont {Shen}}, \bibinfo {author} {\bibfnamefont {N.~A.}\ \bibnamefont {Melosh}},\ and\ \bibinfo {author} {\bibfnamefont {J.}~\bibnamefont {Vu{\v c}kovi{\'c}}},\ }\bibfield  {title} {\bibinfo {title} {Quantum {{Photonic Interface}} for {{Tin-Vacancy Centers}} in {{Diamond}}},\ }\href {https://doi.org/10.1103/PhysRevX.11.031021} {\bibfield  {journal} {\bibinfo  {journal} {Physical Review X}\ }\textbf {\bibinfo {volume} {11}},\ \bibinfo {pages} {031021} (\bibinfo {year} {2021})}\BibitemShut {NoStop}%
\bibitem [{\citenamefont {Debroux}\ \emph {et~al.}(2021)\citenamefont {Debroux}, \citenamefont {Michaels}, \citenamefont {Purser}, \citenamefont {Wan}, \citenamefont {Trusheim}, \citenamefont {Arjona~Mart{\'i}nez}, \citenamefont {Parker}, \citenamefont {Stramma}, \citenamefont {Chen}, \citenamefont {{de Santis}}, \citenamefont {Alexeev}, \citenamefont {Ferrari}, \citenamefont {Englund}, \citenamefont {Gangloff},\ and\ \citenamefont {Atat{\"u}re}}]{debroux2021quantum}%
  \BibitemOpen
  \bibfield  {author} {\bibinfo {author} {\bibfnamefont {R.}~\bibnamefont {Debroux}}, \bibinfo {author} {\bibfnamefont {C.~P.}\ \bibnamefont {Michaels}}, \bibinfo {author} {\bibfnamefont {C.~M.}\ \bibnamefont {Purser}}, \bibinfo {author} {\bibfnamefont {N.}~\bibnamefont {Wan}}, \bibinfo {author} {\bibfnamefont {M.~E.}\ \bibnamefont {Trusheim}}, \bibinfo {author} {\bibfnamefont {J.}~\bibnamefont {Arjona~Mart{\'i}nez}}, \bibinfo {author} {\bibfnamefont {R.~A.}\ \bibnamefont {Parker}}, \bibinfo {author} {\bibfnamefont {A.~M.}\ \bibnamefont {Stramma}}, \bibinfo {author} {\bibfnamefont {K.~C.}\ \bibnamefont {Chen}}, \bibinfo {author} {\bibfnamefont {L.}~\bibnamefont {{de Santis}}}, \bibinfo {author} {\bibfnamefont {E.~M.}\ \bibnamefont {Alexeev}}, \bibinfo {author} {\bibfnamefont {A.~C.}\ \bibnamefont {Ferrari}}, \bibinfo {author} {\bibfnamefont {D.}~\bibnamefont {Englund}}, \bibinfo {author} {\bibfnamefont {D.~A.}\ \bibnamefont {Gangloff}},\ and\ \bibinfo {author} {\bibfnamefont {M.}~\bibnamefont {Atat{\"u}re}},\
  }\bibfield  {title} {\bibinfo {title} {Quantum {{Control}} of the {{Tin-Vacancy Spin Qubit}} in {{Diamond}}},\ }\href {https://doi.org/10.1103/PhysRevX.11.041041} {\bibfield  {journal} {\bibinfo  {journal} {Physical Review X}\ }\textbf {\bibinfo {volume} {11}},\ \bibinfo {pages} {041041} (\bibinfo {year} {2021})}\BibitemShut {NoStop}%
\bibitem [{\citenamefont {Guo}\ \emph {et~al.}(2023)\citenamefont {Guo}, \citenamefont {Stramma}, \citenamefont {Li}, \citenamefont {Roth}, \citenamefont {Huang}, \citenamefont {Jin}, \citenamefont {Parker}, \citenamefont {Arjona~Mart{\'i}nez}, \citenamefont {Shofer}, \citenamefont {Michaels}, \citenamefont {Purser}, \citenamefont {Appel}, \citenamefont {Alexeev}, \citenamefont {Liu}, \citenamefont {Ferrari}, \citenamefont {Awschalom}, \citenamefont {Delegan}, \citenamefont {Pingault}, \citenamefont {Galli}, \citenamefont {Heremans}, \citenamefont {Atat{\"u}re},\ and\ \citenamefont {High}}]{guo2023microwave}%
  \BibitemOpen
  \bibfield  {author} {\bibinfo {author} {\bibfnamefont {X.}~\bibnamefont {Guo}}, \bibinfo {author} {\bibfnamefont {A.~M.}\ \bibnamefont {Stramma}}, \bibinfo {author} {\bibfnamefont {Z.}~\bibnamefont {Li}}, \bibinfo {author} {\bibfnamefont {W.~G.}\ \bibnamefont {Roth}}, \bibinfo {author} {\bibfnamefont {B.}~\bibnamefont {Huang}}, \bibinfo {author} {\bibfnamefont {Y.}~\bibnamefont {Jin}}, \bibinfo {author} {\bibfnamefont {R.~A.}\ \bibnamefont {Parker}}, \bibinfo {author} {\bibfnamefont {J.}~\bibnamefont {Arjona~Mart{\'i}nez}}, \bibinfo {author} {\bibfnamefont {N.}~\bibnamefont {Shofer}}, \bibinfo {author} {\bibfnamefont {C.~P.}\ \bibnamefont {Michaels}}, \bibinfo {author} {\bibfnamefont {C.~P.}\ \bibnamefont {Purser}}, \bibinfo {author} {\bibfnamefont {M.~H.}\ \bibnamefont {Appel}}, \bibinfo {author} {\bibfnamefont {E.~M.}\ \bibnamefont {Alexeev}}, \bibinfo {author} {\bibfnamefont {T.}~\bibnamefont {Liu}}, \bibinfo {author} {\bibfnamefont {A.~C.}\ \bibnamefont {Ferrari}}, \bibinfo {author} {\bibfnamefont
  {D.~D.}\ \bibnamefont {Awschalom}}, \bibinfo {author} {\bibfnamefont {N.}~\bibnamefont {Delegan}}, \bibinfo {author} {\bibfnamefont {B.}~\bibnamefont {Pingault}}, \bibinfo {author} {\bibfnamefont {G.}~\bibnamefont {Galli}}, \bibinfo {author} {\bibfnamefont {F.~J.}\ \bibnamefont {Heremans}}, \bibinfo {author} {\bibfnamefont {M.}~\bibnamefont {Atat{\"u}re}},\ and\ \bibinfo {author} {\bibfnamefont {A.~A.}\ \bibnamefont {High}},\ }\bibfield  {title} {\bibinfo {title} {Microwave-{{Based Quantum Control}} and {{Coherence Protection}} of {{Tin-Vacancy Spin Qubits}} in a {{Strain-Tuned Diamond-Membrane Heterostructure}}},\ }\href {https://doi.org/10.1103/PhysRevX.13.041037} {\bibfield  {journal} {\bibinfo  {journal} {Physical Review X}\ }\textbf {\bibinfo {volume} {13}},\ \bibinfo {pages} {041037} (\bibinfo {year} {2023})}\BibitemShut {NoStop}%
\bibitem [{\citenamefont {Pasini}\ \emph {et~al.}(2024)\citenamefont {Pasini}, \citenamefont {Codreanu}, \citenamefont {Turan}, \citenamefont {Riera~Moral}, \citenamefont {Primavera}, \citenamefont {De~Santis}, \citenamefont {Beukers}, \citenamefont {Brevoord}, \citenamefont {Waas}, \citenamefont {Borregaard},\ and\ \citenamefont {Hanson}}]{pasini2024nonlinear}%
  \BibitemOpen
  \bibfield  {author} {\bibinfo {author} {\bibfnamefont {M.}~\bibnamefont {Pasini}}, \bibinfo {author} {\bibfnamefont {N.}~\bibnamefont {Codreanu}}, \bibinfo {author} {\bibfnamefont {T.}~\bibnamefont {Turan}}, \bibinfo {author} {\bibfnamefont {A.}~\bibnamefont {Riera~Moral}}, \bibinfo {author} {\bibfnamefont {C.~F.}\ \bibnamefont {Primavera}}, \bibinfo {author} {\bibfnamefont {L.}~\bibnamefont {De~Santis}}, \bibinfo {author} {\bibfnamefont {H.~K.~C.}\ \bibnamefont {Beukers}}, \bibinfo {author} {\bibfnamefont {J.~M.}\ \bibnamefont {Brevoord}}, \bibinfo {author} {\bibfnamefont {C.}~\bibnamefont {Waas}}, \bibinfo {author} {\bibfnamefont {J.}~\bibnamefont {Borregaard}},\ and\ \bibinfo {author} {\bibfnamefont {R.}~\bibnamefont {Hanson}},\ }\bibfield  {title} {\bibinfo {title} {Nonlinear {{Quantum Photonics}} with a {{Tin-Vacancy Center Coupled}} to a {{One-Dimensional Diamond Waveguide}}},\ }\href {https://doi.org/10.1103/PhysRevLett.133.023603} {\bibfield  {journal} {\bibinfo  {journal} {Physical Review Letters}\
  }\textbf {\bibinfo {volume} {133}},\ \bibinfo {pages} {023603} (\bibinfo {year} {2024})}\BibitemShut {NoStop}%
\bibitem [{\citenamefont {Brevoord}\ \emph {et~al.}(2024)\citenamefont {Brevoord}, \citenamefont {De~Santis}, \citenamefont {Yamamoto}, \citenamefont {Pasini}, \citenamefont {Codreanu}, \citenamefont {Turan}, \citenamefont {Beukers}, \citenamefont {Waas},\ and\ \citenamefont {Hanson}}]{brevoord2024heralded}%
  \BibitemOpen
  \bibfield  {author} {\bibinfo {author} {\bibfnamefont {J.~M.}\ \bibnamefont {Brevoord}}, \bibinfo {author} {\bibfnamefont {L.}~\bibnamefont {De~Santis}}, \bibinfo {author} {\bibfnamefont {T.}~\bibnamefont {Yamamoto}}, \bibinfo {author} {\bibfnamefont {M.}~\bibnamefont {Pasini}}, \bibinfo {author} {\bibfnamefont {N.}~\bibnamefont {Codreanu}}, \bibinfo {author} {\bibfnamefont {T.}~\bibnamefont {Turan}}, \bibinfo {author} {\bibfnamefont {H.~K.~C.}\ \bibnamefont {Beukers}}, \bibinfo {author} {\bibfnamefont {C.}~\bibnamefont {Waas}},\ and\ \bibinfo {author} {\bibfnamefont {R.}~\bibnamefont {Hanson}},\ }\bibfield  {title} {\bibinfo {title} {Heralded initialization of charge state and optical-transition frequency of diamond tin-vacancy centers},\ }\href {https://doi.org/10.1103/PhysRevApplied.21.054047} {\bibfield  {journal} {\bibinfo  {journal} {Physical Review Applied}\ }\textbf {\bibinfo {volume} {21}},\ \bibinfo {pages} {054047} (\bibinfo {year} {2024})}\BibitemShut {NoStop}%
\bibitem [{\citenamefont {Rosenthal}\ \emph {et~al.}(2024)\citenamefont {Rosenthal}, \citenamefont {Biswas}, \citenamefont {Scuri}, \citenamefont {Lee}, \citenamefont {Stein}, \citenamefont {Kleidermacher}, \citenamefont {Grzesik}, \citenamefont {Rugar}, \citenamefont {Aghaeimeibodi}, \citenamefont {Riedel}, \citenamefont {Titze}, \citenamefont {Bielejec}, \citenamefont {Choi}, \citenamefont {Anderson},\ and\ \citenamefont {Vuckovic}}]{rosenthal2024singleshot}%
  \BibitemOpen
  \bibfield  {author} {\bibinfo {author} {\bibfnamefont {E.~I.}\ \bibnamefont {Rosenthal}}, \bibinfo {author} {\bibfnamefont {S.}~\bibnamefont {Biswas}}, \bibinfo {author} {\bibfnamefont {G.}~\bibnamefont {Scuri}}, \bibinfo {author} {\bibfnamefont {H.}~\bibnamefont {Lee}}, \bibinfo {author} {\bibfnamefont {A.~J.}\ \bibnamefont {Stein}}, \bibinfo {author} {\bibfnamefont {H.~C.}\ \bibnamefont {Kleidermacher}}, \bibinfo {author} {\bibfnamefont {J.}~\bibnamefont {Grzesik}}, \bibinfo {author} {\bibfnamefont {A.~E.}\ \bibnamefont {Rugar}}, \bibinfo {author} {\bibfnamefont {S.}~\bibnamefont {Aghaeimeibodi}}, \bibinfo {author} {\bibfnamefont {D.}~\bibnamefont {Riedel}}, \bibinfo {author} {\bibfnamefont {M.}~\bibnamefont {Titze}}, \bibinfo {author} {\bibfnamefont {E.~S.}\ \bibnamefont {Bielejec}}, \bibinfo {author} {\bibfnamefont {J.}~\bibnamefont {Choi}}, \bibinfo {author} {\bibfnamefont {C.~P.}\ \bibnamefont {Anderson}},\ and\ \bibinfo {author} {\bibfnamefont {J.}~\bibnamefont {Vuckovic}},\ }\href@noop {} {\bibinfo
  {title} {Single-{{Shot Readout}} and {{Weak Measurement}} of a {{Tin-Vacancy Qubit}} in {{Diamond}}}} (\bibinfo {year} {2024}),\ \Eprint {https://arxiv.org/abs/2403.13110} {arXiv:2403.13110} \BibitemShut {NoStop}%
\bibitem [{\citenamefont {{de Lange}}\ \emph {et~al.}(2010)\citenamefont {{de Lange}}, \citenamefont {Wang}, \citenamefont {Rist{\`e}}, \citenamefont {Dobrovitski},\ and\ \citenamefont {Hanson}}]{delange2010universal}%
  \BibitemOpen
  \bibfield  {author} {\bibinfo {author} {\bibfnamefont {G.}~\bibnamefont {{de Lange}}}, \bibinfo {author} {\bibfnamefont {Z.~H.}\ \bibnamefont {Wang}}, \bibinfo {author} {\bibfnamefont {D.}~\bibnamefont {Rist{\`e}}}, \bibinfo {author} {\bibfnamefont {V.~V.}\ \bibnamefont {Dobrovitski}},\ and\ \bibinfo {author} {\bibfnamefont {R.}~\bibnamefont {Hanson}},\ }\bibfield  {title} {\bibinfo {title} {Universal {{Dynamical Decoupling}} of a {{Single Solid-State Spin}} from a {{Spin Bath}}},\ }\href {https://doi.org/10.1126/science.1192739} {\bibfield  {journal} {\bibinfo  {journal} {Science}\ }\textbf {\bibinfo {volume} {330}},\ \bibinfo {pages} {60} (\bibinfo {year} {2010})}\BibitemShut {NoStop}%
\bibitem [{\citenamefont {Karapatzakis}\ \emph {et~al.}(2024)\citenamefont {Karapatzakis}, \citenamefont {Resch}, \citenamefont {Schrodin}, \citenamefont {Fuchs}, \citenamefont {Kieschnick}, \citenamefont {Heupel}, \citenamefont {Kussi}, \citenamefont {S{\"u}rgers}, \citenamefont {Popov}, \citenamefont {Meijer}, \citenamefont {Becher}, \citenamefont {Wernsdorfer},\ and\ \citenamefont {Hunger}}]{karapatzakis2024microwave}%
  \BibitemOpen
  \bibfield  {author} {\bibinfo {author} {\bibfnamefont {I.}~\bibnamefont {Karapatzakis}}, \bibinfo {author} {\bibfnamefont {J.}~\bibnamefont {Resch}}, \bibinfo {author} {\bibfnamefont {M.}~\bibnamefont {Schrodin}}, \bibinfo {author} {\bibfnamefont {P.}~\bibnamefont {Fuchs}}, \bibinfo {author} {\bibfnamefont {M.}~\bibnamefont {Kieschnick}}, \bibinfo {author} {\bibfnamefont {J.}~\bibnamefont {Heupel}}, \bibinfo {author} {\bibfnamefont {L.}~\bibnamefont {Kussi}}, \bibinfo {author} {\bibfnamefont {C.}~\bibnamefont {S{\"u}rgers}}, \bibinfo {author} {\bibfnamefont {C.}~\bibnamefont {Popov}}, \bibinfo {author} {\bibfnamefont {J.}~\bibnamefont {Meijer}}, \bibinfo {author} {\bibfnamefont {C.}~\bibnamefont {Becher}}, \bibinfo {author} {\bibfnamefont {W.}~\bibnamefont {Wernsdorfer}},\ and\ \bibinfo {author} {\bibfnamefont {D.}~\bibnamefont {Hunger}},\ }\href@noop {} {\bibinfo {title} {Microwave {{Control}} of the {{Tin-Vacancy Spin Qubit}} in {{Diamond}} with a {{Superconducting Waveguide}}}} (\bibinfo {year} {2024}),\
  \Eprint {https://arxiv.org/abs/2403.00521} {arXiv:2403.00521} \BibitemShut {NoStop}%
\bibitem [{\citenamefont {Nizovtsev}\ \emph {et~al.}(2018)\citenamefont {Nizovtsev}, \citenamefont {Kilin}, \citenamefont {Pushkarchuk}, \citenamefont {Pushkarchuk}, \citenamefont {Kuten}, \citenamefont {Zhikol}, \citenamefont {Schmitt}, \citenamefont {Unden},\ and\ \citenamefont {Jelezko}}]{nizovtsev2018nonflipping}%
  \BibitemOpen
  \bibfield  {author} {\bibinfo {author} {\bibfnamefont {A.~P.}\ \bibnamefont {Nizovtsev}}, \bibinfo {author} {\bibfnamefont {S.~Y.}\ \bibnamefont {Kilin}}, \bibinfo {author} {\bibfnamefont {A.~L.}\ \bibnamefont {Pushkarchuk}}, \bibinfo {author} {\bibfnamefont {V.~A.}\ \bibnamefont {Pushkarchuk}}, \bibinfo {author} {\bibfnamefont {S.~A.}\ \bibnamefont {Kuten}}, \bibinfo {author} {\bibfnamefont {O.~A.}\ \bibnamefont {Zhikol}}, \bibinfo {author} {\bibfnamefont {S.}~\bibnamefont {Schmitt}}, \bibinfo {author} {\bibfnamefont {T.}~\bibnamefont {Unden}},\ and\ \bibinfo {author} {\bibfnamefont {F.}~\bibnamefont {Jelezko}},\ }\bibfield  {title} {\bibinfo {title} {Non-flipping {{13C}} spins near an {{NV}} center in diamond: Hyperfine and spatial characteristics by density functional theory simulation of the {{C510}}[{{NV}}]{{H252}} cluster},\ }\href {https://doi.org/10.1088/1367-2630/aaa910} {\bibfield  {journal} {\bibinfo  {journal} {New Journal of Physics}\ }\textbf {\bibinfo {volume} {20}},\ \bibinfo {pages}
  {023022} (\bibinfo {year} {2018})}\BibitemShut {NoStop}%
\bibitem [{\citenamefont {Neumann}\ \emph {et~al.}(2008)\citenamefont {Neumann}, \citenamefont {Mizuochi}, \citenamefont {Rempp}, \citenamefont {Hemmer}, \citenamefont {Watanabe}, \citenamefont {Yamasaki}, \citenamefont {Jacques}, \citenamefont {Gaebel}, \citenamefont {Jelezko},\ and\ \citenamefont {Wrachtrup}}]{neumann2008multipartite}%
  \BibitemOpen
  \bibfield  {author} {\bibinfo {author} {\bibfnamefont {P.}~\bibnamefont {Neumann}}, \bibinfo {author} {\bibfnamefont {N.}~\bibnamefont {Mizuochi}}, \bibinfo {author} {\bibfnamefont {F.}~\bibnamefont {Rempp}}, \bibinfo {author} {\bibfnamefont {P.}~\bibnamefont {Hemmer}}, \bibinfo {author} {\bibfnamefont {H.}~\bibnamefont {Watanabe}}, \bibinfo {author} {\bibfnamefont {S.}~\bibnamefont {Yamasaki}}, \bibinfo {author} {\bibfnamefont {V.}~\bibnamefont {Jacques}}, \bibinfo {author} {\bibfnamefont {T.}~\bibnamefont {Gaebel}}, \bibinfo {author} {\bibfnamefont {F.}~\bibnamefont {Jelezko}},\ and\ \bibinfo {author} {\bibfnamefont {J.}~\bibnamefont {Wrachtrup}},\ }\bibfield  {title} {\bibinfo {title} {Multipartite {{Entanglement Among Single Spins}} in {{Diamond}}},\ }\href {https://doi.org/10.1126/science.1157233} {\bibfield  {journal} {\bibinfo  {journal} {Science}\ }\textbf {\bibinfo {volume} {320}},\ \bibinfo {pages} {1326} (\bibinfo {year} {2008})}\BibitemShut {NoStop}%
\bibitem [{\citenamefont {Grimm}\ \emph {et~al.}(2024)\citenamefont {Grimm}, \citenamefont {Senkalla}, \citenamefont {Vetter}, \citenamefont {Frey}, \citenamefont {Gundlapalli}, \citenamefont {Calarco}, \citenamefont {Genov}, \citenamefont {M{\"u}ller},\ and\ \citenamefont {Jelezko}}]{grimm2024coherent}%
  \BibitemOpen
  \bibfield  {author} {\bibinfo {author} {\bibfnamefont {N.}~\bibnamefont {Grimm}}, \bibinfo {author} {\bibfnamefont {K.}~\bibnamefont {Senkalla}}, \bibinfo {author} {\bibfnamefont {P.~J.}\ \bibnamefont {Vetter}}, \bibinfo {author} {\bibfnamefont {J.}~\bibnamefont {Frey}}, \bibinfo {author} {\bibfnamefont {P.}~\bibnamefont {Gundlapalli}}, \bibinfo {author} {\bibfnamefont {T.}~\bibnamefont {Calarco}}, \bibinfo {author} {\bibfnamefont {G.}~\bibnamefont {Genov}}, \bibinfo {author} {\bibfnamefont {M.~M.}\ \bibnamefont {M{\"u}ller}},\ and\ \bibinfo {author} {\bibfnamefont {F.}~\bibnamefont {Jelezko}},\ }\href@noop {} {\bibinfo {title} {Coherent {{Control}} of a {{Long-Lived Nuclear Memory Spin}} in a {{Germanium-Vacancy Multi-Qubit Node}}}} (\bibinfo {year} {2024}),\ \Eprint {https://arxiv.org/abs/2409.06313} {arXiv:2409.06313} \BibitemShut {NoStop}%
\bibitem [{\citenamefont {Kolkowitz}\ \emph {et~al.}(2012)\citenamefont {Kolkowitz}, \citenamefont {Unterreithmeier}, \citenamefont {Bennett},\ and\ \citenamefont {Lukin}}]{kolkowitz2012sensing}%
  \BibitemOpen
  \bibfield  {author} {\bibinfo {author} {\bibfnamefont {S.}~\bibnamefont {Kolkowitz}}, \bibinfo {author} {\bibfnamefont {Q.~P.}\ \bibnamefont {Unterreithmeier}}, \bibinfo {author} {\bibfnamefont {S.~D.}\ \bibnamefont {Bennett}},\ and\ \bibinfo {author} {\bibfnamefont {M.~D.}\ \bibnamefont {Lukin}},\ }\bibfield  {title} {\bibinfo {title} {Sensing {{Distant Nuclear Spins}} with a {{Single Electron Spin}}},\ }\href {https://doi.org/10.1103/PhysRevLett.109.137601} {\bibfield  {journal} {\bibinfo  {journal} {Physical Review Letters}\ }\textbf {\bibinfo {volume} {109}},\ \bibinfo {pages} {137601} (\bibinfo {year} {2012})}\BibitemShut {NoStop}%
\bibitem [{\citenamefont {Zhao}\ \emph {et~al.}(2012)\citenamefont {Zhao}, \citenamefont {Honert}, \citenamefont {Schmid}, \citenamefont {Klas}, \citenamefont {Isoya}, \citenamefont {Markham}, \citenamefont {Twitchen}, \citenamefont {Jelezko}, \citenamefont {Liu}, \citenamefont {Fedder},\ and\ \citenamefont {Wrachtrup}}]{zhao2012sensing}%
  \BibitemOpen
  \bibfield  {author} {\bibinfo {author} {\bibfnamefont {N.}~\bibnamefont {Zhao}}, \bibinfo {author} {\bibfnamefont {J.}~\bibnamefont {Honert}}, \bibinfo {author} {\bibfnamefont {B.}~\bibnamefont {Schmid}}, \bibinfo {author} {\bibfnamefont {M.}~\bibnamefont {Klas}}, \bibinfo {author} {\bibfnamefont {J.}~\bibnamefont {Isoya}}, \bibinfo {author} {\bibfnamefont {M.}~\bibnamefont {Markham}}, \bibinfo {author} {\bibfnamefont {D.}~\bibnamefont {Twitchen}}, \bibinfo {author} {\bibfnamefont {F.}~\bibnamefont {Jelezko}}, \bibinfo {author} {\bibfnamefont {R.-B.}\ \bibnamefont {Liu}}, \bibinfo {author} {\bibfnamefont {H.}~\bibnamefont {Fedder}},\ and\ \bibinfo {author} {\bibfnamefont {J.}~\bibnamefont {Wrachtrup}},\ }\bibfield  {title} {\bibinfo {title} {Sensing single remote nuclear spins},\ }\href {https://doi.org/10.1038/nnano.2012.152} {\bibfield  {journal} {\bibinfo  {journal} {Nature Nanotechnology}\ }\textbf {\bibinfo {volume} {7}},\ \bibinfo {pages} {657} (\bibinfo {year} {2012})}\BibitemShut {NoStop}%
\bibitem [{\citenamefont {Abobeih}(2021)}]{abobeih2021thesis}%
  \BibitemOpen
  \bibfield  {author} {\bibinfo {author} {\bibfnamefont {M.}~\bibnamefont {Abobeih}},\ }\emph {\bibinfo {title} {From Atomic-Scale Imaging to Quantum Fault-Tolerance with Spins in Diamond}},\ \href {https://doi.org/10.4233/UUID:CCE8DBCB-CFC2-4FA2-B78B-99C803DEE02D} {Ph.D. thesis},\ \bibinfo  {school} {Delft University of Technology} (\bibinfo {year} {2021})\BibitemShut {NoStop}%
\bibitem [{\citenamefont {{van Ommen}}\ \emph {et~al.}(2024)\citenamefont {{van Ommen}}, \citenamefont {{van de Stolpe}}, \citenamefont {Demetriou}, \citenamefont {Beukers}, \citenamefont {Yun}, \citenamefont {Fortuin}, \citenamefont {Iuliano}, \citenamefont {Montblanch}, \citenamefont {Hanson},\ and\ \citenamefont {Taminiau}}]{vanommen2024ddrf}%
  \BibitemOpen
  \bibfield  {author} {\bibinfo {author} {\bibfnamefont {H.~B.}\ \bibnamefont {{van Ommen}}}, \bibinfo {author} {\bibfnamefont {G.~L.}\ \bibnamefont {{van de Stolpe}}}, \bibinfo {author} {\bibfnamefont {N.}~\bibnamefont {Demetriou}}, \bibinfo {author} {\bibfnamefont {H.~K.~C.}\ \bibnamefont {Beukers}}, \bibinfo {author} {\bibfnamefont {J.}~\bibnamefont {Yun}}, \bibinfo {author} {\bibfnamefont {T.~R.~J.}\ \bibnamefont {Fortuin}}, \bibinfo {author} {\bibfnamefont {M.}~\bibnamefont {Iuliano}}, \bibinfo {author} {\bibfnamefont {A.~R.-P.}\ \bibnamefont {Montblanch}}, \bibinfo {author} {\bibfnamefont {R.}~\bibnamefont {Hanson}},\ and\ \bibinfo {author} {\bibfnamefont {T.~H.}\ \bibnamefont {Taminiau}},\ }\href@noop {} {\bibinfo {title} {Improved {{Electron-Nuclear Quantum Gates}} for {{Spin Sensing}} and {{Control}}}} (\bibinfo {year} {2024}),\ \Eprint {https://arxiv.org/abs/2409.13610} {arXiv:2409.13610} \BibitemShut {NoStop}%
\bibitem [{\citenamefont {Uhrig}(2007)}]{uhrig2007udd}%
  \BibitemOpen
  \bibfield  {author} {\bibinfo {author} {\bibfnamefont {G.~S.}\ \bibnamefont {Uhrig}},\ }\bibfield  {title} {\bibinfo {title} {Keeping a {{Quantum Bit Alive}} by {{Optimized}} {$\pi$}-{{Pulse Sequences}}},\ }\href {https://doi.org/10.1103/PhysRevLett.98.100504} {\bibfield  {journal} {\bibinfo  {journal} {Physical Review Letters}\ }\textbf {\bibinfo {volume} {98}},\ \bibinfo {pages} {100504} (\bibinfo {year} {2007})}\BibitemShut {NoStop}%
\bibitem [{\citenamefont {Loudon}(2000)}]{loudon2000quantum}%
  \BibitemOpen
  \bibfield  {author} {\bibinfo {author} {\bibfnamefont {R.}~\bibnamefont {Loudon}},\ }\href@noop {} {\emph {\bibinfo {title} {The Quantum Theory of Light}}},\ \bibinfo {edition} {3rd}\ ed.,\ Oxford Science Publications\ (\bibinfo  {publisher} {Oxford University Press},\ \bibinfo {address} {Oxford ; New York},\ \bibinfo {year} {2000})\BibitemShut {NoStop}%
\bibitem [{\citenamefont {Astner}\ \emph {et~al.}(2024)\citenamefont {Astner}, \citenamefont {Koller}, \citenamefont {Gilardoni}, \citenamefont {Hendriks}, \citenamefont {Son}, \citenamefont {Ivanov}, \citenamefont {Hassan}, \citenamefont {van~der Wal},\ and\ \citenamefont {Trupke}}]{astner2024vanadium}%
  \BibitemOpen
  \bibfield  {author} {\bibinfo {author} {\bibfnamefont {T.}~\bibnamefont {Astner}}, \bibinfo {author} {\bibfnamefont {P.}~\bibnamefont {Koller}}, \bibinfo {author} {\bibfnamefont {C.~M.}\ \bibnamefont {Gilardoni}}, \bibinfo {author} {\bibfnamefont {J.}~\bibnamefont {Hendriks}}, \bibinfo {author} {\bibfnamefont {N.~T.}\ \bibnamefont {Son}}, \bibinfo {author} {\bibfnamefont {I.~G.}\ \bibnamefont {Ivanov}}, \bibinfo {author} {\bibfnamefont {J.~U.}\ \bibnamefont {Hassan}}, \bibinfo {author} {\bibfnamefont {C.~H.}\ \bibnamefont {van~der Wal}},\ and\ \bibinfo {author} {\bibfnamefont {M.}~\bibnamefont {Trupke}},\ }\bibfield  {title} {\bibinfo {title} {Vanadium in silicon carbide: Telecom-ready spin centres with long relaxation lifetimes and hyperfine-resolved optical transitions},\ }\href {https://doi.org/10.1088/2058-9565/ad48b1} {\bibfield  {journal} {\bibinfo  {journal} {Quantum Science and Technology}\ }\textbf {\bibinfo {volume} {9}},\ \bibinfo {pages} {035038} (\bibinfo {year} {2024})}\BibitemShut {NoStop}%
\bibitem [{\citenamefont {Casanova}\ \emph {et~al.}(2015)\citenamefont {Casanova}, \citenamefont {Wang}, \citenamefont {Haase},\ and\ \citenamefont {Plenio}}]{casanova2015robust}%
  \BibitemOpen
  \bibfield  {author} {\bibinfo {author} {\bibfnamefont {J.}~\bibnamefont {Casanova}}, \bibinfo {author} {\bibfnamefont {Z.-Y.}\ \bibnamefont {Wang}}, \bibinfo {author} {\bibfnamefont {J.~F.}\ \bibnamefont {Haase}},\ and\ \bibinfo {author} {\bibfnamefont {M.~B.}\ \bibnamefont {Plenio}},\ }\bibfield  {title} {\bibinfo {title} {Robust dynamical decoupling sequences for individual-nuclear-spin addressing},\ }\href {https://doi.org/10.1103/PhysRevA.92.042304} {\bibfield  {journal} {\bibinfo  {journal} {Physical Review A}\ }\textbf {\bibinfo {volume} {92}},\ \bibinfo {pages} {042304} (\bibinfo {year} {2015})}\BibitemShut {NoStop}%
\bibitem [{\citenamefont {Stas}\ \emph {et~al.}(2022)\citenamefont {Stas}, \citenamefont {Huan}, \citenamefont {Machielse}, \citenamefont {Knall}, \citenamefont {Suleymanzade}, \citenamefont {Pingault}, \citenamefont {Sutula}, \citenamefont {Ding}, \citenamefont {Knaut}, \citenamefont {Assumpcao}, \citenamefont {Wei}, \citenamefont {Bhaskar}, \citenamefont {Riedinger}, \citenamefont {Sukachev}, \citenamefont {Park}, \citenamefont {Lon{\v c}ar}, \citenamefont {Levonian},\ and\ \citenamefont {Lukin}}]{stas2022robust}%
  \BibitemOpen
  \bibfield  {author} {\bibinfo {author} {\bibfnamefont {P.-J.}\ \bibnamefont {Stas}}, \bibinfo {author} {\bibfnamefont {Y.~Q.}\ \bibnamefont {Huan}}, \bibinfo {author} {\bibfnamefont {B.}~\bibnamefont {Machielse}}, \bibinfo {author} {\bibfnamefont {E.~N.}\ \bibnamefont {Knall}}, \bibinfo {author} {\bibfnamefont {A.}~\bibnamefont {Suleymanzade}}, \bibinfo {author} {\bibfnamefont {B.}~\bibnamefont {Pingault}}, \bibinfo {author} {\bibfnamefont {M.}~\bibnamefont {Sutula}}, \bibinfo {author} {\bibfnamefont {S.~W.}\ \bibnamefont {Ding}}, \bibinfo {author} {\bibfnamefont {C.~M.}\ \bibnamefont {Knaut}}, \bibinfo {author} {\bibfnamefont {D.~R.}\ \bibnamefont {Assumpcao}}, \bibinfo {author} {\bibfnamefont {Y.-C.}\ \bibnamefont {Wei}}, \bibinfo {author} {\bibfnamefont {M.~K.}\ \bibnamefont {Bhaskar}}, \bibinfo {author} {\bibfnamefont {R.}~\bibnamefont {Riedinger}}, \bibinfo {author} {\bibfnamefont {D.~D.}\ \bibnamefont {Sukachev}}, \bibinfo {author} {\bibfnamefont {H.}~\bibnamefont {Park}}, \bibinfo {author}
  {\bibfnamefont {M.}~\bibnamefont {Lon{\v c}ar}}, \bibinfo {author} {\bibfnamefont {D.~S.}\ \bibnamefont {Levonian}},\ and\ \bibinfo {author} {\bibfnamefont {M.~D.}\ \bibnamefont {Lukin}},\ }\bibfield  {title} {\bibinfo {title} {Robust multi-qubit quantum network node with integrated error detection},\ }\href {https://doi.org/10.1126/science.add9771} {\bibfield  {journal} {\bibinfo  {journal} {Science}\ }\textbf {\bibinfo {volume} {378}},\ \bibinfo {pages} {557} (\bibinfo {year} {2022})}\BibitemShut {NoStop}%
\bibitem [{\citenamefont {Harris}\ \emph {et~al.}(2023)\citenamefont {Harris}, \citenamefont {Michaels}, \citenamefont {Chen}, \citenamefont {Parker}, \citenamefont {Titze}, \citenamefont {Arjona~Mart{\'i}nez}, \citenamefont {Sutula}, \citenamefont {Christen}, \citenamefont {Stramma}, \citenamefont {Roth}, \citenamefont {Purser}, \citenamefont {Appel}, \citenamefont {Li}, \citenamefont {Trusheim}, \citenamefont {Palmer}, \citenamefont {Markham}, \citenamefont {Bielejec}, \citenamefont {Atat{\"u}re},\ and\ \citenamefont {Englund}}]{harris2023hyperfine}%
  \BibitemOpen
  \bibfield  {author} {\bibinfo {author} {\bibfnamefont {I.~B.~W.}\ \bibnamefont {Harris}}, \bibinfo {author} {\bibfnamefont {C.~P.}\ \bibnamefont {Michaels}}, \bibinfo {author} {\bibfnamefont {K.~C.}\ \bibnamefont {Chen}}, \bibinfo {author} {\bibfnamefont {R.~A.}\ \bibnamefont {Parker}}, \bibinfo {author} {\bibfnamefont {M.}~\bibnamefont {Titze}}, \bibinfo {author} {\bibfnamefont {J.}~\bibnamefont {Arjona~Mart{\'i}nez}}, \bibinfo {author} {\bibfnamefont {M.}~\bibnamefont {Sutula}}, \bibinfo {author} {\bibfnamefont {I.~R.}\ \bibnamefont {Christen}}, \bibinfo {author} {\bibfnamefont {A.~M.}\ \bibnamefont {Stramma}}, \bibinfo {author} {\bibfnamefont {W.}~\bibnamefont {Roth}}, \bibinfo {author} {\bibfnamefont {C.~M.}\ \bibnamefont {Purser}}, \bibinfo {author} {\bibfnamefont {M.~H.}\ \bibnamefont {Appel}}, \bibinfo {author} {\bibfnamefont {C.}~\bibnamefont {Li}}, \bibinfo {author} {\bibfnamefont {M.~E.}\ \bibnamefont {Trusheim}}, \bibinfo {author} {\bibfnamefont {N.~L.}\ \bibnamefont {Palmer}}, \bibinfo {author}
  {\bibfnamefont {M.~L.}\ \bibnamefont {Markham}}, \bibinfo {author} {\bibfnamefont {E.~S.}\ \bibnamefont {Bielejec}}, \bibinfo {author} {\bibfnamefont {M.}~\bibnamefont {Atat{\"u}re}},\ and\ \bibinfo {author} {\bibfnamefont {D.}~\bibnamefont {Englund}},\ }\bibfield  {title} {\bibinfo {title} {Hyperfine {{Spectroscopy}} of {{Isotopically Engineered Group-IV Color Centers}} in {{Diamond}}},\ }\href {https://doi.org/10.1103/PRXQuantum.4.040301} {\bibfield  {journal} {\bibinfo  {journal} {PRX Quantum}\ }\textbf {\bibinfo {volume} {4}},\ \bibinfo {pages} {040301} (\bibinfo {year} {2023})}\BibitemShut {NoStop}%
\bibitem [{\citenamefont {Beukers}\ \emph {et~al.}(2024{\natexlab{b}})\citenamefont {Beukers}, \citenamefont {Waas}, \citenamefont {Pasini}, \citenamefont {{van Ommen}}, \citenamefont {Codreanu}, \citenamefont {Brevoord}, \citenamefont {Turan}, \citenamefont {Iuliano}, \citenamefont {Ademi}, \citenamefont {Taminiau},\ and\ \citenamefont {Hanson}}]{beukers2024dataset4tu}%
  \BibitemOpen
  \bibfield  {author} {\bibinfo {author} {\bibfnamefont {H.~K.~C.}\ \bibnamefont {Beukers}}, \bibinfo {author} {\bibfnamefont {C.}~\bibnamefont {Waas}}, \bibinfo {author} {\bibfnamefont {M.}~\bibnamefont {Pasini}}, \bibinfo {author} {\bibfnamefont {H.~B.}\ \bibnamefont {{van Ommen}}}, \bibinfo {author} {\bibfnamefont {N.}~\bibnamefont {Codreanu}}, \bibinfo {author} {\bibfnamefont {J.~M.}\ \bibnamefont {Brevoord}}, \bibinfo {author} {\bibfnamefont {T.}~\bibnamefont {Turan}}, \bibinfo {author} {\bibfnamefont {M.}~\bibnamefont {Iuliano}}, \bibinfo {author} {\bibfnamefont {Z.}~\bibnamefont {Ademi}}, \bibinfo {author} {\bibfnamefont {T.~H.}\ \bibnamefont {Taminiau}},\ and\ \bibinfo {author} {\bibfnamefont {R.}~\bibnamefont {Hanson}},\ }\href@noop {} {\bibinfo {title} {Data and simulations underlying the research article "{{Control}} of solid-state nuclear spin qubits using an electron spin-1/2"}},\ \bibinfo {howpublished} {4TU.ResearchData} (\bibinfo {year} {2024}{\natexlab{b}})\BibitemShut {NoStop}%
\bibitem [{\citenamefont {Raa}\ \emph {et~al.}(2023)\citenamefont {Raa}, \citenamefont {Ervasti}, \citenamefont {Botma}, \citenamefont {Visser}, \citenamefont {Budhrani}, \citenamefont {{van Rantwijk}}, \citenamefont {Cadot}, \citenamefont {Vermeltfoort}, \citenamefont {Pompili},\ and\ \citenamefont {Stolk}}]{raa2023qmi}%
  \BibitemOpen
  \bibfield  {author} {\bibinfo {author} {\bibfnamefont {I.~T.}\ \bibnamefont {Raa}}, \bibinfo {author} {\bibfnamefont {H.}~\bibnamefont {Ervasti}}, \bibinfo {author} {\bibfnamefont {P.}~\bibnamefont {Botma}}, \bibinfo {author} {\bibfnamefont {L.}~\bibnamefont {Visser}}, \bibinfo {author} {\bibfnamefont {R.}~\bibnamefont {Budhrani}}, \bibinfo {author} {\bibfnamefont {J.}~\bibnamefont {{van Rantwijk}}}, \bibinfo {author} {\bibfnamefont {S.}~\bibnamefont {Cadot}}, \bibinfo {author} {\bibfnamefont {J.}~\bibnamefont {Vermeltfoort}}, \bibinfo {author} {\bibfnamefont {M.}~\bibnamefont {Pompili}},\ and\ \bibinfo {author} {\bibfnamefont {A.}~\bibnamefont {Stolk}},\ }\href {https://doi.org/10.4121/6D39C6DB-2F50-4A49-AD60-5BB08F40CB52} {\bibinfo {title} {{{QMI}} - {{Quantum Measurement Infrastructure}}, a {{Python}} 3 framework for controlling laboratory equipment}},\ \bibinfo {howpublished} {4TU.ResearchData} (\bibinfo {year} {2023})\BibitemShut {NoStop}%
\bibitem [{\citenamefont {Taminiau}\ \emph {et~al.}(2014)\citenamefont {Taminiau}, \citenamefont {Cramer}, \citenamefont {{van der Sar}}, \citenamefont {Dobrovitski},\ and\ \citenamefont {Hanson}}]{taminiau2014universal}%
  \BibitemOpen
  \bibfield  {author} {\bibinfo {author} {\bibfnamefont {T.~H.}\ \bibnamefont {Taminiau}}, \bibinfo {author} {\bibfnamefont {J.}~\bibnamefont {Cramer}}, \bibinfo {author} {\bibfnamefont {T.}~\bibnamefont {{van der Sar}}}, \bibinfo {author} {\bibfnamefont {V.~V.}\ \bibnamefont {Dobrovitski}},\ and\ \bibinfo {author} {\bibfnamefont {R.}~\bibnamefont {Hanson}},\ }\bibfield  {title} {\bibinfo {title} {Universal control and error correction in multi-qubit spin registers in diamond},\ }\href {https://doi.org/10.1038/nnano.2014.2} {\bibfield  {journal} {\bibinfo  {journal} {Nature Nanotechnology}\ }\textbf {\bibinfo {volume} {9}},\ \bibinfo {pages} {171} (\bibinfo {year} {2014})}\BibitemShut {NoStop}%
\bibitem [{\citenamefont {Wang}\ \emph {et~al.}(2012)\citenamefont {Wang}, \citenamefont {{de Lange}}, \citenamefont {Rist{\`e}}, \citenamefont {Hanson},\ and\ \citenamefont {Dobrovitski}}]{wang2012comparison}%
  \BibitemOpen
  \bibfield  {author} {\bibinfo {author} {\bibfnamefont {Z.-H.}\ \bibnamefont {Wang}}, \bibinfo {author} {\bibfnamefont {G.}~\bibnamefont {{de Lange}}}, \bibinfo {author} {\bibfnamefont {D.}~\bibnamefont {Rist{\`e}}}, \bibinfo {author} {\bibfnamefont {R.}~\bibnamefont {Hanson}},\ and\ \bibinfo {author} {\bibfnamefont {V.~V.}\ \bibnamefont {Dobrovitski}},\ }\bibfield  {title} {\bibinfo {title} {Comparison of dynamical decoupling protocols for a nitrogen-vacancy center in diamond},\ }\href {https://doi.org/10.1103/PhysRevB.85.155204} {\bibfield  {journal} {\bibinfo  {journal} {Physical Review B}\ }\textbf {\bibinfo {volume} {85}},\ \bibinfo {pages} {155204} (\bibinfo {year} {2012})}\BibitemShut {NoStop}%
\bibitem [{\citenamefont {Johansson}\ \emph {et~al.}(2012)\citenamefont {Johansson}, \citenamefont {Nation},\ and\ \citenamefont {Nori}}]{johansson2012qutip}%
  \BibitemOpen
  \bibfield  {author} {\bibinfo {author} {\bibfnamefont {J.~R.}\ \bibnamefont {Johansson}}, \bibinfo {author} {\bibfnamefont {P.~D.}\ \bibnamefont {Nation}},\ and\ \bibinfo {author} {\bibfnamefont {F.}~\bibnamefont {Nori}},\ }\bibfield  {title} {\bibinfo {title} {{{QuTiP}}: {{An}} open-source {{Python}} framework for the dynamics of open quantum systems},\ }\href {https://doi.org/10.1016/j.cpc.2012.02.021} {\bibfield  {journal} {\bibinfo  {journal} {Computer Physics Communications}\ }\textbf {\bibinfo {volume} {183}},\ \bibinfo {pages} {1760} (\bibinfo {year} {2012})}\BibitemShut {NoStop}%
\bibitem [{\citenamefont {{van de Stolpe}}\ \emph {et~al.}(2024)\citenamefont {{van de Stolpe}}, \citenamefont {Kwiatkowski}, \citenamefont {Bradley}, \citenamefont {Randall}, \citenamefont {Abobeih}, \citenamefont {Breitweiser}, \citenamefont {Bassett}, \citenamefont {Markham}, \citenamefont {Twitchen},\ and\ \citenamefont {Taminiau}}]{vandestolpe2024mapping}%
  \BibitemOpen
  \bibfield  {author} {\bibinfo {author} {\bibfnamefont {G.~L.}\ \bibnamefont {{van de Stolpe}}}, \bibinfo {author} {\bibfnamefont {D.~P.}\ \bibnamefont {Kwiatkowski}}, \bibinfo {author} {\bibfnamefont {C.~E.}\ \bibnamefont {Bradley}}, \bibinfo {author} {\bibfnamefont {J.}~\bibnamefont {Randall}}, \bibinfo {author} {\bibfnamefont {M.~H.}\ \bibnamefont {Abobeih}}, \bibinfo {author} {\bibfnamefont {S.~A.}\ \bibnamefont {Breitweiser}}, \bibinfo {author} {\bibfnamefont {L.~C.}\ \bibnamefont {Bassett}}, \bibinfo {author} {\bibfnamefont {M.}~\bibnamefont {Markham}}, \bibinfo {author} {\bibfnamefont {D.~J.}\ \bibnamefont {Twitchen}},\ and\ \bibinfo {author} {\bibfnamefont {T.~H.}\ \bibnamefont {Taminiau}},\ }\bibfield  {title} {\bibinfo {title} {Mapping a 50-spin-qubit network through correlated sensing},\ }\href {https://doi.org/10.1038/s41467-024-46075-4} {\bibfield  {journal} {\bibinfo  {journal} {Nature Communications}\ }\textbf {\bibinfo {volume} {15}},\ \bibinfo {pages} {2006} (\bibinfo {year}
  {2024})}\BibitemShut {NoStop}%
\bibitem [{\citenamefont {Pompili}(2021)}]{pompili2021thesis}%
  \BibitemOpen
  \bibfield  {author} {\bibinfo {author} {\bibfnamefont {M.}~\bibnamefont {Pompili}},\ }\emph {\bibinfo {title} {Multi-{{Node Quantum Networks}} with {{Diamond Qubits}}}},\ \href {https://doi.org/10.4233/UUID:B125EC2D-E2AF-4708-BCCC-0A2357A533B1} {Ph.D. thesis},\ \bibinfo  {school} {Delft University of Technology} (\bibinfo {year} {2021})\BibitemShut {NoStop}%
\end{thebibliography}%

\end{document}